\documentclass[english, a4paper, 11pt, DIV=12, numbers=noenddot]{scrartcl}
\usepackage{blindtext}
\usepackage{amsmath}
\usepackage{amssymb}
\usepackage{bbm}
\usepackage{slashed}
\usepackage{cite}
\usepackage{setspace}
\usepackage[bottom]{footmisc}
\usepackage{hyperref}
\usepackage{catchfile}
\usepackage{graphicx}
\usepackage[labelfont=bf]{caption}
\usepackage{subcaption}
\usepackage{braket}
\usepackage{chngcntr}
\usepackage{booktabs}
\usepackage{color, colortbl, xcolor}
\usepackage{multirow}
\numberwithin{equation}{section} 
\counterwithin{figure}{section} 
\counterwithin{table}{section} 
\hypersetup{linktocpage,
    colorlinks,
    linkcolor={red!70!black},
    citecolor={green!50!blue},
    urlcolor={blue!50!black}
}
\newcommand*{\GtrSim}{\smallrel\gtrsim}
\newcommand*{\LessSim}{\smallrel\lesssim}

\makeatletter
\newcommand*{\smallrel}[2][.8]{%
  \mathrel{\mathpalette{\smallrel@{#1}}{#2}}%
}
\newcommand*{\smallrel@}[3]{%
  \sbox0{$#2\vcenter{}$}%
  \dimen@=\ht0 %
  \raise\dimen@\hbox{%
    \scalebox{#1}{%
      \raise-\dimen@\hbox{$#2#3\m@th$}%
    }%
  }%
}
\makeatother

\def\EmissT{\,{}/ \hspace{-1.5ex}  E_{T}}
\begin{document}
\begin{titlepage}
\begin{flushright}
TTP21-031\\
P3H-21-065
\end{flushright}
\vskip2cm
\begin{center}
{\LARGE \bfseries Tasting Flavoured Majorana Dark Matter}
\vskip1.0cm
{\large Harun Acaro\u{g}lu$^a$, Monika Blanke$^{a,b}$}
\vskip0.5cm
 \textit{$^a$Institut f\"ur Theoretische Teilchenphysik,
  Karlsruhe Institute of Technology, \\
Engesserstra\ss e 7,
  D-76128 Karlsruhe, Germany}
 \vspace{3mm}\\
  \textit{$^b$Institut f\"ur Astroteilchenphysik, Karlsruhe Institute of Technology,\\
  Hermann-von-Helmholtz-Platz 1,
  D-76344 Eggenstein-Leopoldshafen, Germany}

\vskip1cm


\vskip1cm

{\large \bfseries Abstract\\[10pt]} \parbox[t]{.9\textwidth}{
We study a simplified model of flavoured Majorana dark matter in the Dark Minimal Flavour Violation framework. The model extends the Standard Model by a dark matter flavour triplet and a scalar mediator, through which the new dark fermions couple to right-handed up-type quarks. This interaction is governed by a new coupling matrix $\lambda$ which is assumed to constitute the only new source of flavour and CP violation. We analyse the parameter space of this model by using constraints from collider searches, $D^0-\bar{D}^0$ mixing, cosmology and direct dark matter searches. Throughout our study, we point out crucial differences between the Majorana and Dirac dark matter cases. After performing a combined analysis within the context of all the experimental constraints mentioned above, we analyse which flavour for the dark matter particle is preferred by experimental data. We further investigate if this model is capable of explaining the large measured value of the direct CP asymmetry $\Delta A^\text{dir}_{CP}$ in charm decays. We find that significant enhancements with respect to the Standard Model expectation are compatible with all constraints, and even the central value of the measurement can be reached.  We also advertise the flavour-violating final state with two same-sign top quarks produced in association with missing transverse energy as a smoking-gun signature for flavoured Majorana dark matter at the LHC. 
}

\end{center}
\end{titlepage}

\tableofcontents 
\newpage

\section{Introduction}
\label{sec:intro}

Despite the overwhelming astrophysical and cosmological evidence for the existence of dark matter (DM), its particle nature remains obscure. With the non-observation of DM particles in direct detection experiments and at the LHC, the WIMP paradigm -- i.\,e.\ a weakly-coupled DM particle with weak-scale mass -- has been put under severe pressure \cite{Arcadi:2017kky}. As a consequence, alternatives to the simple WIMP scenario  have become increasingly popular.

One possibility to reconcile the WIMP paradigm with experimental data is the introduction of an extended dark sector in which the DM candidate carries charge under a flavour symmetry \cite{Kile:2011mn,Kamenik:2011nb,Batell:2011tc,Agrawal:2011ze,
Batell:2013zwa,Kile:2013ola,Lopez-Honorez:2013wla,Kumar:2013hfa,
Zhang:2012da}. While the first studies of such flavoured DM models were confined to the Minimal Flavour Violation (MFV) \cite{mfv3,mfv4,mfv5,mfv1,mfv2} hypothesis, more recently the concept of flavoured DM with a non-trivial flavour structure has been put forward \cite{dmfv,tfdm,Jubb:2017rhm,ldm,Chen:2015jkt}. In this case the coupling between the DM flavour multiplet and the Standard Model (SM) fermions constitutes a new source of flavour and CP violation, so that the phenomenology vastly changes with respect to the MFV-type models.

In order to systematically study the phenomenological implications of such a non-MFV coupling between DM and the SM, the concept of Dark Minimal Flavour Violation (DMFV) has been introduced \cite{dmfv}. Conceptually similar to the MFV hypothesis, in DMFV models the coupling between the DM flavour triplet and the SM quark or lepton flavour triplet is assumed to be the only new source of flavour and CP violation beyond the SM Yukawa couplings. The assumption of DMFV has the advantage of keeping the number of parameters in the simplified models minimal and thus allowing for an efficient  study of their rich phenomenology. 

Previous analyses of simplified models with DMFV introduced DM as a Dirac fermion transforming as a triplet under a new $U(3)_\chi$ flavour symmetry, either coupling to the SM down-type quarks \cite{dmfv}, up-type quarks \cite{tfdm,Jubb:2017rhm}, left-handed quark doublets \cite{ldm} or charged leptons \cite{Chen:2015jkt} via a scalar mediator.  
In the present paper we instead explore the possibility to introduce the DM flavour triplet as a Majorana fermion. Since the Majorana nature requires the DM to transform under a real representation, we take the flavour symmetry in the dark sector to be $O(3)_\chi$, instead of $U(3)_\chi$ in the Dirac case. 
 
As a benchmark case, we consider the case of flavoured Majorana DM in which the dark flavour triplet $\chi$ couples to right-handed up-type quarks, in analogy to the Dirac DM model studied in \cite{tfdm,Jubb:2017rhm}.
To allow for a straightforward comparison of the phenomenology of the two models, let us briefly recall the main results of \cite{tfdm}:
\begin{itemize}
\item
The stringent constraints from $D^0-\bar D^0$ mixing required a strong suppression of flavour changing neutral current (FCNC) processes, thereby forbidding interesting non-standard effects in rare and CP-violating $D$ meson decays.
\item
The LHC experiments ATLAS and CMS put lower bounds on the mass of the scalar mediator, depending on the DM mass and the coupling strength to SM quarks. The strongest constraints arise from searches for final states with jets plus missing transverse energy $\EmissT$.
\item
Similarly to flavour-violating supersymmetric models \cite{fvsquark2,Chakraborty:2018rpn}, the model predicts the new benchmark signatures $tj+\EmissT$ and monotop$+\EmissT$ at a significant rate. A dedicated experimental search has been proposed in \cite{Blanke:2020bsf}.
\item
The limits from direct detection experiments severely constrain the allowed ranges of DM--quark couplings, requiring a destructive interference between the various tree- and loop-level contributions to the DM--nucleon scattering cross section. In combination with the constraints from the thermal freeze-out condition, the direct detection limits translate into a lower bound on the DM mass. 
\end{itemize}

Analysing the  constraints from collider searches, flavour physics, cosmology and direct detection experiments, we shall see that the phenomenology changes drastically when, instead of a Dirac fermion, DM is introduced as a Majorana fermion.

\section{Flavoured Majorana Dark Matter in DMFV}
\label{sec::theory}
The subject of this paper is a simplified model based on the DMFV hypothesis \cite{dmfv}. This section briefly introduces the DMFV framework in general and then presents the explicit DMFV model to be studied.

	\subsection{A Simplified DMFV Model}
	\label{subsec::dmfv}
	In the DMFV framework the SM is extended by a new flavour symmetry, generically denoted as {$\mathcal{G}(3)_\chi$}, and a DM field $\chi$, that transforms under the fundamental representation of this new symmetry. This results in a global 
	\begin{equation}
	\mathcal{G}_\text{DMFV} = 
	U(3)_q \times U(3)_u \times U(3)_d \times \mathcal{G}(3)_\chi\,,
	\label{eq::Gdmfv}
	\end{equation}
	flavour symmetry, where we focus on DM interacting with SM quarks. Depending on whether the field $\chi$ is a Dirac or Majorana fermion the symmetry group $\mathcal{G}(3)_\chi$ is either a $U(3)_\chi$ or an $O(3)_\chi$ group, respectively. In DMFV the SM Yukawa couplings $Y_u$, $Y_d$, and $\lambda$ -- the  coupling of $\chi$ to quarks -- constitute the only sources that break the flavour symmetry $\mathcal{G}_\text{DMFV}$. In this sense, the DMFV framework goes beyond the scope of MFV \cite{mfv1,mfv2,mfv3,mfv4,mfv5}, as it extends the SM flavour symmetry by  $\mathcal{G}(3)_\chi$ and includes one new source of flavour and CP violation, namely $\lambda$. DMFV models can be classified by the type of flavour to which the DM couples  and the fermion nature of the DM particle. The cases of $\chi$ being a Dirac fermion and coupling to right-handed up quarks $u_R$ \cite{tfdm,Jubb:2017rhm}, right-handed down quarks $d_R$ \cite{dmfv}, left-handed quark doublets $q_L$ \cite{ldm} and right-handed charged leptons \cite{Chen:2015jkt} have already been studied. \par
	In this paper we present a simplified model that constitutes the first realization of the DMFV ansatz where the new DM field $\chi$ is assumed to be a Majorana fermion. In analogy to the model analysed in \cite{tfdm}, it is assumed that $\chi$ couples to right-handed up quarks through the exchange of a scalar boson $\phi$. The Lagrangian of this model reads: 
	\begin{align}
	\nonumber
	\mathcal{L}=\,&\mathcal{L}_\text{SM} + \frac{1}{2} \left(i\bar{\chi}\slashed{\partial}\chi - M_\chi \bar{\chi}\chi\right) - \left(\lambda_{ij}\,\bar{u}_{Ri}\chi_j\,\phi + \mathrm{h.c.}\right)+(D_\mu\phi)^\dagger (D^\mu \phi) \\
	& - m_\phi^2\, \phi^\dagger \phi + \lambda_{H\phi}\, \phi^\dagger \phi\, H^\dagger H + \lambda_{\phi\phi} \left(\phi^\dagger \phi\right)^2.
	\label{eq::lagrangian}
	\end{align}
	Here we have introduced $\chi$ as a four-component Majorana spinor $\chi = (\chi_L, i\sigma_2 \chi_L^*)^T$, with $\chi_L$ being a two-component Weyl spinor.
	Note that due to its Majorana nature the kinetic term and the mass term of $\chi$ include a factor of $1/2$. The field $\chi$ transforms as a singlet under the SM gauge group and as a triplet under a global $O(3)_\chi$ symmetry, i.e.\@ it comes in three generations. Its lightest generation is assumed to constitute the observed DM in the universe. Interactions between SM quarks and this new DM field $\chi$ are parametrised by the coupling $\lambda$, a general complex $3 \times 3$ matrix. These interactions are mediated by the scalar boson $\phi$ which carries hypercharge and colour. 
	
	In passing we note that the coupling $\lambda_{H\phi}$ contributes to the loop-induced effective couplings of the SM Higgs to gluons and photons. However these contributions are suppressed by the square of the mediator mass, $m_\phi^2$, and hence expected to be small for the mediator masses allowed by LHC searches, see Section \ref{sec::collider}. Furthermore, the coupling $\lambda_{H\phi}$ contributes to the $\bar\chi_3\chi_3 h$ coupling at the one-loop level, potentially relevant for DM direct detection experiments. We will return to this topic in Section \ref{sec:dd}. However let us stress already here that the main goal of our analysis is to shed light on the possible flavour structure of Majorana DM, governed by the coupling matrix $\lambda$. We hence neglect the impact of the couplings $\lambda_{H\phi}$ and $\lambda_{\phi\phi}$ in the rest of our paper.

	\subsection[Parametrisation of the Dark-Matter--Quark Coupling \texorpdfstring{$\lambda$}{Lambda}]{Parametrisation of the Dark-Matter--Quark Coupling \texorpdfstring{\boldmath $\lambda$\unboldmath}{Lambda}}
	
	\label{subsec::parametrization}
	In order to analyse the constraints on our model we need to first parametrise the physical degrees of freedom in the flavour sector. For the Yukawa couplings $Y_u$ and $Y_d$ one can proceed as usual and use the SM flavour symmetry in order to remove unphysical parameters to finally end up with six quark masses and the CKM matrix.\par 
	A similar procedure has to be performed for the new coupling $\lambda$. According to the DMFV ansatz $\lambda$ is an arbitrary complex $3 \times 3$ matrix. Therefore it contains 18 parameters, consisting of 9 real parameters and 9 complex phases. In the following we want to make use of the flavour symmetry $\mathcal{G}_\text{DMFV}$ and the symmetry $O(3)_\chi$ in particular in order to remove unphysical degrees of freedom from $\lambda$.\par 
	As a first step $\lambda$ is expressed through a singular value decomposition, yielding 
	\begin{equation}
	\lambda = U_\lambda D_\lambda V_\lambda\,.
	\label{eq::svd}
	\end{equation}
	Here, $U_\lambda$ and $V_\lambda$ are unitary matrices and $D_\lambda$ is a diagonal matrix with positive and real entries. 
Taking into account that 	eq.\ \eqref{eq::svd} is invariant under the diagonal rephasing
\begin{align}
	\nonumber
	U_\lambda^\prime=\,&U_\lambda\, \text{diag}(e^{i\alpha_1},e^{i\alpha_2},e^{i\alpha_3})\,, \\
	V_\lambda^\prime=\,&\text{diag}(e^{-i\alpha_1},e^{-i\alpha_2},e^{-i\alpha_3})\,V_\lambda\,, \label{eq::rephasing}
	\end{align}	
	this parametrization indeed contains 9 real parameters and 9 complex phases.

	One can now use the flavour symmetry $\mathcal{G}_\text{DMFV}$ in order to remove more unphysical degrees of freedom from $\lambda$. Note that for the case of $\chi$ being a Dirac fermion $\mathcal{G}_\text{DMFV}$ contains a $U(3)_\chi$ symmetry. In such models the flavour symmetry of $\chi$ can thus be used to fully remove the unitary matrix $V_\lambda$ and end up with $\lambda = U_\lambda D_\lambda$ \cite{dmfv,tfdm,ldm}.  
	For our case of a Majorana fermion $\chi$ the Lagrangian is no longer invariant under such $U(3)_\chi$ transformations, since Majorana fields can only transform under real representations. Indeed the bilinear
	\begin{equation}
	\bar{\chi}\chi = i(\chi^\dagger_L \sigma_2 \chi_L^*-\chi_L^T \sigma_2 \chi_L)\,.
	\label{eq::majoranamass}
	\end{equation}
	  is only invariant under orthogonal transformations. Thus, one is left with an $O(3)_\chi$ symmetry that can be used to remove 3 real degrees of freedom from $V_\lambda$.
	  
	   According to \cite{decomp} the unitary matrix $V_\lambda$ can be decomposed as
	\begin{equation}
	V_\lambda = O_V\, d_V\, P_V\,.
	\end{equation}
	Here, $O_V$ and $P_V$ are orthogonal matrices with 3 real degrees of freedom and $d_V$ is a diagonal unitary matrix with 3 complex phases. The orthogonal matrix $P_V$ can now be removed from $V_\lambda$ by using the transformation 
	\begin{equation}
	\chi \rightarrow P_V^{-1} \chi\,.
	\end{equation}
	This finally yields 
	\begin{equation}
	\lambda = U\, D\, O\, d\,\,,
	\label{eq::lambdaparam}
	\end{equation}
	where we have omitted the indices for brevity of notation.\par
	We adopt the parametrisation for $U$ from \cite{tfdm,Blanke:2006xr}, which reads
	\begin{eqnarray}
	U &=& U_{23}\, U_{13}\, U_{12}\, \nonumber\\
	&=&
 	\begin{pmatrix}
  	1 & 0 & 0\\
 	 0 & c_{23}^\theta & s_{23}^\theta e^{- i\delta_{23}}\\
 	 0 & -s_{23}^\theta e^{i\delta_{23}} & c_{23}^\theta\\
	\end{pmatrix}
	\begin{pmatrix}
 	 c_{13}^\theta & 0 & s_{13}^\theta e^{- i\delta_{13}}\\
 	 0 & 1 & 0\\
 	 -s_{13}^\theta e^{ i\delta_{13}} & 0 & c_{13}^\theta\\
	\end{pmatrix}
	\begin{pmatrix}
 	 c_{12}^\theta & s_{12}^\theta e^{- i\delta_{12}} & 0\\
 	 -s_{12}^\theta e^{i\delta_{12}} & c_{12}^\theta & 0\\
 	 0 & 0 & 1\\
	\end{pmatrix},\nonumber\\
	&&
	\end{eqnarray}
	where we have used the shorthand notation $c_{ij}^\theta = \cos\theta_{ij}$ and $s_{ij}^\theta = \sin\theta_{ij}$. Note that in this parametrization the rephasing of eq.\ \eqref{eq::rephasing} has already been used to remove three complex phases from $U$. The matrix $O$ is parametrised as
		\begin{eqnarray}
	O &=& O_{23}\, O_{13}\, O_{12} \nonumber\\
	&=&
 	\begin{pmatrix}
  	1 & 0 & 0\\
 	 0 & c_{23}^\phi & s_{23}^\phi\\
 	 0 & -s_{23}^\phi & c_{23}^\phi\\
	\end{pmatrix}
	\begin{pmatrix}
 	 c_{13}^\phi & 0 & s_{13}^\phi\\
 	 0 & 1 & 0\\
 	 -s_{13}^\phi & 0 & c_{13}^\phi\\
	\end{pmatrix}
	\begin{pmatrix}
 	 c_{12}^\phi & s_{12}^\phi & 0\\
 	 -s_{12}^\phi & c_{12}^\phi & 0\\
 	 0 & 0 & 1\\
	\end{pmatrix},\qquad\label{eq::O}
	\end{eqnarray}
	with $c_{ij}^\phi = \cos\phi_{ij}$ and $s_{ij}^\phi = \sin\phi_{ij}$, and the diagonal matrices $D$ and $d$ are parametrised as 
	\begin{equation}\label{eq::Dd}
	D=\text{diag}(D_1,D_2,D_3)\,,\qquad \text{and} \qquad  d=\text{diag}(e^{i\gamma_1},e^{i\gamma_2},e^{i\gamma_3})\,.
	\end{equation}
	 In total the coupling matrix $\lambda$ then has the following 15 physical  parameters:
	\begin{equation}
	\theta_{23},\,\theta_{13},\,\theta_{12},\,\phi_{23},\,\phi_{13},\,\phi_{12},\,\delta_{23},\,\delta_{13},\,\delta_{12},\,\gamma_1,\,\gamma_2,\,\gamma_3,\,D_1,\,D_2,\,D_3\,. 
	\end{equation}
	In order to avoid a double-counting of the parameter space in our numerical analysis, we restrict the parameters of the coupling matrix $\lambda$ to the following ranges:
	\begin{equation}
	\theta_{ij} \in [0,\frac{\pi}{4}], \quad \phi_{ij} \in [0,\frac{\pi}{4}], \quad \delta_{ij} \in [0,2\pi), \quad \gamma_i \in [0,2\pi), \quad D_i >0\,.
	\end{equation}\par
	

	\subsection{Mass Spectrum and Dark Matter Stability}
	\label{subsec::mspl}

Following the DMFV assumption the mass term $M_\chi$ cannot be a generic $3\times 3$ matrix, as this would constitute an additional source of flavour violation. Instead, similarly to the MFV spurion expansion \cite{mfv4}, we can expand the DM mass matrix in powers of the flavour violating coupling $\lambda$:
\begin{equation}
	M_{\chi,ij}=m_\chi\,\left\{\mathbbm{1}+\frac{\eta}{2}\, (\lambda^\dagger\lambda +\lambda^T\lambda^*)+ \mathcal{O}(\lambda^4)\right\}_{ij}\,.
	\label{eq::msplnew}
	\end{equation}
	The expansion parameter $\eta$ is an additional parameter of the simplified model that accounts for our ignorance of the UV completion of the theory. In order to ensure that the mass corrections always reduce the DM mass with respect to the leading-order mass parameter $m_\chi$, we choose $\eta < 0$.
	
	The expression in eq.\ \eqref{eq::msplnew} differs from the one employed in the case of Dirac DM \cite{tfdm} -- the second summand in the round brackets is required in order to render the mass matrix symmetric, as required for Majorana fermions. 	
	Inserting our parametrization for $\lambda$ we find
	\begin{equation}
	M_{\chi,ij}=m_\chi\,\left\{\mathbbm{1}-\frac{|\eta|}{2}\, (d^*\,O^T\,D^2\,O\,d+ d\,O^T\,D^2\,O\,d^*)+ \mathcal{O}(\lambda^4)\right\}_{ij}\,.
	\label{eq::massspl}
	\end{equation}
	 Note that in contrast to the Dirac fermion case the mass matrix $M_\chi$ is not diagonal per se, i.e.\@ there is a misalignment between the flavour and the mass eigenstates for $\chi$. The diagonalization of $M_\chi$ can be achieved through an Autonne-Takagi factorization \cite{autonne,takagi}, where 
	 \begin{equation}
	 M_\chi = W^T M^D_\chi W\,.
	 \label{eq::takagi}
	 \end{equation}
	 Here, $M^D_\chi$ is a diagonal matrix with real positive entries and $W$ is an orthogonal matrix, since the mass matrix $M_\chi$ is real. The necessary field redefinition  $\chi_L \rightarrow W \chi_L$ then transforms the coupling of the DM field $\chi$ to the SM quarks $u_R$ to
	 \begin{equation}\label{eq::lambda-tilde}
	 \tilde{\lambda} = \lambda W^T\,.
	 \end{equation}
	 We further arrange the rows of $W$ in such a way that we always have 
	\begin{equation} 
	M_\chi^D=\text{diag}(m_{\chi_1},m_{\chi_2},m_{\chi_3})\,,
	\end{equation} 
	with $m_{\chi_1}>m_{\chi_2}>m_{\chi_3}$, i.\,e.\ the third dark generation is the lightest state, and we assume it to form the DM of the universe. Due to the complexity of the mass corrections it is not possible to provide an analytical expression for $W$ and thus, calculating analytical conditions that the model parameters have to fulfil in order to generate a particular mass splitting, as was done in \cite{tfdm}, is not possible. 
	
			In DMFV models with Dirac fermions DM stability is guaranteed by an unbroken $\mathbb{Z}_3$ symmetry which follows from the flavour symmetry \cite{dmfv, tfdm, ldm}\footnote{An analogous residual symmetry had previously been found in the case of DM models with MFV \cite{Batell:2011tc}.}. However, for Majorana DM this symmetry cannot be present due to its non-trivial representation being complex. Hence, we impose a $\mathbb{Z}_2$ symmetry under which only the new particles $\chi$ and $\phi$ are charged in order to forbid the decay of $\chi$ and $\phi$ into final states with SM particles only. The lightest flavour of the DM triplet $\chi$ is then rendered stable, as long as its mass is smaller than the mass of the coloured scalar boson $\phi$. Thus, we additionally choose 
	\begin{equation}
	m_\chi < m_\phi\,.
	\label{eq::mchismallermphi}
\end{equation}

\section{Collider Phenomenology}
\label{sec::collider}

Searches for new particles at the LHC lead to stringent constraints on the model presented above. We discuss these constraints in this section to determine the experimentally excluded regions in the parameter space of our model.

It was shown in \cite{monojet} that for models with a coloured $t$-channel mediator the most stringent limits arise in general from the pair-production of the mediator, subsequently decaying into quarks and missing transverse energy ($\slashed{E}_T$), as opposed to searches for monojet$+\slashed{E}_T$. The authors of \cite{dmfv} found very similar results for a DMFV model where $\chi$ is a Dirac fermion and couples to right-handed down-type quarks. We expect that this behaviour also applies to our model and therefore focus on $\phi$ pair production. 

The relevant final states constrained by the LHC experiments are those of searches for supersymmetric top squarks ($t\bar t+\slashed{E}_T$)  and squarks of the first two generations (jets$+\slashed{E}_T$).  Note that the limits used in \cite{tfdm} to constrain the Dirac version of top-flavoured DM were based on LHC run 1 data with $\sqrt{s}=8\,\mathrm{TeV}$. Updated bounds resulting from a recast of 136$\,\text{fb}^{-1}$ of LHC run 2 data at $\sqrt{s}=13\,\mathrm{TeV}$ were presented in \cite{Blanke:2020bsf}.
In this study we will use the same run 2 limits as in \cite{Blanke:2020bsf}  to constrain our model of flavoured Majorana DM.

	\subsection{LHC Signatures from Mediator Pair-Production}
	\label{subsec::lhcsign}
	
	Just like squarks in SUSY, the scalar boson $\phi$ is a colour triplet that is odd under a $\mathbb{Z}_2$ symmetry and is therefore produced in pairs through QCD interactions. Note that this production channel neither depends on the DM mass $m_\chi$ nor on the coupling $\lambda$.
	
\begin{figure}[b!]
	\centering
		\begin{subfigure}[t]{0.32\textwidth}
		\includegraphics[width=\textwidth]{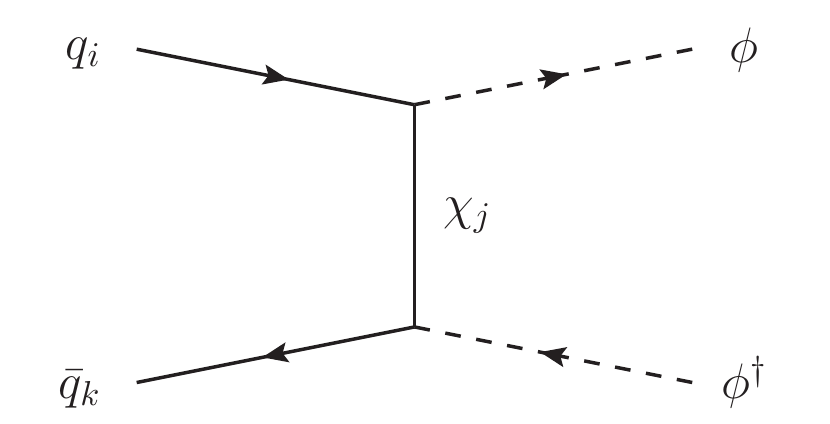}
		\caption{$\phi\phi^\dagger$ production}
		\end{subfigure}
		\hfill
		\begin{subfigure}[t]{0.32\textwidth}
		\includegraphics[width=\textwidth]{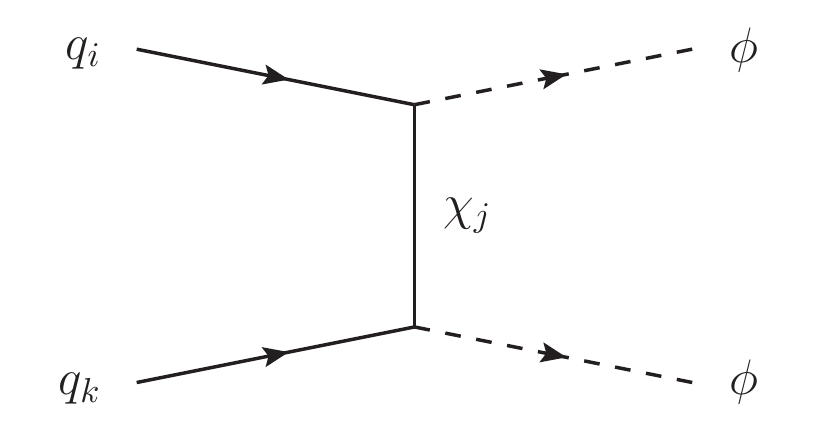}
		\caption{$\phi\phi$ production}
		\end{subfigure}
		\hfill
		\begin{subfigure}[t]{0.32\textwidth}
		\includegraphics[width=\textwidth]{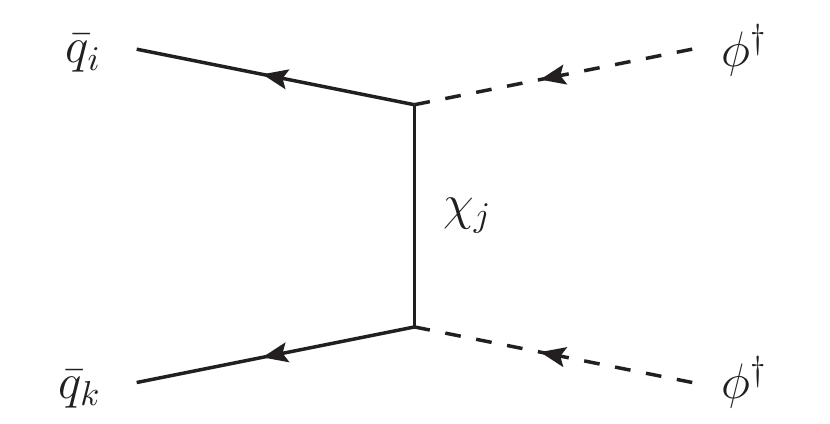}
		\caption{$\phi^\dagger\phi^\dagger$ production}
		\end{subfigure}
	\caption{Feynman diagrams of the $t$-channel $\chi$ exchange production modes of $\phi$.}
	\label{fig::phiproduction}
	\end{figure}
	 However, $\phi$ can also be pair-produced through processes involving the $t$-channel exchange of $\chi$, as shown in Figure \ref{fig::phiproduction}. These diagrams are proportional to $\lambda^2$. As $\chi$ is a Majorana fermion the production modes for $\phi$ also include same-sign production channels, i.e.\@ the mediator can be produced as $\phi\phi^\dagger, \phi\phi$ and $\phi^\dagger\phi^\dagger$ pairs. 
	Note that the production of a $\phi\phi$ pair is enhanced by the up quark parton distribution function (PDF) \cite{pdf}, due to the pair of valence up quarks in the proton.
	 As it is the mass term of eq.\@ \eqref{eq::majoranamass} that mixes $\chi^\dagger_L$ and $\chi_L$, these production channels are additionally proportional to the mass parameter $m_\chi$ of the Lagrangian. 
	 
\begin{figure}[t!]
	\centering
		\includegraphics[width=.35\textwidth]{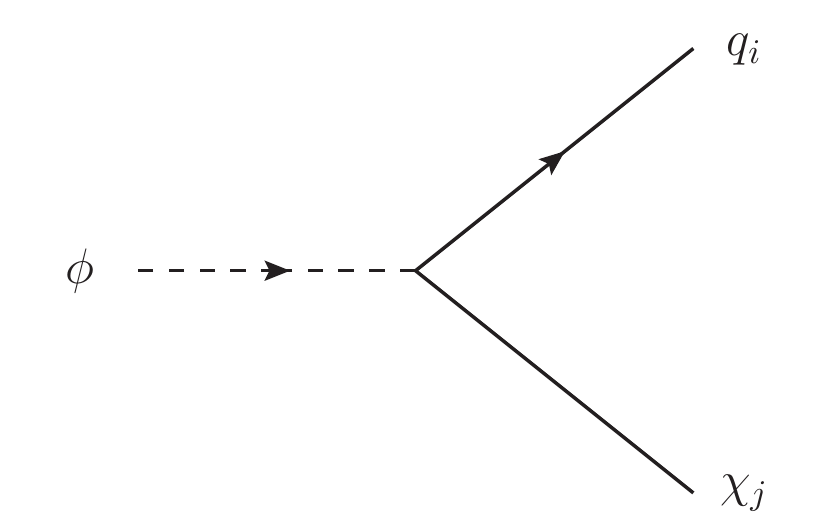}
	\caption{Feynman diagram of the mediator decay.}
	\label{fig::decaychannel}
	\end{figure}

	Due to the $\mathbb{Z}_2$ symmetry introduced in section \ref{subsec::mspl} 	
	 the scalar boson $\phi$ can only decay into final states that include $\chi$. 
	 The by far dominant decay modes are then the ones into a single quark $q_i$ accompanied by $\chi_j$, where $i,j$ are flavour indices, as depicted in Figure \ref{fig::decaychannel}.

	Combining the pair-production and decay of the mediator boson $\phi$, we find the following parton level processes that are relevant for the LHC analysis:
	\begin{equation}
	\begin{array}{ccccc}
	pp &\rightarrow & \phi\,\phi^\dagger &\rightarrow& \chi_i\, \chi_j\, q_k\, \bar{q}_l\,,\\
	pp &\rightarrow & \phi\,\phi &\rightarrow& \chi_i\, \chi_j\, q_k\, q_l\,,\\
	pp &\rightarrow & \phi^\dagger\,\phi^\dagger &\rightarrow& \chi_i\, \chi_j\, \bar{q}_k\, \bar{q}_l\,.
	\end{array}
	\label{eq::decaysigns}
	\end{equation}
	Here, $i,j,k$ and $l$ are flavour indices. The explicit constellation of the final states at the LHC depends on the SM flavour indices $k$ and $l$. An (anti)quark of the first or second generation leads to a jet in the final state which is very difficult to distinguish in flavour. Top and antitop quarks, however, can be experimentally distinguished in their semileptonic decay channels, by measuring the charge of the final-state lepton. 
The dark flavours $\chi_i$, on the other hand, are indistinguishable at the LHC, as they appear as missing transverse energy $\slashed{E}_T$ -- due to the small mass splitting possible decay products of the heavy flavours are too soft to be detected.
	
 The relevant final state signatures for the processes discussed above are then $t\bar{t}+\slashed{E}_T$, $tt+\slashed{E}_T$, $\bar{t}\bar{t}+\slashed{E}_T$, and $jj+\slashed{E}_T$. The former three are not distinguished experimentally if hadronically decaying top quarks are considered. In that case we use the shorthand notation $\text{tops}+\slashed{E}_T$ for the sum of these three processes.
 	The signatures $\text{tops}+\slashed{E}_T$ and $jj+\slashed{E}_T$ are the same as in SUSY searches for a pair of squarks decaying into SM quarks and a neutralino.  As the spin-statistics in the SUSY case and in our model are the same, the final-state kinematics is the same as far as the QCD production channel is concerned. The $t$-channel production process is expected to yield significantly larger contributions than in the case of SUSY neutralino or gluino $t$-channel exchange, as the latter are suppressed by the smallness of the neutralino couplings and the experimental constraints on the gluino mass, $m_{\tilde g} \gg 1\,\text{TeV}$.\footnote{Typically the gluino contribution is not included in the simplified models on which the LHC squark searches are based.} Hence, the final-state kinematics in our model could in principle differ from the SUSY case, however we do not expect a relevant impact on our results. Thus, we can straightforwardly transfer the limits obtained by such SUSY searches to our case.

  It is also possible that the two intermediate $\phi$ bosons decay into two $\chi$ particles, one top quark and one light quark, generating the signature $tj+\slashed{E}_T$, which is also generated in SUSY models with flavour violating squark decays \cite{fvsquark1,fvsquark2,fvsquark3,fvsquark4,fvsquark5,fvsquark6}. While no dedicated searches for this signature currently exist, it has been shown in \cite{Blanke:2020bsf} to have significant discovery potential in the case of Dirac DM. We expect that similar conclusions also hold  in our model, but leave a detailed analysis for future work.

Lastly,  the enhanced same-sign signature $tt+\slashed{E}_T$ is a unique feature of Majorana DM models which, to the best of our knowledge, has not received attention in the literature. We return to this signature in Section \ref{subsec::tt}.

	\subsection{Recast of LHC Constraints}
	\label{subsec::lhcresults}
	
	For the analysis of constraints that LHC searches for SUSY squarks pose on our model, we  use the bounds obtained by the CMS collaboration \cite{lhclimits} using the full run 2 data set with an integrated luminosity of $136\, \mathrm{fb}^{-1}$. Their analysis places limits on coloured SUSY particles from final states with multiple hadronic jets associated by $\slashed{E}_T$. In particular, the final states of interest $jj+\slashed{E}_T$ and $\text{tops}+\slashed{E}_T$ are directly addressed in that analysis, and cross section limits are provided in tabular form. Note that the results of \cite{lhclimits} are not distinctive with respect to the charge of the final state top quarks.

	In order to recast the experimental limits we calculate the leading order (LO) signal cross section using \texttt{MadGraph 5} \cite{madgraph}. The simplified model is implemented in \texttt{FeynRules} \cite{feynrules} using the Lagrangian in eq.\@ \eqref{eq::lagrangian}. Note that for simplicity we assume a degenerate mass spectrum for the DM triplet $\chi$ and neglect the small mass splitting that was discussed in Section \ref{subsec::mspl}. This approximation is justified as the mass splitting would only lead to additional soft decay products that are difficult to detect and therefore do not qualitatively influence the results of this analysis \cite{tfdm}. As we are primarily interested in the constraints that the LHC searches impose on the mass parameters $m_\phi$ and $m_\chi$, we set the mixing angles and phases in the coupling matrix $\lambda$ to zero. Non-zero mixing angles tend to reduce the branching ratio of a given flavour-conserving final state in comparison to the case with vanishing mixing angles, and therefore lead to a smaller signal cross section. We also assume $D_1=D_2$ for simplicity.
	
\begin{figure}
		\centering
		\begin{subfigure}[t]{0.49\textwidth}
		\includegraphics[width=\textwidth]{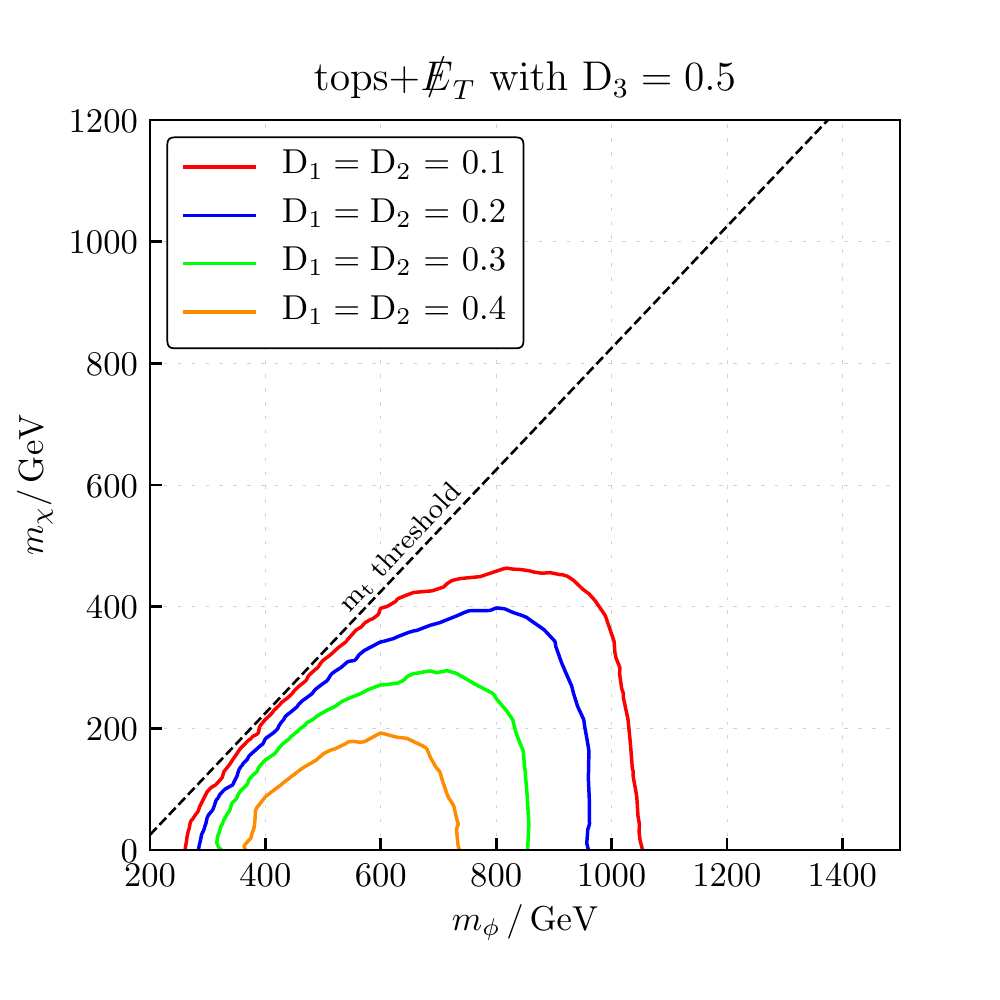}
		\caption{$D_3=0.5$ and varying $D_1=D_2$}
		\label{fig::LHCttMETa}
		\end{subfigure}
		\hfill
		\begin{subfigure}[t]{0.49\textwidth}
		\includegraphics[width=\textwidth]{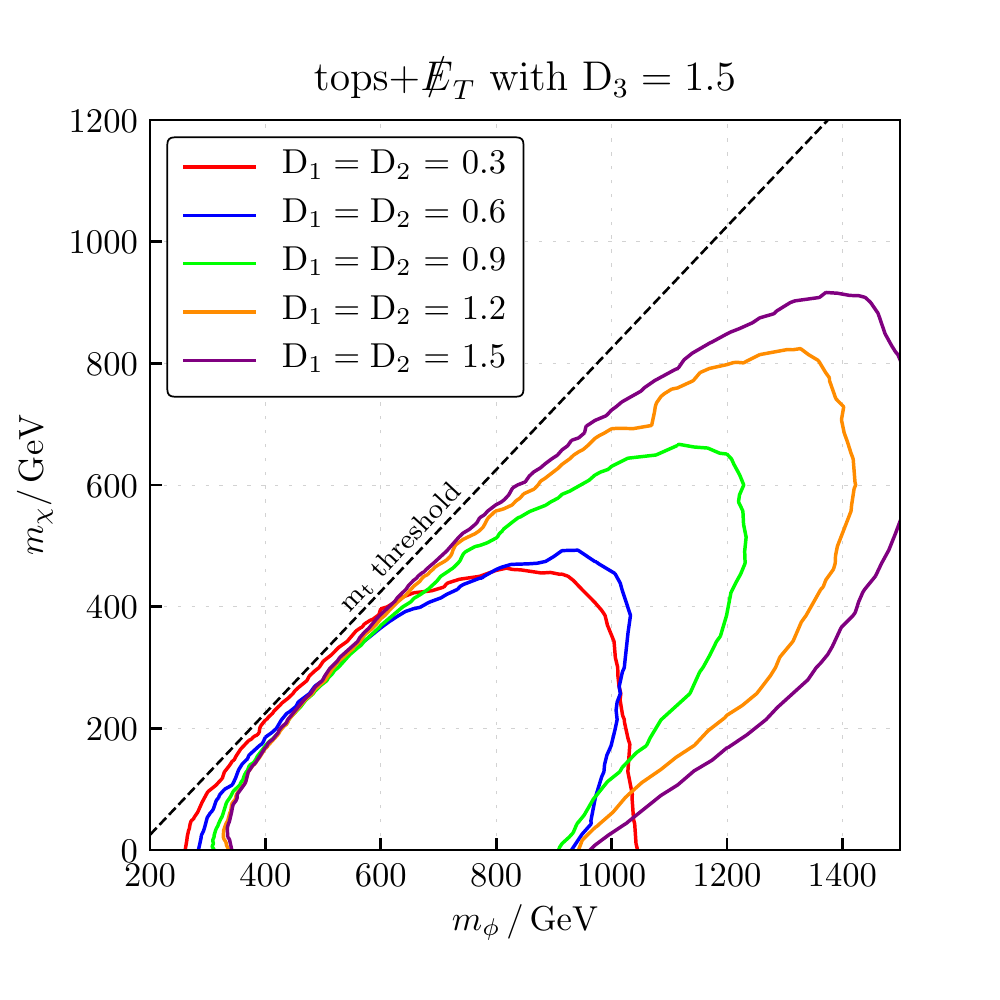}
		\caption{$D_3=1.5$ and varying $D_1=D_2$}
		\label{fig::LHCttMETb}
		\end{subfigure}
	\caption{Constraints on the final states $t\bar{t}+\slashed{E}_T$, $tt+\slashed{E}_T$ and $\bar{t}\bar{t}+\slashed{E}_T$ obtained from \cite{lhclimits}. The area under the curve is excluded.}
	\label{fig::LHCttMET}
	\end{figure}

	In Figure \ref{fig::LHCttMET} we show the exclusion contours resulting from the final state $\text{tops}+\slashed{E}_T$ in the $m_\phi-m_\chi$ plane, obtained by recasting the cross section limits of \cite{lhclimits}. The value of $D_3$ is fixed whereas $D_1=D_2$ is varied. In Figure \ref{fig::LHCttMETa} the excluded region shrinks with growing couplings $D_1=D_2$. This is due to the fact that increasing the couplings to up and charm quarks reduces the branching ratio of the mediator $\phi$ decaying into final states with top flavour. 
	
	At the same time a growing coupling $D_1$ enhances the $t$-channel production process. In Figure \ref{fig::LHCttMETb} one can see that due to this reason for couplings $D_1 > 0.5$ the excluded area grows for increasing values of $D_1=D_2$. Note that in this case contributions of the $t$-channel production of $\phi$ proportional to $\lambda^2$ grow larger than the QCD contributions to the overall cross section. In particular, a large coupling $D_1$ also enhances the production of the same sign final state $tt+\slashed{E}_T$, enhanced by the PDFs of two up quarks in the initial state. 
	This can be seen explicitly in Figure \ref{fig::LHCttMETb}. The excluded region grows quickly for increasing $D_1$ and non-vanishing DM mass $m_\chi$. As explained above, this dependence on $m_\chi$ originates from the Majorana nature of $\chi$ necessary for this contribution. Thus, even for the maximally constraining case of $D_1=D_2=D_3=1.5$ regions with a small $m_\chi$ and $m_\phi \gtrsim 1\,\mathrm{TeV}$ are not excluded.

	\begin{figure}
		\centering
		\begin{subfigure}[t]{0.49\textwidth}
		\includegraphics[width=\textwidth]{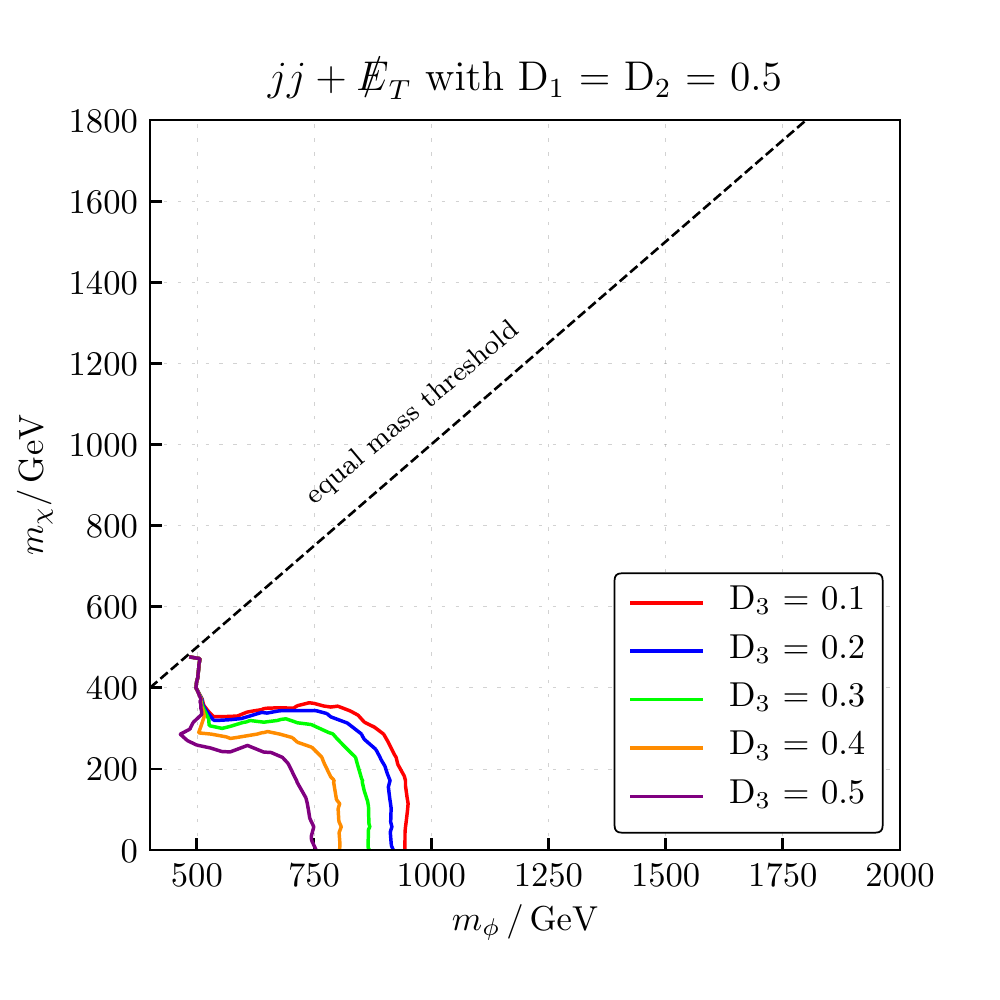}
		\caption{$D_1=D_2=0.5$ and varying $D_3$}
		\label{fig::LHCjjMETa}
		\end{subfigure}
		\hfill
		\begin{subfigure}[t]{0.49\textwidth}
		\includegraphics[width=\textwidth]{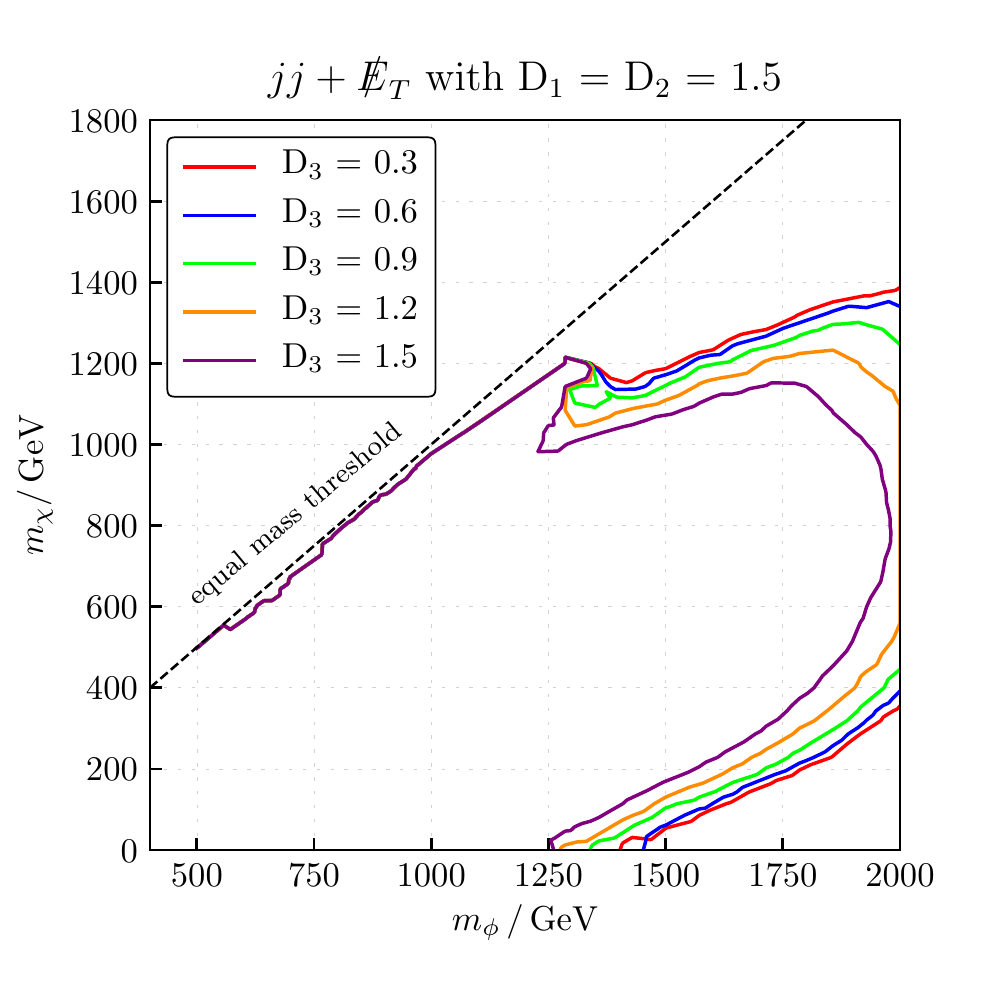}
		\caption{$D_1=D_2=1.5$ and varying $D_3$}
		\label{fig::LHCjjMETb}
		\end{subfigure}
	\caption{Constraints on the final state $jj+\slashed{E}_T$ obtained from \cite{lhclimits}.}
	\label{fig::LHCjjMET}
	\end{figure}
	The results of recasting the $jj+\slashed{E}_T$ limits are shown in Figure \ref{fig::LHCjjMET}. In this case we fix the value of $D_1=D_2$ and vary the value of $D_3$. In contrast to the final states with top flavour, increasing the value of $D_3$ reduces the branching ratio into this final state. At the same time, both the mediator pair-production cross section and the final state branching ratio grow with increasing $D_1 = D_2$.
	 The pattern in Figure \ref{fig::LHCjjMET} matches this expectation.
	  We observe that an increasing value of $D_3$ shrinks the excluded area. When comparing Figure \ref{fig::LHCjjMETa} and Figure \ref{fig::LHCjjMETb}, we also see that the excluded region grows sizeably when the values of $D_1$ and $D_2$ are increased. While this in general is due to an increased production cross section and branching ratio, it partially originates -- in analogy to the final states with top flavour -- from the production rate of the same-sign intermediate  $\phi\phi$ state that grows for an increasing value of $D_1$ and is again governed by the mass parameter $m_\chi$. Just as for the final states with top flavour, regions with a small $m_\chi$ are therefore not excluded for sufficiently high mediator masses $m_\phi \GtrSim 1.5\,\mathrm{TeV}$. Overall, comparing Figure \ref{fig::LHCttMET} and Figure \ref{fig::LHCjjMET} shows that the limits for final states with two jets are significantly more constraining for large $D_1=D_2$ than the limits for final states with top flavour. As there is no interplay between a decreased branching ratio into the final state and a concurrent increased production of the intermediate state when increasing the value of $D_1$, this was to be expected. Another reason is that for degenerate couplings $D_i$ the branching ratio of the final state with $jj+\slashed{E}_T$ is larger, due to the larger multiplicity of possible parton-level final states ($u$- and $c$-jets).
	  
In total, we conclude that for both signatures the same-sign contributions present only in the Majorana DM model constrain a significant part of the $m_\phi-m_\chi$ plane. In order to avoid large exclusion limits on the masses, we require $D_1$ and $D_2$ to be small. As this choice ensures the $t$-channel production of $\phi$ to be small compared to its QCD production, it is safe for $D_3$ to have larger values. Requiring the couplings to lie in the ranges  
	\begin{gather}
	\nonumber
	0< D_3 \leq 1.5\,,\\
	0< D_1,D_2 \leq 0.5\,,
	\end{gather}
it is possible to fulfil the LHC constraints for masses $m_\phi \GtrSim 1\,\mathrm{TeV}$ and arbitrary $m_\chi$. As we will see later on, the limits from flavour and DM phenomenology support this choice of ranges especially in the case of top-flavoured DM. Note that even for larger values $D_1,D_2 \GtrSim 1.0$ it is still possible to fulfil the LHC constraints by choosing $m_\phi \GtrSim 1.5\,\mathrm{TeV}$ and $m_\chi \LessSim 200\,\mathrm{GeV}$.

\subsection{Same-Sign Tops at the LHC}
\label{subsec::tt}

As discussed above, the $t$-channel exchange of a Majorana fermion $\chi$ leads to the production of same-sign mediator pairs $\phi\phi$, with the cross section being enhanced by two powers of the up quark PDF of the proton. The subsequent decay of both $\phi$ scalars into a top quark and a DM flavour then induces the same-sign di-top final state $tt+\slashed{E}_T$. Since this signature is absent in the case of Dirac DM and strongly suppressed by the smallness of the relevant couplings in SUSY, we regard this final state as a smoking-gun signature of our model. Experimentally, this final state can be distinguished from the more common $t\bar t+\slashed{E}_T$ one by measuring the lepton charge of semileptonic top decays.

Therefore we also present a prediction for the $tt+\slashed{E}_T$ production cross section for the upcoming LHC runs, i.e.\@ at a centre of mass energy of $\sqrt{s}=14\,\mathrm{TeV}$. 
Note that experimental NP searches dedicated to same-sign top signatures exist, see  \cite{samesignlimits}. The limits obtained in the latter study generally also pose constraints on our model. However, the NP processes under consideration are different from our case, and hence the final-state kinematics are not the same. A proper recasting would therefore be in order to derive meaningful bounds on the parameter space of our model, which is beyond the scope of the present paper.

\begin{figure}
		\centering
		\includegraphics[width=.5\textwidth]{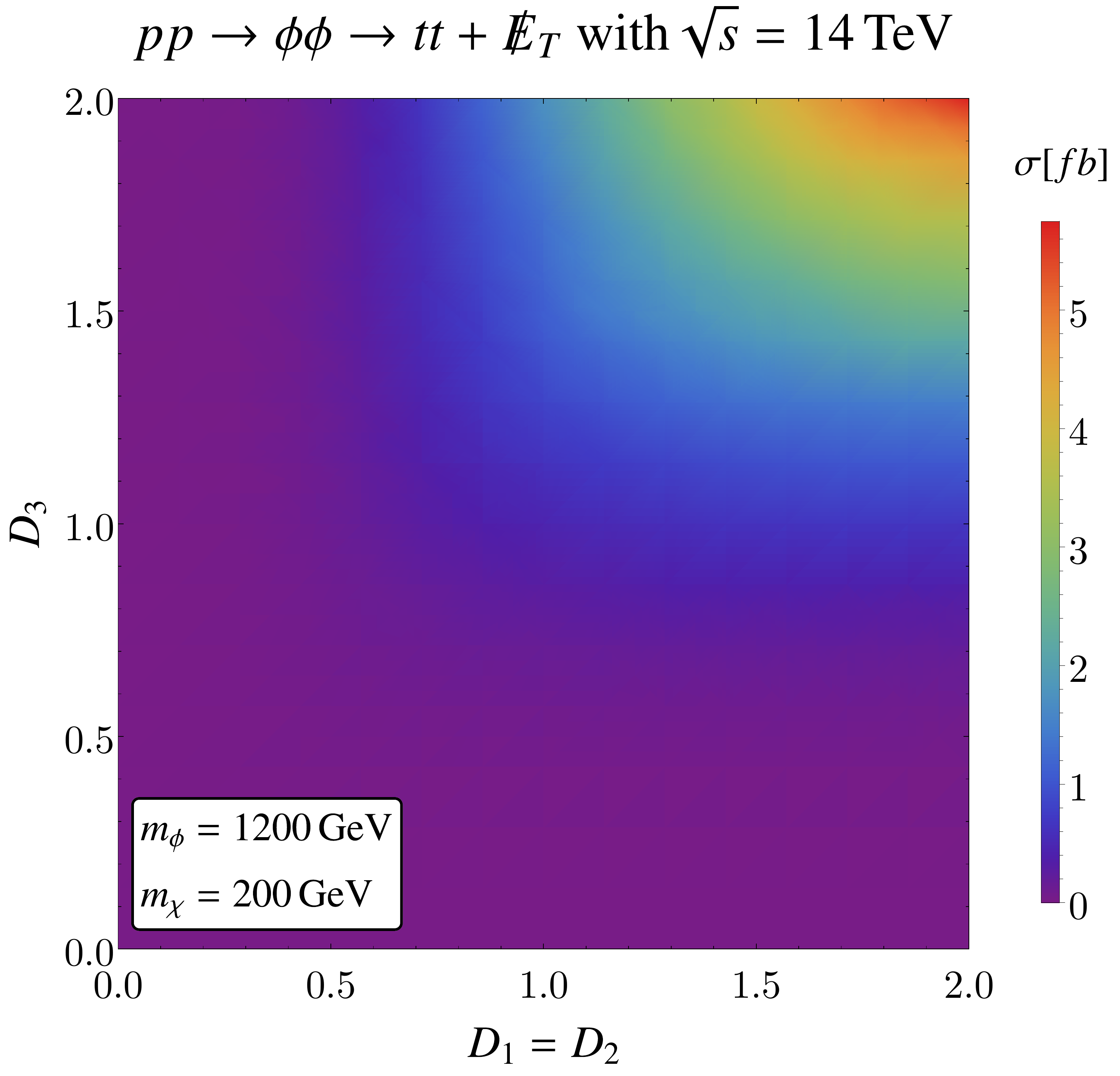}
	\caption{LO cross section of $pp \rightarrow \phi\phi \rightarrow tt+\slashed{E}_T$ in $14\,\mathrm{TeV}$ collisions for $m_\phi=1200\,\mathrm{GeV}$ and $m_\chi = 200\,\mathrm{GeV}$.}
	\label{fig::LHCpredic}
	\end{figure}

	Figure \ref{fig::LHCpredic} illustrates the LO cross section for the same-sign signature $tt+\slashed{E}_T$ for a centre-of-mass energy of $\sqrt{s} = 14\,\mathrm{TeV}$. The cross section is calculated for varying coupling strengths $D_1=D_2$ as well as $D_3$, whereas the masses are fixed to $m_\phi = 1200\,\mathrm{GeV}$ and $m_\chi = 200\, \mathrm{GeV}$, in agreement with the LHC constraints derived in the previous section. As expected, the cross section grows with an increasing coupling $D_1$ since the same-sign intermediate state is mainly produced through the process $uu\rightarrow \phi\phi$. At the same time, the branching ratio into top final states grows for increasing $D_3$, yielding the largest $tt+\slashed{E}_T$ in the part of the parameter space where all couplings $D_i$ are large. In this case, rates of the order of several fb are predicted, well in the reach of future LHC studies.


\section{Flavour Physics Phenomenology}
\label{sec::flavour}

A central aspect of the DMFV ansatz is the introduction of the new flavour and CP-violating coupling $\lambda$ which can generally lead to large FCNC effects and therefore is very sensitive to constraints stemming from flavour observables. Often, the most stringent constraints on flavoured NP interactions stem from $\Delta F = 2$ observables, i.\,e.\ observables measuring the oscillation of neutral mesons. Since in our model the new particles couple to the right-handed up-type quarks, the relevant process is $D^0-\bar{D}^0$ mixing ($\Delta C=2$), to which we dedicate this section.
We derive the relevant expressions for the contributions of our flavoured Majorana DM model  and restrict the structure of the coupling matrix $\lambda$ through existing experimental bounds on $D^0-\bar{D}^0$ mixing observables.
 A detailed introduction into the underlying formalism  of effective Hamiltonians describing flavour violating processes can be found in \cite{df21,df22}.
 
 Concerning $\Delta F = 1$ processes, the constraints on NP contributions to rare $D$ decays are generally weaker than the ones from $D^0-\bar{D}^0$ mixing, and we do not consider them here. Following \cite{tfdm} we also ignore rare flavour violating top decays, as their constraints are not stringent enough yet. 
 However, after combining all available constraints on the parameter space of our model, in Section \ref{sec::CPV} we return to flavour physics and 
  study the contributions to directly CP-violating charm decays, measured in the asymmetry $\Delta A^\text{dir}_{CP}$.   
  
	\subsection[Neutral \texorpdfstring{$D$}{D} Meson Mixing]{Neutral \texorpdfstring{\boldmath $D$\unboldmath}{D} Meson Mixing}
	
\begin{figure}[b!]
		\centering
		\begin{subfigure}[t]{0.49\textwidth}
		\includegraphics[width=\textwidth]{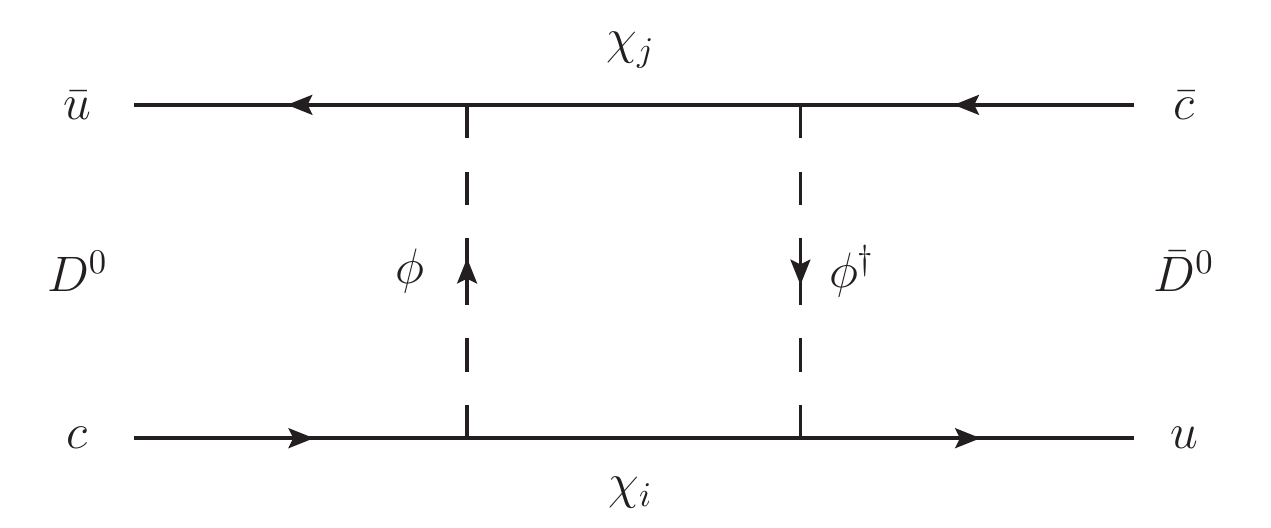}
		\caption{standard box diagram}
		\label{fig::mixinga}
		\end{subfigure}
		\hfill
		\begin{subfigure}[t]{0.49\textwidth}
		\includegraphics[width=\textwidth]{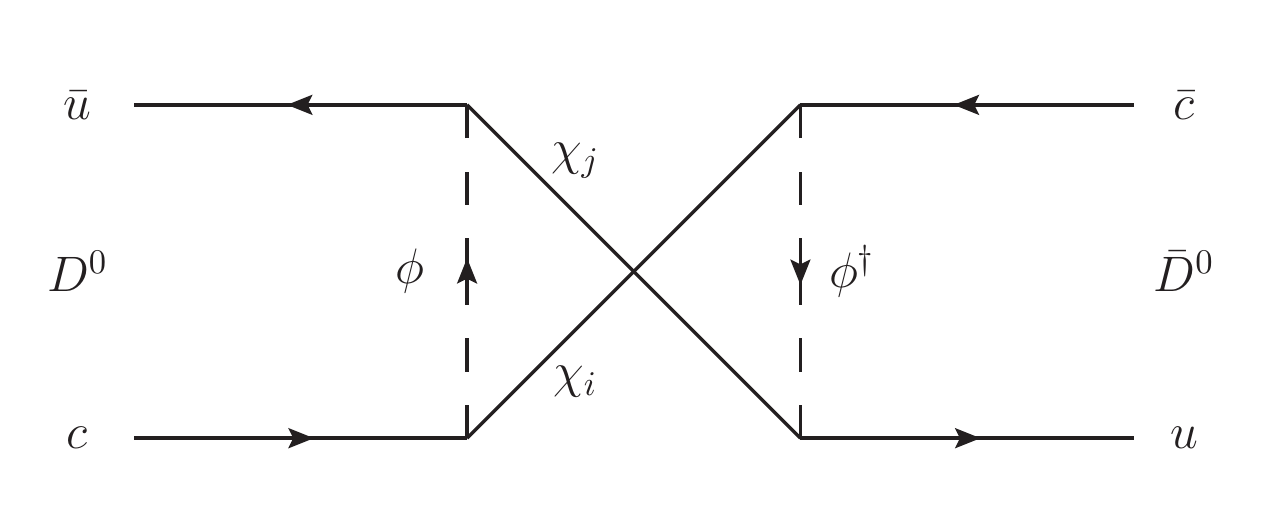}
		\caption{crossed box diagram}
		\label{fig::mixingb}
		\end{subfigure}
	\caption{Feynman diagrams for the NP contributions to $D^0-\bar{D}^0$ mixing at LO.}
	\label{fig::mixing}
	\end{figure}

	The flavour violating interaction term in eq.\@ \eqref{eq::lagrangian}  gives rise to NP contributions to $D^0-\bar{D}^0$ mixing. This $\Delta C=2$ process is first generated at one-loop order, i.e.\@ $\mathcal{O}(\lambda^4)$. The relevant diagrams are shown in Figure \ref{fig::mixing}. 
	Note that the crossed diagram shown in Figure \ref{fig::mixingb} only exists if $\chi$ is a Majorana fermion. In models where $\chi$ is a Dirac fermion one only has the standard box diagram shown in Figure \ref{fig::mixinga}. Using the techniques presented in \cite{df21,df22} and evaluating the diagrams one arrives at the following effective Hamiltonian:
	\begin{equation}
	\mathcal{H}^{\Delta C = 2,\,\text{NP}}_\text{eff}=
	\frac{1}{128 \pi^2 m_\phi^2} \sum_{ij} \lambda_{uj}\lambda_{ci}^*\Bigl[\lambda_{ui}\lambda_{cj}^*\cdot F(x_i,x_j) - 2\lambda_{uj}\lambda_{ci}^*\cdot G(x_i,x_j)\Bigr] \times Q_{uc}^{VRR}+\mathrm{h.c.}\,,
	\label{eq::effham}
	\end{equation}
	where we have used the short-hand notation
	\begin{equation}
	Q_{uc}^{VRR} = (\bar{u}_\alpha \gamma_\mu P_R c_\alpha)(\bar{u}_\beta \gamma^\mu P_R c_\beta)\,,
	\label{eq::effop}
	\end{equation}
	for the relevant  effective four-fermion operator.
	The loop functions read
	\begin{align}
	\nonumber
	F(x_i,x_j)=\phantom{-}&\left(\frac{x_i^2 \log(x_i)}{(x_i-x_j)(1-x_i)^2}+\frac{x_j^2 \log(x_j)}{(x_j-x_i)(1-x_j)^2}+\frac{1}{(1-x_i)(1-x_j)}\right)\,,\\
	G(x_i,x_j)=-&\left(\frac{\sqrt{x_i x_j}x_i \log(x_i)}{(x_i-x_j)(1-x_i)^2}+\frac{\sqrt{x_i x_j}x_j \log(x_j)}{(x_j-x_i)(1-x_j)^2}+\frac{\sqrt{x_i x_j}}{(1-x_i)(1-x_j)}\right)\,,
	\label{eq::loopfunctions}
	\end{align}
	with $x_i=m_{\chi_i}^2/m_\phi^2$. In eq.\@ \eqref{eq::effop} we imply a summation over the colour indices $\alpha$ and $\beta$. In the interval $x_i \in \left[0,1\right]$ the loop functions of eq.\@ \eqref{eq::loopfunctions} have the same sign, i.e.\@ the two diagrams of Figure \ref{fig::mixing} can interfere destructively due to the relative minus sign in the effective Hamiltonian in eq.\@ \eqref{eq::effham}. This is a well known effect from SUSY where the box diagram contains a squark and a gluino \cite{susymixing}. For $x_i=x_j=1$ we even have $2G(1,1)=F(1,1)$, but as the coupling $\lambda$ is not flavour-universal a destructive interference can only occur in eq.\@ \eqref{eq::effham} if the different prefactors of $F(x_i,x_j)$ and $G(x_i,x_j)$ have the same sign.
	
	Using the effective Hamiltonian from above we can now calculate the off-diagonal element of the $D^0-\bar{D}^0$ mass matrix as
	\begin{align}
	 \nonumber
	M_{12}^{D,\,\text{NP}}&=\frac{1}{2 m_D}\bra{\bar{D}^0}\mathcal{H}^{\Delta C = 2,\,\text{NP}}_\text{eff}\ket{D^0}^*\\
	&=\frac{\eta_D m_D f_D^2 \hat{B}_D }{384\pi^2 m_\phi^2} \sum_{ij} \lambda_{uj}^*\lambda_{ci}\Bigl[\lambda_{ui}^*\lambda_{cj}\cdot F(x_i,x_j)- 2\lambda_{uj}^*\lambda_{ci}\cdot G(x_i,x_j)\Bigr]\,,
	\label{eq::mixingoffdiag}
	\end{align}
	where we have used 
	\begin{equation}
	\bra{\bar{D}^0}Q^{VRR}_{uc}(\mu)\ket{D^0} = \frac{2}{3} m_D^2 f_D^2 \hat{B}_D\,,
	\end{equation}
	for the hadronic matrix element at the low-energy scale $\mu = 3\,\text{GeV}$, at which the parameters $f_D$ and $\hat{B}_D$ are obtained through lattice QCD calculations \cite{lattice1,lattice2}. In eq.\@ \eqref{eq::mixingoffdiag} the parameter $\eta_D$ accounts for NLO contributions from the renormalization group running   between the weak scale $\mu = M_W$ and the low-energy scale $\mu = 3\,\mathrm{GeV}$ \cite{mesonscale}. It also serves as a parametrisation of threshold corrections for the matching of the SM to the effective theory. We neglect the additional threshold matching corrections  between the SM and our DMFV model which, following \cite{dmfv}, we expect to be small.
	Also note that there is no NP contribution to the absorptive part of the mixing amplitude $\Gamma_{12}^D$, since the NP scale is above the $D^0$ meson mass scale and therefore cannot contribute on-shell.
	 
	The expression in eq.\@ \eqref{eq::mixingoffdiag} describes the general case with a possible mass splitting between different DM flavours $\chi_i$. As the mass corrections described in Section \ref{subsec::mspl} are NLO in the DMFV expansion, plugging them into the expressions from above will generate a higher-order DMFV correction that we assume to be small. Further, we have checked numerically that $x_i \neq x_j$ only causes corrections of a few percent for the loop functions $F(x_i,x_j)$ and $G(x_i,x_j)$. Thus, the mass splitting among the fields $\chi_i$ is neglected in this section. We can then evaluate the sum in eq.\@ \eqref{eq::mixingoffdiag} and find 
	\begin{equation}
	M_{12}^{D,\,\text{NP}} =\frac{\eta_D m_D f_D^2 \hat{B}_D }{384\pi^2 m_\phi^2} \Bigl[\xi_f \cdot f(x)- 2\, \xi_g\cdot g(x)\Bigr]\,,
	\end{equation}
	where we have introduced
	\begin{align}
	\nonumber
	\xi_f &=\sum_{ij} \lambda_{ui}^*\lambda_{ci} \lambda_{uj}^*\lambda_{cj}=\left(\lambda\lambda^\dagger\right)_{cu}^2\,,\\
	\xi_g &=\sum_{ij} \lambda_{ci}\lambda_{ci} \lambda_{uj}^*\lambda_{uj}^*=\left(\lambda\lambda^T\right)_{cc}\left(\lambda\lambda^T\right)_{uu}^*\,.\label{eq::xi}
	\end{align}
	For the loop functions we find 
	\begin{align}
	f(x) &= \lim_{y \rightarrow x} F(x,y)= \frac{1+x}{(1-x)^2}+ \frac{2x \log(x)}{(1-x)^3}\,,\\
	g(x) &= \lim_{y \rightarrow x} G(x,y)= -\frac{2x}{(1-x)^2}- \frac{x(1+x)}{(1-x)^3}\log(x)\,.
	\label{eq::mixingxi}
	\end{align}
	We can now calculate the NP contribution to $x^{D}_{12}$ according to 
	\begin{equation}
	x^{D,\text{NP}}_{12} = 2\, \tau_{D^0}\, |M^{D,\text{NP}}_{12}|\,.
	\end{equation}
	As $\Gamma^{D,\text{SM}}_{12}$ is real to an excellent approximation in the standard parametrisation of the CKM matrix, the CP-violating phase simply reduces to $\phi_{12}^D=\text{Arg}\left(M^D_{12}\right)$. 
	
	\subsection[Application of the \texorpdfstring{$D$}{D} Meson Mixing Constraints]{Application of the \texorpdfstring{\boldmath$D$\unboldmath}{D} Meson Mixing Constraints}
	
	In Table \ref{tab::mixingvalues} we show the relevant values of parameters and the model-independent limits for $x_{12}^D$ as well as $\phi_{12}^D$ that we use in the numerical analysis.
	\begin{table}[b!]
	\centering
	\begin{tabular}{cc} \toprule
    \multicolumn{2}{c}{\textbf{Numerical Values and Limits}} \\ \bottomrule \toprule
	$\hat{B}_D$	& $0.75 \pm 0.02$ \\
	$f_D$	& $209.0 \pm 2.4\,\mathrm{MeV}$ \\
	$\eta_D$	& $0.772$ \\
	$m_{D^0}$	& $1864.83 \pm 0.05 \,\mathrm{MeV}$ \\
	$\tau_{D^0}$	& $410.1 \pm 1.5 \,\mathrm{fs}$ \\ \midrule
	$x_{12}^D$	& $\left[0.21\%,0.63\%\right]$ \\
	$\phi_{12}^D$	& $\left[-2.8^\circ,1.7^\circ\right]$ \\\bottomrule
	\end{tabular}
	\caption{Numerical values and limits used for the analysis of the $D^0-\bar{D}^0$ mixing constraints \cite{lattice1,lattice2,mesonscale,pdg,mixinglimits}. The limits on $x_{12}^D$ and $\phi_{12}^D$ are given at $95\%$ C.L. and were obtained from the {Heavy Flavour Averaging Group}'s (HFLAV) website \cite{mixinglimits}.}
	\label{tab::mixingvalues}
	\end{table}	
	 
\begin{figure}
		\centering
		\begin{subfigure}[t]{0.49\textwidth}
		\includegraphics[width=\textwidth]{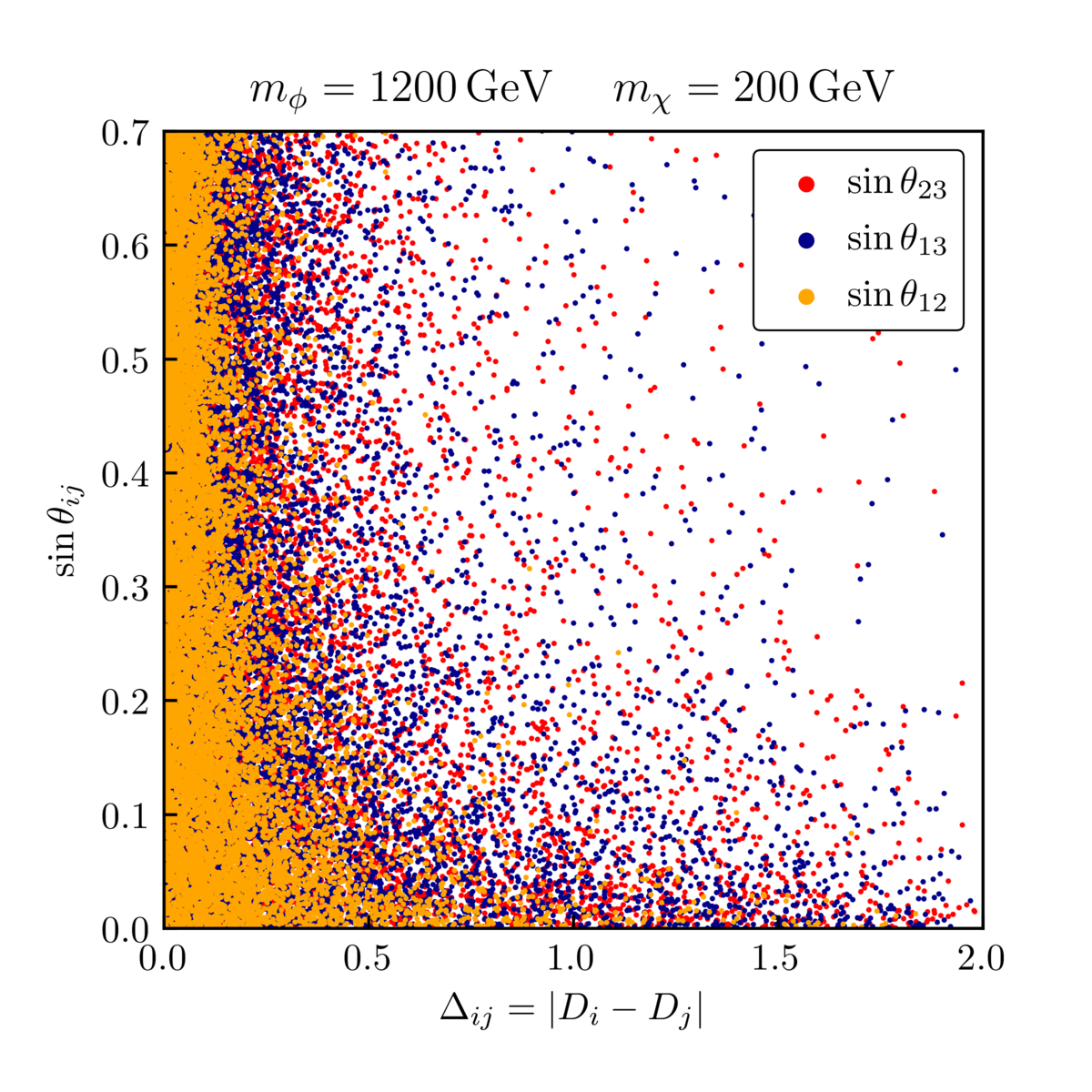}
		\caption{$m_\phi=1200\,\mathrm{GeV}$ and $m_\chi=200\,\mathrm{GeV}$}
		\label{fig::deltavsthetaa}
		\end{subfigure}
		\hfill
		\begin{subfigure}[t]{0.49\textwidth}
		\includegraphics[width=\textwidth]{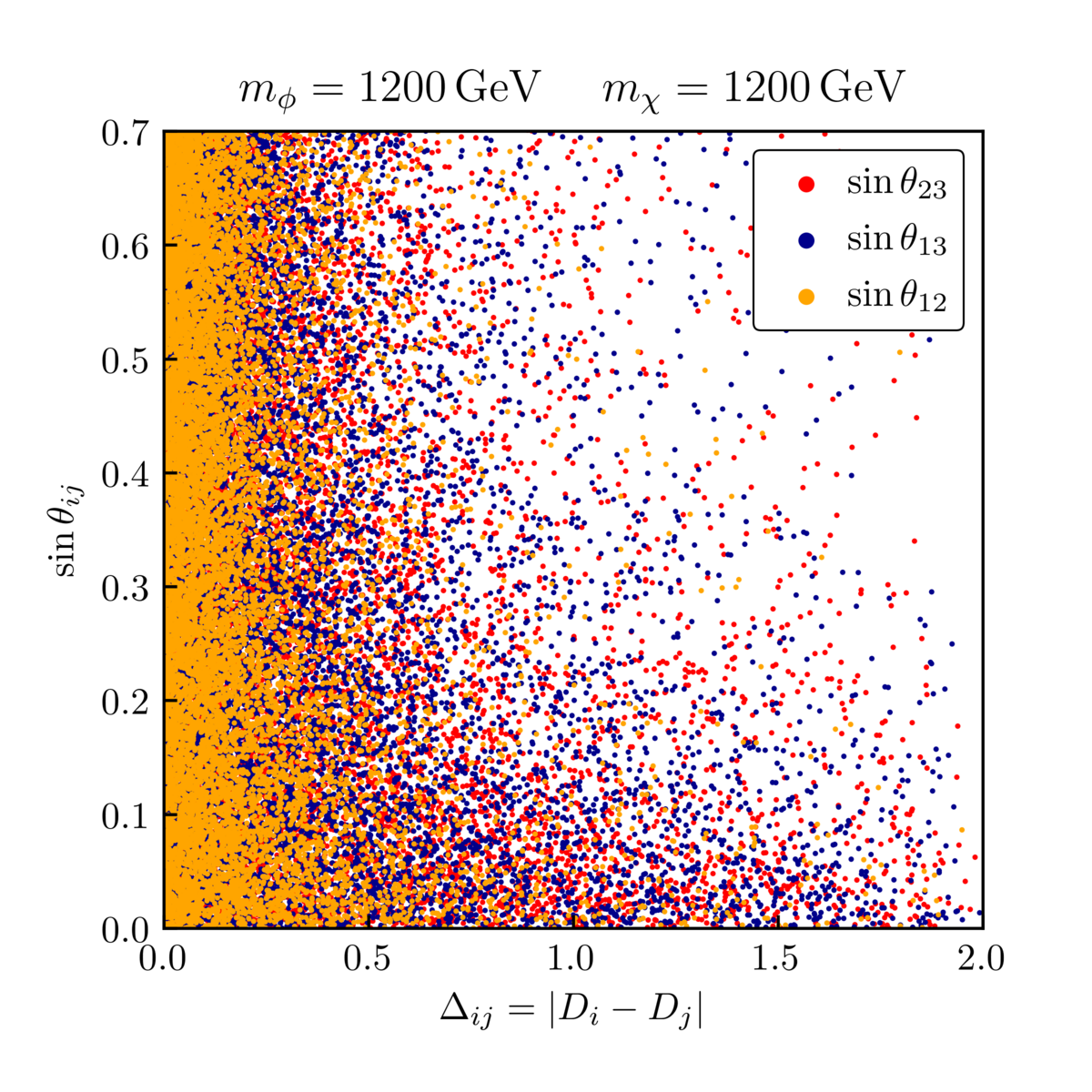}
		\caption{$m_\phi=1200\,\mathrm{GeV}$ and $m_\chi=1200\,\mathrm{GeV}$}
		\label{fig::deltavsthetab}
		\end{subfigure}
	\caption{Allowed mixing angles $\theta_{ij}$ as a function of the splitting $\Delta_{ij}=|D_i-D_j|$ for two choices of $m_\phi$ and $m_\chi$.}
	\label{fig::deltavstheta}
	\end{figure}

\begin{figure}
		\centering
		\begin{subfigure}[t]{0.49\textwidth}
		\includegraphics[width=\textwidth]{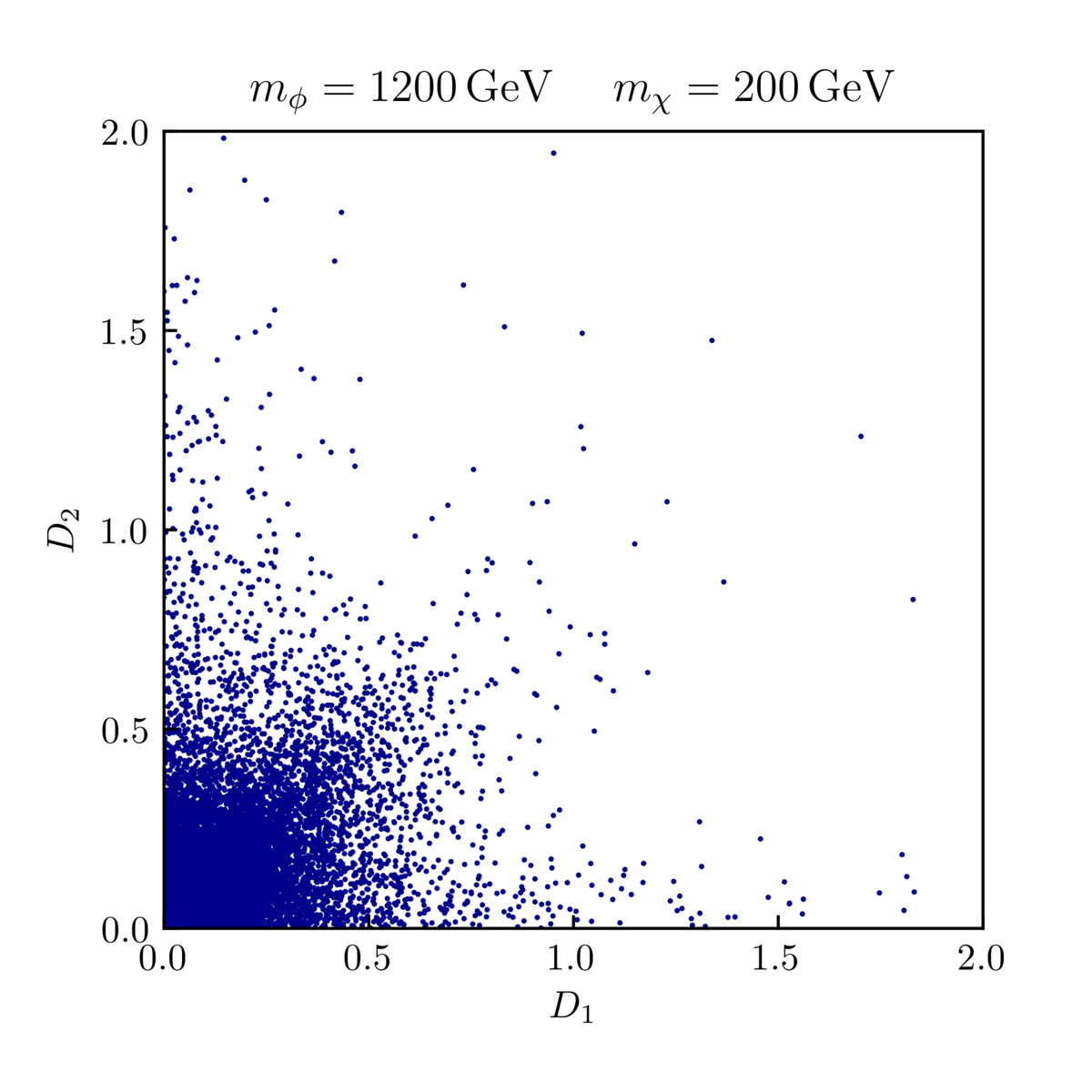}
		\caption{$D_1-D_2$ plane}
		\label{fig::d1vsd2}
		\end{subfigure}
		\hfill
		\begin{subfigure}[t]{0.49\textwidth}
		\includegraphics[width=\textwidth]{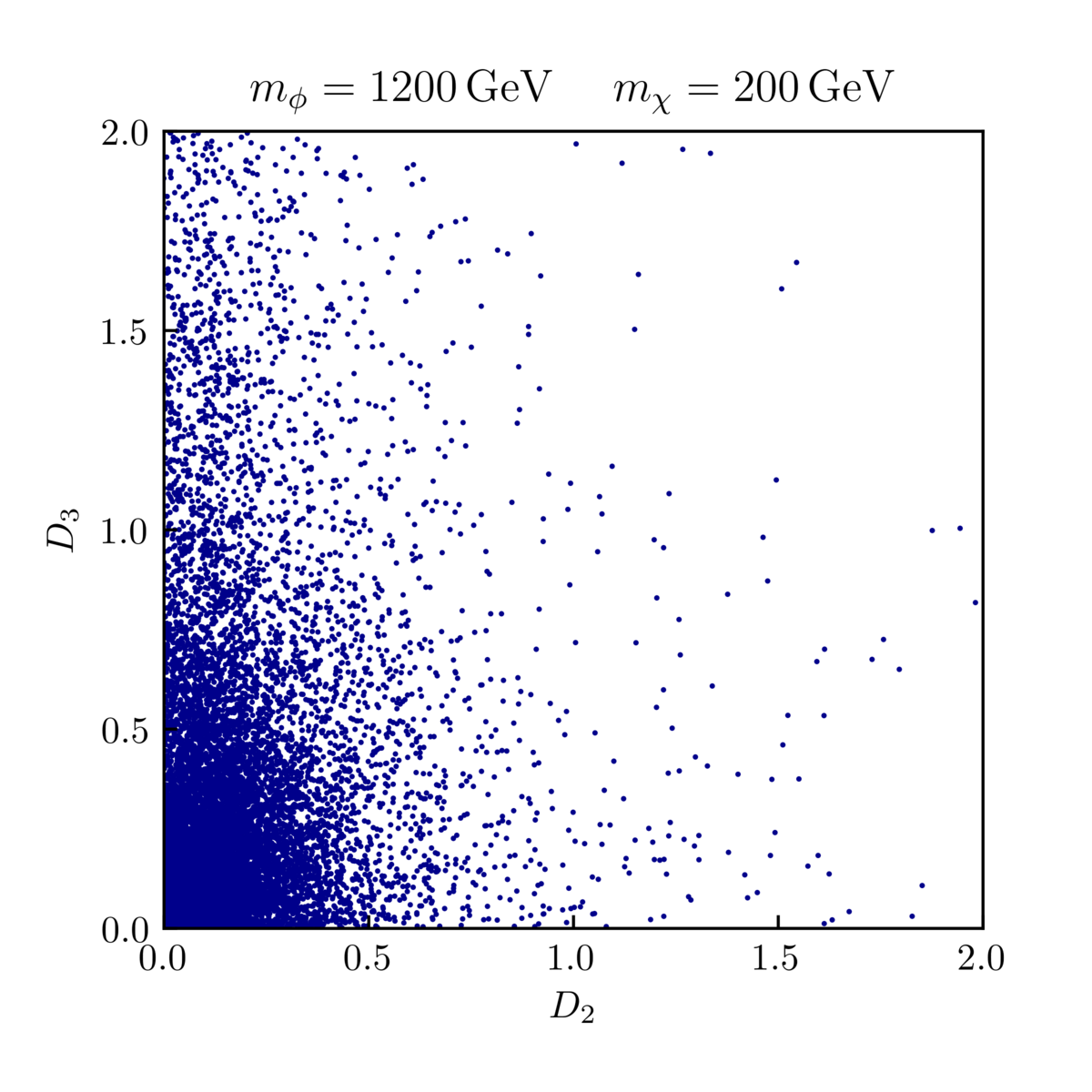}
		\caption{$D_2-D_3$ plane}
		\label{fig::d2vsd3}
		\end{subfigure}
	\caption{Allowed parameter space in the $D_i-D_j$ plane for $m_\phi=1200\,\mathrm{GeV}$ and $m_\chi=200\,\mathrm{GeV}$.}
	\label{fig::divsdj}
	\end{figure}

\begin{figure}[b!]
		\centering
		\begin{subfigure}[t]{0.49\textwidth}
		\includegraphics[width=\textwidth]{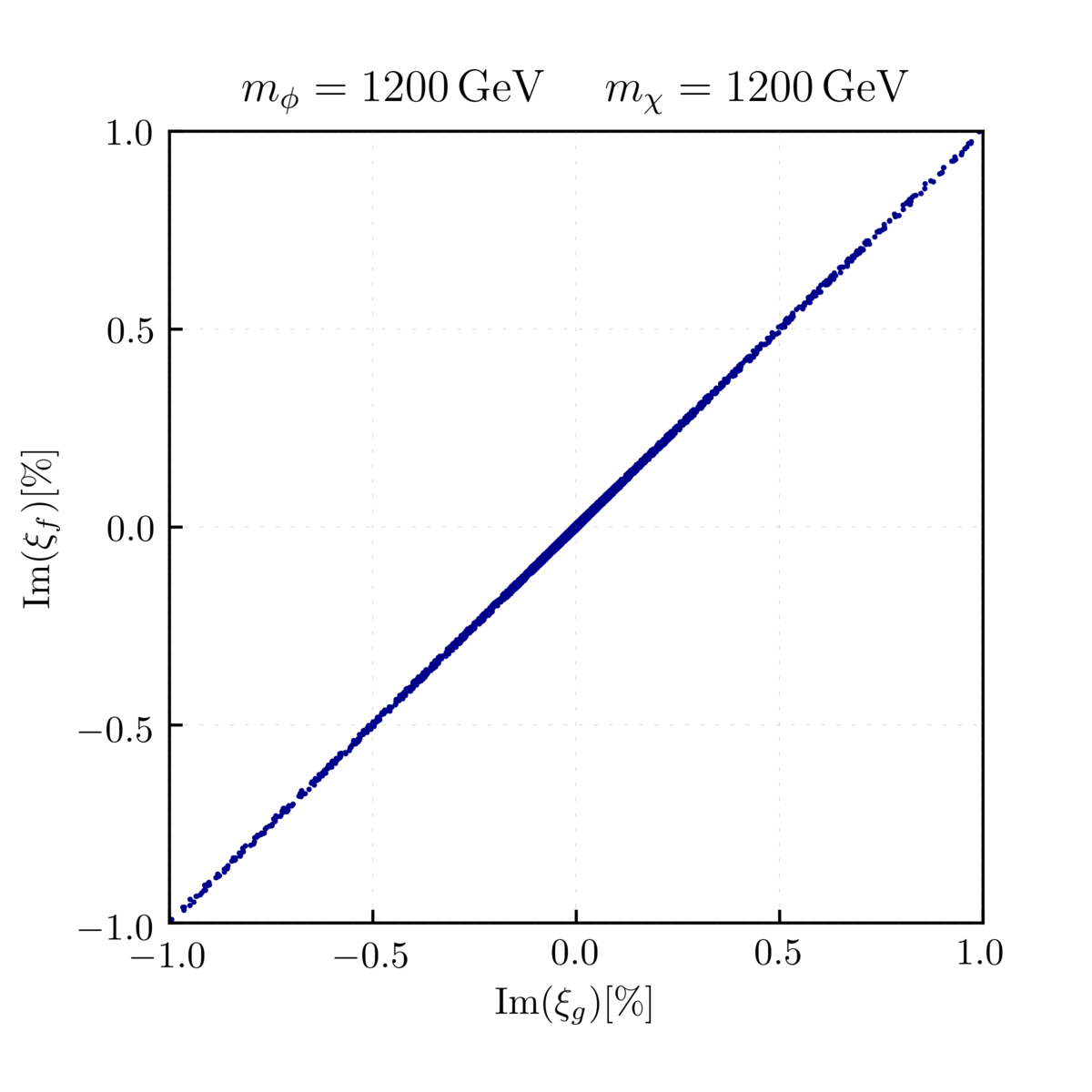}
		\caption{$m_\phi=1200\,\mathrm{GeV}$ and $m_\chi=1200\,\mathrm{GeV}$}
		\label{fig::xia}
		\end{subfigure}
		\hfill
		\begin{subfigure}[t]{0.49\textwidth}
		\includegraphics[width=\textwidth]{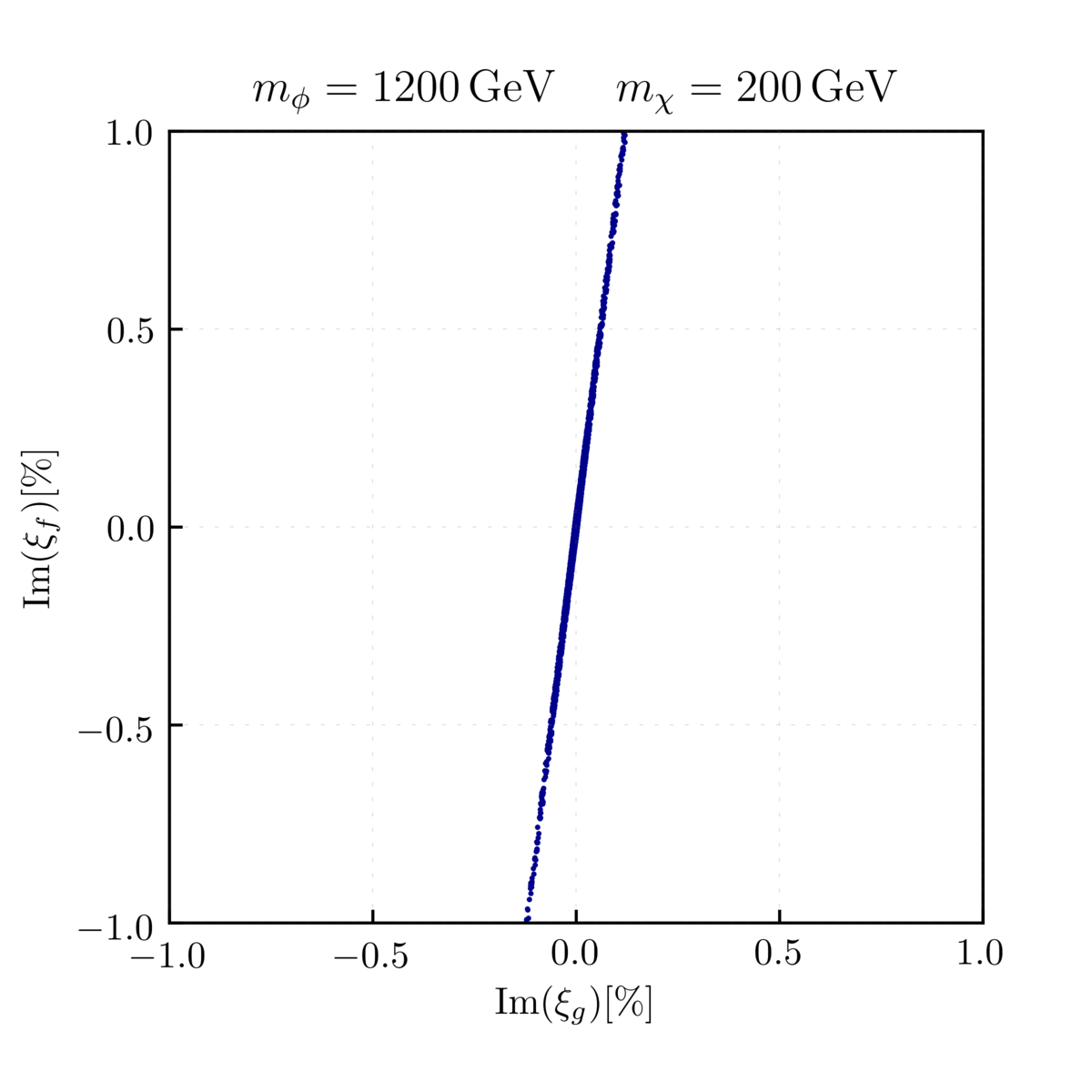}
		\caption{$m_\phi=1200\,\mathrm{GeV}$ and $m_\chi=200\,\mathrm{GeV}$}
		\label{fig::xib}
		\end{subfigure}
	\caption{Imaginary part of the two factors $\xi_f$ and $\xi_g$ in percent.}
	\label{fig::xi}
	\end{figure}

	For a proper treatment of the restrictions these constraints pose on $\lambda$ it is necessary to also consider the SM contribution to $x_{12}^D$ and $\phi_{12}^D$. As the CP-violating phase is expected to be of order $\mathcal{O}(10^{-3})$ in the SM, we neglect this small contribution and assume $M_{12}^{D,\text{SM}}$ to be real.
	 As for the SM contributions to $x_{12}^D$, they are dominated by long-distance effects and hence suffer from large theory uncertainties. Nevertheless, estimates of $x_{12}^{D,\text{SM}}$ find it of the order $\mathcal{O}(10^{-2})$ \cite{Petrov:2013usa}. Given these uncertainties, we conservatively assume the SM contribution to lie in the range 
	\begin{equation}
	x_{12}^{D,\text{SM}} \in [-3\%,3\%]\,.
	\label{eq::mixingxdsm}
	\end{equation}
	Using the results provided in the previous section we then demand that the total values for $x_{12}^D$ and $\phi_{12}^D$ lie in their $95\%$ C.L. intervals given in Table \ref{tab::mixingvalues} to restrict the parameters of our model. The results are shown in Figure  \ref{fig::deltavstheta}, Figure \ref{fig::divsdj} and Figure \ref{fig::xi}.\par 
	Figure \ref{fig::deltavstheta} illustrates the allowed mixing angles $\theta_{ij}$ in dependence of the difference $\Delta_{ij}=|D_i - D_j|$ for $m_\chi < m_\phi$ and $m_\phi = m_\chi$.\footnote{The latter case is excluded as it would render $\phi$ stable, in conflict with cosmological observations.
	 Nevertheless, we also discuss the case of equal masses here in order to provide a better understanding of the flavour physics phenomenology and the interference between the diagrams shown in \ref{fig::mixing}.} We see that in both cases the strongest restrictions are placed on $\theta_{12}$ and that the experimental constraints become especially effective for large $\Delta_{ij}$. For nearly degenerate couplings $D_i \approx D_j$ the corresponding angle $\theta_{ij}$ can be chosen freely. The reason is that the factor $\xi_f$ defined in eq.\@ \eqref{eq::mixingxi} approaches zero in the limit of degenerate $D_i$ as $\lambda\lambda^\dagger = UD^2U^\dagger$ becomes diagonal. Thus, the contribution of the diagram in Figure \ref{fig::mixinga} vanishes.  Figure \ref{fig::deltavsthetab} shows that choosing $m_\phi = m_\chi$ only slightly loosens the bounds on $\Delta_{ij}$.\par 
	
	As the crossed box diagram in Figure \ref{fig::mixingb} is proportional to the diagonal entries of $\lambda\lambda^T$ and its complex conjugate, it does not approach zero for vanishing mixing angles and phases. In order to gain insight on the constraints this places on the parameters of $\lambda$, we analyse the $D_i - D_j$ plane of the parameter space. Figure \ref{fig::d1vsd2} shows that the constraints require the couplings $D_1$ and $D_2$ to roughly lie in the interval $D_1,D_2 \in [0.0,0.5]$. This is due to the fact that $\xi_g$ mainly depends on these two couplings. For zero mixing angles and phases it reduces to $\xi_g = D_1^2 D_2^2$. Thus, combinations of a large $D_1$ with a small $D_2$ or vice versa are also allowed. In Figure \ref{fig::d2vsd3} we see that there are no such constraints on $D_3$ and that the latter coupling can be chosen freely. The same pattern can also be observed for other masses $m_\phi$ and $m_\chi$. 

	We conclude that, barring cancellations,
	\begin{gather}
	\nonumber
	0< D_3 \leq 2.0\,,\\
	0< D_1,D_2 \leq 0.5\,,
	\end{gather}
is required to fulfil the $D^0-\bar D^0$ mixing constraints while allowing for large mixing angles for most of the remaining parameter space. It is interesting to see that this result perfectly matches the LHC constraints identified in Section \ref{subsec::lhcresults}.
Note that while in both Figs.\ \ref{fig::deltavstheta} and \ref{fig::divsdj} the majority of allowed parameter points lies within these boundaries, we find some parameter points significantly outside this region. Those points can be traced back to the  aforementioned destructive interference of the two diagrams shown in Figure \ref{fig::mixing}.
\par  
	
	As we show next, this destructive interference indeed extends the allowed parameter space, relative to the case of Dirac DM. Due to the large non-perturbative uncertainty in $M_{12}^{D,\text{SM}}$ (which is CP-conserving), for illustrative purposes we focus on the constraint from CP violation in $D^0-\bar D^0$ mixing.
	
	To illustrate the interference effect, in Figure \ref{fig::xi} we show the imaginary parts of the coupling combinations $\xi_i$, defined in eq.\ \eqref{eq::xi},  for two choices for the masses $m_\phi$ and $m_\chi$. One can see that for both choices of masses the allowed imaginary parts scatter around a linear function. Remembering that the limits on $\phi^D_{12}$ are very stringent, i.e.\@ $\phi^D_{12} \approx 0$, this behaviour can be understood analytically. Since $\phi^D_{12}$ is given as 
	\begin{equation}
	\phi^D_{12} =\arctan\frac{\text{Im}(M_{12}^{D,\text{NP}})}{\text{Re}(M_{12}^{D,\text{NP}})+\text{Re}(M_{12}^{D,\text{SM}})}\,,
	\end{equation}	
	small values for $\phi^D_{12}$ require a small imaginary part of $M_{12}^{D,\text{NP}}$. This requirement forces the two coupling factors $\xi_f$ and $\xi_g$ to follow the relation 
	\begin{equation}
	\text{Im}\xi_f \approx \text{Im}\xi_g \frac{2g(x)}{f(x)}\,.
	\label{eq::mixingxilin}
	\end{equation}
		 We confirm this finding in Figure \ref{fig::xia} where we have $x=1$ and $2g(1)=f(1)=1/3$. In this case the slope is basically one and we have $\text{Im}\xi_f = \text{Im}\xi_g$. For any other point in the $m_\phi - m_\chi$ plane the slope is given by the actual value of $2g(x)/f(x)$ as can be seen in Figure \ref{fig::xib}. Recall that in the case of Dirac DM the contribution to $M_{12}^D$ proportional to $\xi_g$ is absent, so that in that case  $\text{Im}\xi_f \approx 0$ is required. Hence indeed the Majorana nature of $\chi_i$ implies larger allowed values of the CP-violating coupling parameters $\text{Im}\xi_{f,g}$. {We note that the valid parameter points found in our scan extend well beyond the range shown in Figur \ref{fig::xi}. While the density of points decreases with increasing values of $\text{Im}\xi_{f,g}$, values as large as $\text{Im}\xi_{f,g}\sim \mathcal{O}(1)$ can be reached.} In Section \ref{sec::CPV} we return to the practical implications of this finding.

	 In the described figures the scattering of the allowed points around this linear function is due to the fact that $\phi_{12}^D$ is not exactly zero but is allowed to lie in the interval given in Table \ref{tab::mixingvalues}. The same interference between the diagrams of Figure \ref{fig::mixing} can also be seen in the real part of $M_{12}^{D,\text{NP}}$, with the sole difference of a larger scattering around the linear function of eq.\@ \eqref{eq::mixingxilin}. This is due to the theoretical uncertainty in the SM contribution to the real part of $M_{12}^{D}$.

\section{Cosmological Implications}
\label{sec::relicabundance}
The constraints from collider and flavour physics experiments analysed so far are generally relevant for any extension of the SM that contains new flavoured particles. As the DMFV ansatz further assumes the lightest flavour of the field $\chi$ to constitute the observed DM in the universe \cite{planck}, it is also necessary to consider constraints from cosmology. Thus, we  use this section to discuss the implications of the observed DM relic abundance on our model.

	\subsection{Dark Matter Annihilation and Freeze-Out Scenarios}
	
	A commonly used approach  to explain the observed amount of DM is to assume a freeze-out of dark particles from thermal equilibrium at a given temperature $T_f \approx m_\chi/ 20$ \cite{wimp}, i.e.\@ at temperatures below $T_f$ the DM production and annihilation rates both approach zero and one is left with a relic of DM. The resulting  relic abundance depends on the annihilation rate of DM into SM matter. We  first define two benchmark freeze-out scenarios to be discussed in this analysis and then  provide the relevant formulae for the annihilation of DM into SM matter.
	
	As our model contains three generations of DM particles it is important to consider the mass hierarchy of the fields $\chi_i$ in order to identify different freeze-out scenarios. This hierarchy is generated by the DMFV mass corrections discussed in Section \ref{subsec::mspl}. If there is only a very small difference between the different DM masses $m_{\chi_i}$, all three flavours will be present at the time of thermal freeze-out since the decay of the slightly heavier flavours into the lightest one is kinematically highly suppressed. The frozen-out heavy flavours will then completely decay into the lightest DM particle at later times or lower temperatures, respectively. If on the other hand we allow for a significant splitting, the lifetime of the heavier DM particles is short compared to the time of freeze-out. Flavour-changing scattering processes of the type $\chi_i u_j \to \chi_k u_l$ still maintain a relative equilibrium between the three DM flavours, however the abundance of the heavier flavours is suppressed by a Boltzmann factor $\exp(-\delta m_\chi/T_f)$. We can thus approximate the latter case by assuming that only the lightest flavour $\chi_3$ contributes to the DM freeze-out.\footnote{To overcome the approximate nature of our ansatz, a full numerical solution of the coupled three-flavour Boltzmann equations would be necessary. As we expect the dominant effects to be captured by our approximation, we leave a more in-depth analysis for future work.} We follow \cite{dmfv,tfdm} and define the following two freeze-out scenarios as benchmarks:
	\begin{itemize}
	\item We call the case of a very small mass splitting  the \textbf{quasi-degenerate freeze-out (QDF)} scenario. In this case the mass difference
	\begin{equation}
	\Delta m_{i3} = \frac{m_{\chi_i}}{m_{\chi_3}} - 1\,,
	\end{equation}
	between the heavier flavours $i\in\{1,2\}$ and the lightest flavour\footnote{Remember that we conventionally choose the third generation to be the lightest of the three DM particles.} is restricted to be below $1\%$. In order to suppress the DMFV correction to the DM mass matrix, we set $\eta=-0.01$ in eq.\@ \eqref{eq::massspl}. Smaller magnitudes of $\eta$ would be implausible, as the contribution is generated at the one-loop level.
	\item In the \textbf{single flavour freeze-out scenario (SFF)} the mass splitting $\Delta m_{i3}$ is assumed to be larger than $10\%$ but still small enough to ensure the convergence of eq.\@ \eqref{eq::massspl}. We choose $\eta=-0.0575$ for this scenario. This yields a maximal splitting of $\Delta m_{i3}^\text{max} \simeq 30\%$ for couplings $D_i \leq 2.0$.
	\end{itemize}

\begin{figure}[t!]
		\centering
		\begin{subfigure}[t]{0.35\textwidth}
		\includegraphics[width=\textwidth]{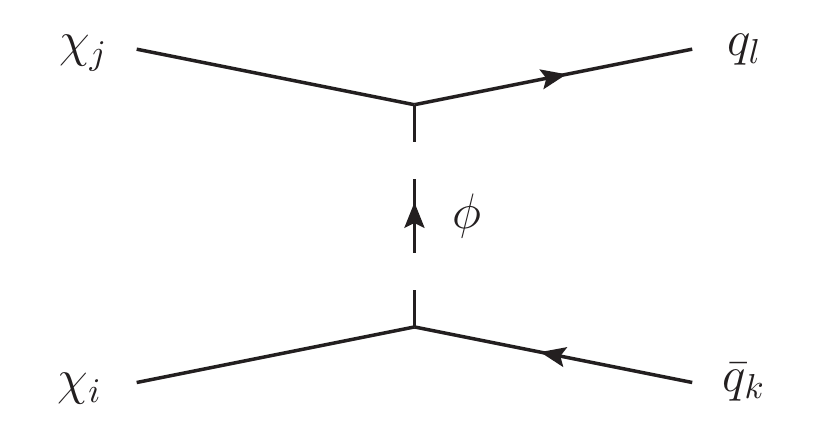}
		\caption{$t$-channel annihilation}
		\label{fig::annihilationa}
		\end{subfigure}
		\hspace{2cm}
		\begin{subfigure}[t]{0.35\textwidth}
		\includegraphics[width=\textwidth]{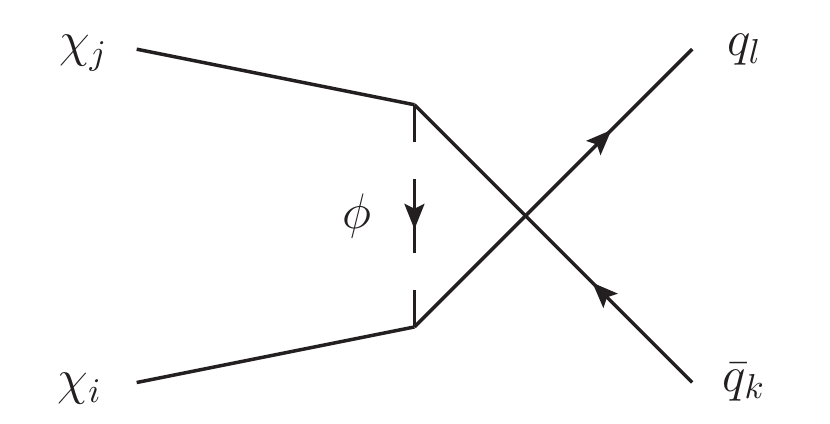}
		\caption{$u$-channel annihilation}
		\label{fig::annihilationb}
		\end{subfigure}
	\caption{Feynman diagrams for the annihilation of two DM particles into two SM quarks at LO. Note that the $u$-channel diagram only exists for Majorana fermions.}
	\label{fig::annihilation}
	\end{figure}

	The relevant diagrams for the DM annihilation process at tree-level are shown in Figure \ref{fig::annihilation}. Evaluating the diagrams we find
	\begin{equation}\label{eq::Mbar2}
	\overline{|M|^2}=\overline{|M_t|^2}+\overline{|M_u|^2}-2\,\overline{M_{tu}}\,,
	\end{equation}	
	for the flavour, colour and spin averaged squared amplitude $\overline{|M|^2}$, where the destructive interference between the $t$- and $u$-channel contributions arises from the crossing of fermion lines.
	The summands in eq.\ \eqref{eq::Mbar2} are given as
	\begin{align}
	\label{eq::amplit}
	\overline{|M_t|^2} &= \frac{1}{9}\cdot \frac{3}{4}\sum_{ij} \sum_{kl} c^t_{ijkl} \frac{(m_{\chi_i}^2+m_k^2-t)(m_{\chi_j}^2+m_l^2-t)}{(t-m_\phi^2)^2}\,,\\
	\label{eq::ampliu}
		\overline{|M_u|^2} &= \frac{1}{9}\cdot \frac{3}{4}\sum_{ij} \sum_{kl} c^u_{ijkl} \frac{(m_{\chi_i}^2+m_l^2-u)(m_{\chi_j}^2+m_k^2-u)}{(u-m_\phi^2)^2}\,,\\
	\label{eq::amplitu}
		\overline{M_{tu}} &=\text{Re}(\overline{M_t^\dagger M_u})=\frac{1}{9}\cdot \frac{3}{4}\sum_{ij} \sum_{kl} c^{tu}_{ijkl} \frac{m_{\chi_i}m_{\chi_j}(s-m_k^2-m_l^2)}{(u-m_\phi^2)(t-m_\phi^2)}\,.
	\end{align} 
	 Here, $s,t$ and $u$ are the Mandelstam variables defined as $s=(p_1+p_2)^2$, $t=(p_1-p_3)^2$ and $u=(p_1-p_4)^2$ and the indices $i,j,k$ and $l$ are flavour indices. The couplings $c^\alpha$ are defined through
	\begin{align}
	c^t_{ijkl} &= |\tilde{\lambda}_{ki}|^2|\tilde{\lambda}_{lj}|^2\,,\\
	c^u_{ijkl} &= |\tilde{\lambda}_{li}|^2|\tilde{\lambda}_{kj}|^2\,,\\
	c^{tu}_{ijkl} &= \text{Re}\left(\tilde{\lambda}_{ki}^* \tilde{\lambda}_{lj}\tilde{\lambda}_{li}^*\tilde{\lambda}_{kj}\right)\,,
	\label{eq::calpha}
	\end{align}	
{with $\tilde{\lambda}$ defined in eq.\ \eqref{eq::lambda-tilde}.}	
	
	Adopting the usual low-velocity expansion for the thermally averaged annihilation cross section \cite{Gondolo:1990dk,Srednicki:1988ce} we write
	\begin{align}
	\langle\sigma v\rangle = a + b\, \langle v^2\rangle + \mathcal{O}\left(\langle v^4\rangle\right)\,,
	\label{eq::cxanni}
	\end{align}
	where $\langle v^2\rangle=6 T_f/m_\chi \approx 0.3$. The coefficients $a$ and $b$ for the s- and p-wave contributions were calculated using the techniques provided in \cite{Gondolo:1990dk,annihilation} and can be found in Appendix \ref{app::partialwave}. In the limit of vanishing final state masses $m_k = m_l =0$ and equal intial state masses $m_{\chi_i}=m_{\chi_j}=m_\chi$ they read
	\begin{align}
	a &= \frac{1}{96 \pi m_\chi^2 \left(1+\mu\right)^2}\sum_{ij} \sum_{kl} \left(c^t_{ijkl}+c^u_{ijkl}-2\,c^{tu}_{ijkl}\right)\,,\\
	\nonumber
	b &=  \frac{1}{1152 \pi m_\chi^2 \left(1+\mu\right)^4} \sum_{ij} \sum_{kl} \biggl(22\, c^{tu}_{ijkl} - 7\, \left(c^t_{ijkl}+c^u_{ijkl}\right)\\
	&{} -18\,\left(c^t_{ijkl}+c^u_{ijkl}-2\,c^{tu}_{ijkl}\right)\mu + \left(c^t_{ijkl}+c^u_{ijkl}+6\,c^{tu}_{ijkl}\right)\mu^2\biggr)\,.
	\label{eq::cxanniab}
	\end{align} 
	where we have additionally defined $\mu = m_\phi^2/m_\chi^2$.\par  
	Using equal initial state masses is evident for the SFF scenario, as only the lightest flavour is present and there is no co-annihilation between different flavours of $\chi$. Note that in this case eq.\@ \eqref{eq::cxanni} does not contain the flavour averaging factor $1/9$ as there is no sum over the initial state flavours $i$ and $j$, and the parameter $m_\chi$ has to be understood as the mass of the lightest DM particle $\chi_3$. It is crucial to also note that in the SFF scenario the couplings $c^\alpha$ reduce to 
	\begin{equation}
	c^t_{33kl} = c^u_{33kl} = c^{tu}_{33kl} = c_{kl} = |\tilde{\lambda}_{k3}|^2|\tilde{\lambda}_{l3}|^2\,.
	\end{equation}
	This leads to a velocity suppresion of the cross section for annihilations into massless final states as one finds 
		\begin{align}
	a &= 0\,,\\
	b &= \sum_{kl} c_{kl}\,\frac{1+\mu^2}{16 \pi m_\chi^2  \left(1+\mu\right)^4}\,,
	\label{eq::cxannibsff}
	\end{align} 
	in this limit. 
	
	In the QDF scenario, on the other hand, all three flavours are present but we demand the mass splittings $\Delta m_{i3}$ to be smaller than $1\%$. Thus, setting $m_{\chi_i}=m_{\chi_j}$ is a very good approximation. We have checked that this in fact only causes a negligibly small difference of order $\mathcal{O}(1\%)$ in the results. It is also important to note that the sum over the final state flavours $k$ and $l$ depends on the value of $m_\chi$. If $m_\chi < m_t$, final states with a top antitop pair are kinematically forbidden. In this case the only allowed final states with top flavour are single-top final states, i.e.\@ the term with $k=l=3$ is excluded from the sum. For even smaller values with $m_\chi<m_t/2$ final states with top flavour are excluded completely and thus one has $k,l\in \{1,2\}$. We also state that eq.\@ \eqref{eq::cxanni} does not contain the additional factor of $1/2$ present in the case of Dirac DM          \cite{tfdm} as the DM particles are Majorana fermions in our model \cite{majoranafactor1,majoranafactor2}.
	\subsection{Analysis of the Relic Abundance Constraints}

	For the numerical analysis of the relic abundance constraints we demand that 
	\begin{equation}
	\langle\sigma v\rangle = 2.2 \times 10^{-26}\,\mathrm{cm}^3/\mathrm{s}\,,
	\end{equation}
	within a $10\%$ tolerance range. This value is adopted from \cite{wimp} and represents the effective annihilation cross section that is necessary to produce the observed relic abundance through a single particle freeze-out for masses $m_\chi> 10\,\mathrm{GeV}$. For both scenarios we include terms up to order $\mathcal{O}\left(\langle v^2\rangle\right)$ in the partial wave expansion of $\langle\sigma v\rangle$. We further set the masses of the light quarks to zero and use $m_t=173.5\,\mathrm{GeV}$.\par
\begin{figure}[!t]
		\centering
		\begin{subfigure}[t]{0.49\textwidth}
		\includegraphics[width=\textwidth]{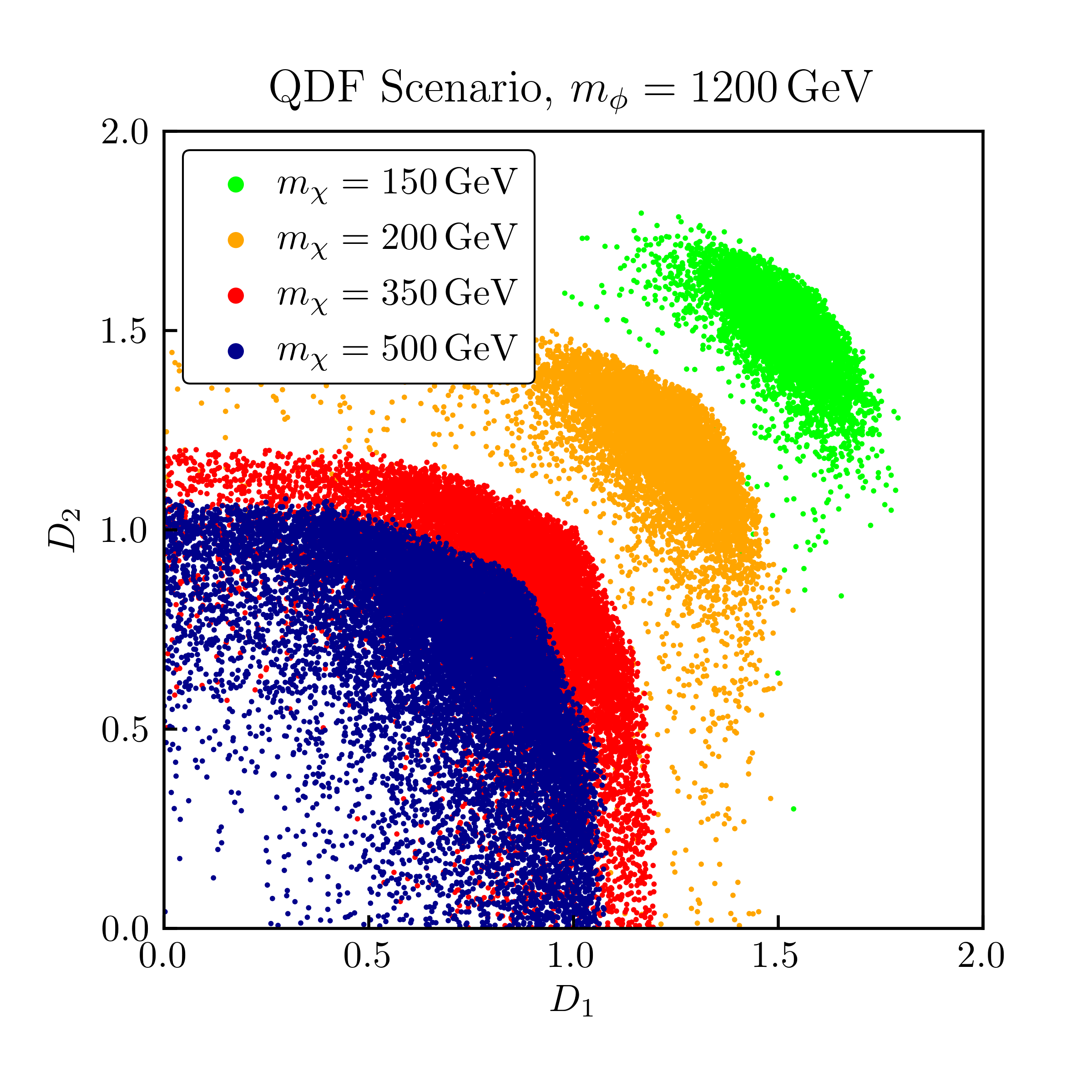}
		\caption{$D_1-D_2$ plane in the QDF scenario for varying $m_\chi$ and $m_\phi =1200\,\mathrm{GeV}$}
		\label{fig::relicabundancea}
		\end{subfigure}
		\hfill
		\begin{subfigure}[t]{0.49\textwidth}
		\includegraphics[width=\textwidth]{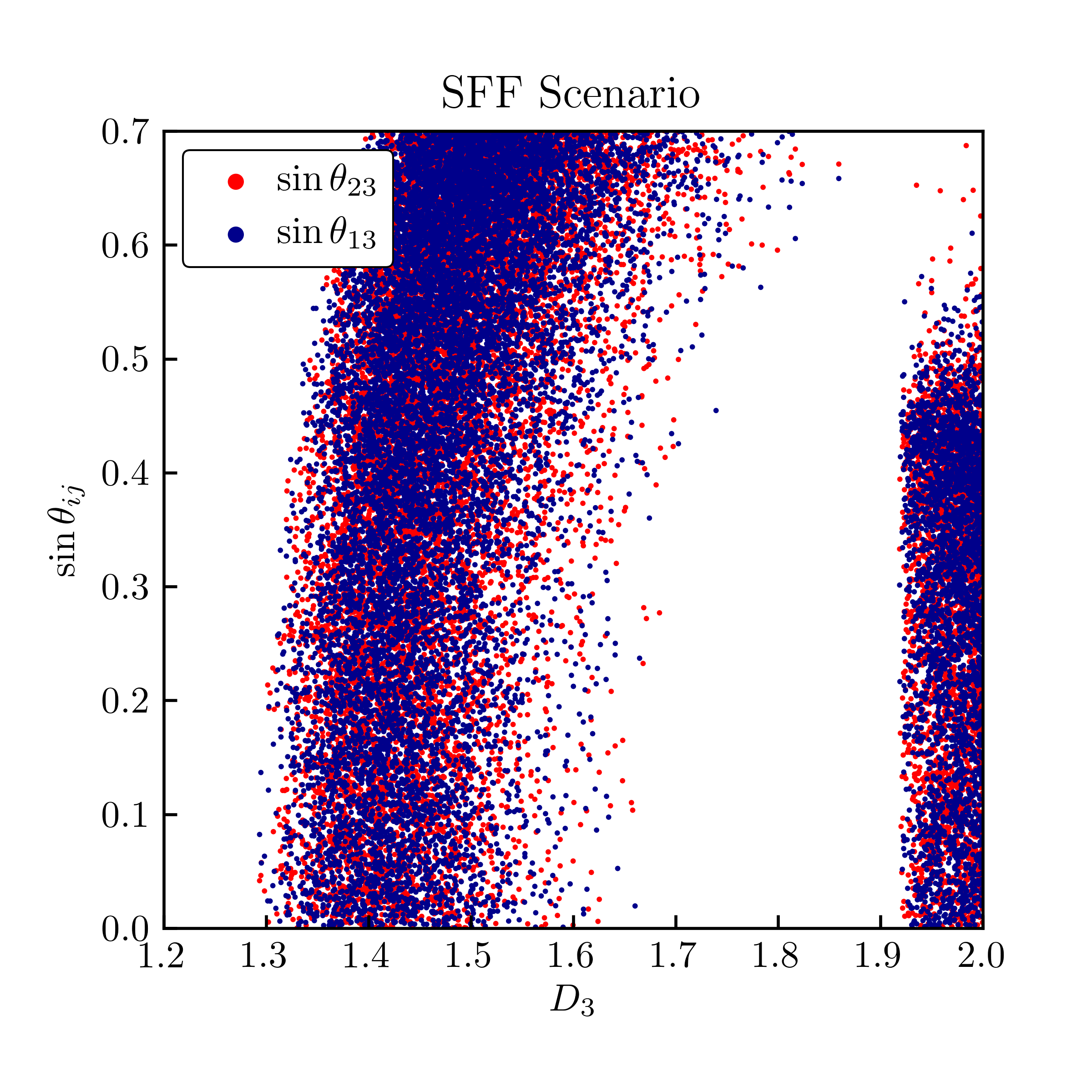}
		\caption{Allowed $\sin\theta_{i3}$ and $D_3$ in the SFF scenario for $m_\chi = 220\,\mathrm{GeV}$ and $m_\phi = 950\,\mathrm{GeV}$}
		\label{fig::relicabundanceb}
		\end{subfigure}
	\caption{Restrictions of the relic abundance constraints on the model parameters for both freeze-out scenarios. } 
	\label{fig::relicabundance}
	\end{figure}

 
	The results are shown in Figure \ref{fig::relicabundance}. The allowed parameter space that we see for the QDF scenario in Figure \ref{fig::relicabundancea} corresponds to the overlap between the allowed parameter space that remains after demanding the QDF mass splitting and after demanding the annihilation cross section to lie in the required range. For masses $m_\chi \gg m_t$ we find
	\begin{align}
	\langle\sigma v\rangle &= \frac{1}{96 \pi}\frac{m_\chi^2}{\left(m_\chi^2+m_\phi^2\right)^2} \sum_{ij} \sum_{kl} \left(c^t_{ijkl}+c^u_{ijkl}-2\,c^{tu}_{ijkl}\right)+ \mathcal{O}\left(\langle v^2\rangle\right)\,,
	\label{eq::cxanniqdfa}
	\end{align}
	where we have only included the leading term of the partial wave expansion.\footnote{Note that this expression is only provided for illustration. For the numerical analysis we have used the full expressions from Appendix \ref{app::partialwave} in both scenarios.} This is due to the fact that in the QDF scenario the s-wave contribution does not vanish as in the SFF scenario. For the sum over the couplings $c^\alpha$ defined in eq.\@ \eqref{eq::calpha} we here find
	\begin{align}
	\sum_{ij}\sum_{kl} c^t_{ijkl}&=\sum_{ij}\sum_{kl} c^u_{ijkl}= \text{Tr} \left[\tilde{\lambda}^\dagger \tilde{\lambda}\right]^2 = \text{Tr}\left[D^2\right]^2\,,\\
	 \sum_{ij}\sum_{kl} c^{tu}_{ijkl}&= \text{Tr} \left[\tilde{\lambda}^T \tilde{\lambda}^* \tilde{\lambda}^\dagger \tilde{\lambda}\right] = \text{Tr}\left[O d^2 O^T D^2 O {d^*}^2 O^T D^2\right]\,.
	\end{align}
	The first trace is trivially found to be
	\begin{equation}
	\text{Tr}\left[D^2\right]^2=\left(D_1^2+D_2^2+D_3^2\right)^2\,,
	\end{equation}	 
	while the second trace is less trivial. It is bounded from above and its maximum is found in the limit of vanishing mixing angles $\phi_{ij}$ and phases $\gamma_i$ (see eqs.\ \eqref{eq::O} and \eqref{eq::Dd} for their definition) to be
	\begin{equation}
	\text{Tr}\left[O d^2 O^T D^2 O {d^*}^2 O^T D^2\right] \leq \text{Tr}\left[D^4\right]=D_1^4+D_2^4+D_3^4\,.
	\end{equation}
	This shows that the leading term in the partial wave expansion given in eq.\@ \eqref{eq::cxanniqdfa} can never vanish due to the destructive interference between the two diagrams shown in Figure \ref{fig::annihilation}, as we find
	\begin{equation}
	\langle\sigma v\rangle \geq \frac{1}{24 \pi}\frac{m_\chi^2}{\left(m_\chi^2+m_\phi^2\right)^2} \left(D_1^2 D_2^2 + D_1^2 D_3^2 + D_2^2 D_3^2\right)+ \mathcal{O}\left(\langle v^2\rangle\right)\,,
	\end{equation}
	for the thermally averaged annihilation cross section.\par  
	In sum, the relic abundance constraint translates to the condition
	\begin{equation}
	\label{eq::relicqdftrace}
	\left(D_1^2+D_2^2+D_3^2\right)^2 - \text{Tr}\left[O d^2 O^T D^2 O {d^*}^2 O^T D^2\right] \approx \text{const.}\,,
	\end{equation}
	for a given point in the $m_\phi - m_\chi$ plane in this scenario.\footnote{We have checked numerically that the trace in eq. \eqref{eq::relicqdftrace} is typically much smaller than the first summand.}
	The circular pattern of Figure \ref{fig::relicabundancea} is due to this condition. The green and orange points in Figure \ref{fig::relicabundancea} show that we encounter the same behaviour in the region $m_\chi \approx m_t$, as well as for $m_\chi < m_t$, where final states with top flavour become inaccessible. The $m_\chi$ dependence of $\langle\sigma v\rangle$ can easily be seen, as small masses $m_\chi$ require large couplings $D_1$ and $D_2$. Another reason for why large couplings are required in the case of small $m_\chi$ is the reduced number of annihilation channels due to the exclusion of final states with top flavour for $m_\chi<m_t$ and $m_\chi< m_t/2$ respectively. Given that we restrict the couplings $D_i$ to $D_i \in \left[0,2\right]$ this poses a lower limit on $m_\chi$. \par
	There is no sum over intial state flavours in eqs.\@ \eqref{eq::amplit}--\eqref{eq::amplitu} in the SFF scenario as the only flavour contributing to the freeze-out is $\chi_3$. In Figure \ref{fig::relicabundanceb} we show the restrictions on the mixing angles $\theta_{13}$ and $\theta_{23}$ in the SFF scenario for the case of top-flavoured DM, i.\,e.\ the lightest flavour $\chi_3$ coupling dominantly to the top quark. We see that for the smallest allowed values of $D_3$ the mixing angles need to be small. Large mixing angles lead to additional contributions to the cross section from annihilations of $\chi_3$ into up- and charm-quarks as well as annihilations into final states with a single top- or antitop-quark, while they at the same time reduce the contribution of annihilations into a top--antitop pair to the total cross section. For annihilations into light quarks we have discussed earlier that the s-wave coefficient $a$ vanishes, and thus these annihilations are velocity-suppressed as they only contribute to the p-wave coefficient $b$. While this is not the case for annihilations into final states with a single top or antitop quark, such annihilations are still suppressed due to the massless quark in the final state. We find these contributions to the s-wave coefficient $a$ to be roughly three orders of magnitude smaller than the contribution of annihilations into a top--antitop pair. As increasing mixing angles $\theta_{i3}$ reduce the latter contributions while they only generate suppressed new contributions, they in total lead to a smaller annihilation cross section. This explains the necessity to have small mixing angles for the smallest allowed values of $D_3$. With a growing coupling $D_3$ larger mixing angles become allowed due to the  reasons explained above, as the total annihilation cross section grows with $D_3$. For values $D_3 \GtrSim 1.65$ the cross section tends to be too large and hence we need large mixing angles in order to reduce it and push it into the tolerance interval. It is important to note that the suppression through mixing angles becomes weaker for growing couplings $D_3$, as the contributions to $a$ from single top/antitop final states and the contributions to $b$ from massless and single top/antitop final states also grow. 	Thus, for values $D_3 \GtrSim 1.80$ the annihilation cross section grows too large and the constraint cannot be satisfied as long as annihilations into a top--antitop pair are possible. 
	Finally, for $D_3 \GtrSim 1.92$ a large range of mixing angles $\theta_{i3}$ becomes allowed again. The explanation can be found in the dependende of the DM mass $m_{\chi_3}$ on the coupling $D_3$. 	In the limit of small mixing angles $\phi_{ij}$ and phases $\gamma_i$ the mass corrections for $\chi_3$ are  $m_{\chi_3} \approx m_\chi \left(1- |\eta| D_3^2\right)$, and thus $m_{\chi_3}$ decreases with increasing $D_3$. For $D_3 \GtrSim 1.92$ final states with a top--antitop pair are kinematically forbidden.  Therefore, only very large mixing angles $\theta_{i3}$ are excluded in this case as they would further enhance the accessible annihilation channels and lead to a too high annihilation rate.\par 
	As we do not fix the values of $\sin\theta_{13}$ while looking at $\sin\theta_{23}$ and vice versa, we see no difference for these two mixing angles in Figure \ref{fig::relicabundanceb}. Recall that we allow for a rearrangement of the diagonalization matrix $W$ of eq.\@ \eqref{eq::takagi} in order to obtain a hierarchy such that the third generation is always the lightest. As this at the same time determines the flavour of $\chi_3$, its mass corrections may also depend on $D_1$ or $D_2$ and the respective mixing angles if $\chi_3$ has up or charm flavour. In these cases the parameters shown in Figure \ref{fig::relicabundanceb} are basically free, with the only restriction of a sufficiently small $D_3$ in order to fulfil the SFF mass splitting condition.\par
\begin{figure}[t]
	\centering
	\includegraphics[width=0.75\textwidth]{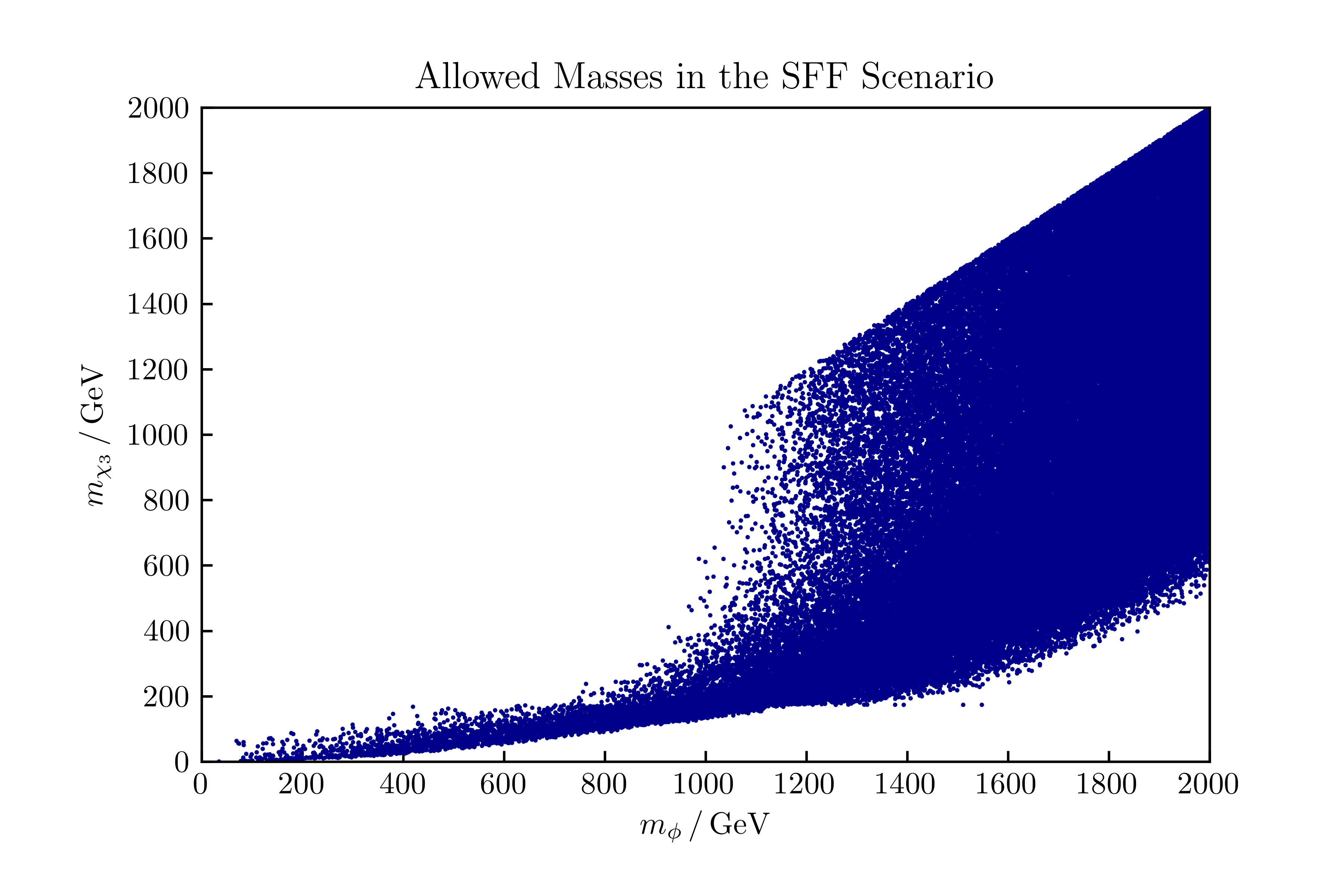}
	\caption{Allowed points in the $m_\phi - m_{\chi_3}$ plane for the SFF scenario.}
	\label{fig::sffmasses}
	\end{figure}

	Like in the QDF scenario, also in the SFF scenario there is a lower bound on $m_{\chi_3}$ due to the upper limit $D_i \leq 2.0$ we chose in our analysis. Additionally, for the SFF scenario we also find an upper bound on $m_{\chi_3}$ as can be seen in Figure \ref{fig::sffmasses}. This upper bound can be explained through eq.\@ \eqref{eq::cxannibsff}. For large values of $m_{\chi_3}$ the couplings $\tilde{\lambda}_{i3}$ have to be correspondingly small in order to keep the annihilation cross section in the allowed interval. At the same time the mass splitting of the SFF scenario forces one $\tilde{\lambda}_{i3}$ to be large. For too large $m_{\chi_3}$ it is not possible to maintain the SFF mass splitting while at the same time fulfilling the relic abundance bound, and thus we observe an upper limit. As the mediator mass $m_\phi$ suppresses the annihilation cross section, the allowed region for $m_{\chi_3}$ grows with $m_\phi$. For values $m_\phi \GtrSim 1000\,\mathrm{GeV}$ this suppression dominates and values up to the equal mass threshold $m_{\chi_3}=m_\phi$ become viable.

\section{Dark Matter Phenomenology}\label{sec:dd}

The Majorana nature of the DM field $\chi$ has profound implications on its signatures in direct detection experiments.
 The DM--nucleon scattering  generally splits up into a spin-dependent and spin-independent part in the non-relativistic limit, where constraints on the latter are typically stronger due to a coherent scattering off all nucleons in the nucleus. In the spin-dependent case there is no such enhancement, since the DM particle couples to the modulus of the total spin \cite{agrawalspin} and the nucleon spins cancel in pairs.
 Hence,  it is sufficient to only consider the spin-independent part of the scattering cross section, unless it is suppressed, and neglect the spin-dependent contributions \cite{dmfv,tfdm,ldm}. However, for our case of DM with Majorana nature and a chiral coupling the leading dimension-six operators for spin-independent scattering $\bar{\chi}\chi\bar{q} q$ and $(\bar{\chi}\gamma^\mu\chi)(\bar{q} \gamma_\mu q)$ are identically zero due to the coupling structure or since for Majorana particles bilinears which are anti-symmetric under $C$ parity vanish, respectively \cite{ddvogl,taitdd,agrawalspin}. Thus, it is not only necessary to also include the spin-dependent cross section into the analysis, but also to go beyond leading order and include loop-induced scattering between DM and gluons for the spin-independent part. In Figure \ref{fig::diagramsdd} we show representative Feynman diagrams for the relevant processes. 
 
 In addition to the contributions discussed here,  there is also a one-loop Higgs penguin contribution to spin-independent scattering. The impact of such a contribution has been discussed in the context of neutralino DM in the MSSM in \cite{Drees:1996pk,Djouadi:2001kba}. The size of the effective $\bar \chi_3\chi_3 h $ vertex is determined by the coupling parameter $\lambda_{H\phi}$ in eq.\ \eqref{eq::lagrangian} as well as the SM Yukawa coupling of the quark flavour running in the loop. Hence, while in general the Higgs penguin can yield a relevant contribution to the spin-independent DM scattering off nuclei, it is always possible to suppress this contribution by the proper choice of $\lambda_{H\phi}$. Since the latter coupling is not constrained by the rest of our analysis, we use this freedom to assume the Higgs penguin contribution to be negligible.

\subsection{Dark-Matter--Nucleon Scattering Processes}
As already mentioned above, the DM--nucleon scattering cross section splits up into a spin-dependent part \cite{ddsd,taitdd,ddvogl}
\begin{equation}
\sigma_\text{SD}^N = \frac{3}{16 \pi} \frac{m_N^2 m_\chi^2}{(m_N+m_\chi)^2} \left(\sum_{q=u,d,s} \Delta q^N a_q\right)^2\,,
\end{equation}
and a spin-independent part \cite{taitdd,ddvogl,agrawalspin}
\begin{equation}
\sigma_\text{SI}^N = \frac{4}{\pi} \frac{m_N^2 m_\chi^2}{(m_N+m_\chi)^2}\, |f_N|^2\,,
\label{eq::spinindep}
\end{equation}
where $a_q$ is the Wilson coefficient of the spin-dependent quark-nucleon interaction, $\Delta q^N$ is the spin content of the nucleon $N=\{p,n\}$ in terms of the quark $q$ and $m_N$ is its mass. The factor $f_N$ is the matrix element of the spin-independent quark-nucleon interaction. \par

\begin{figure}[!t]
	\centering
		\begin{subfigure}[t]{0.3\textwidth}
		\includegraphics[width=\textwidth]{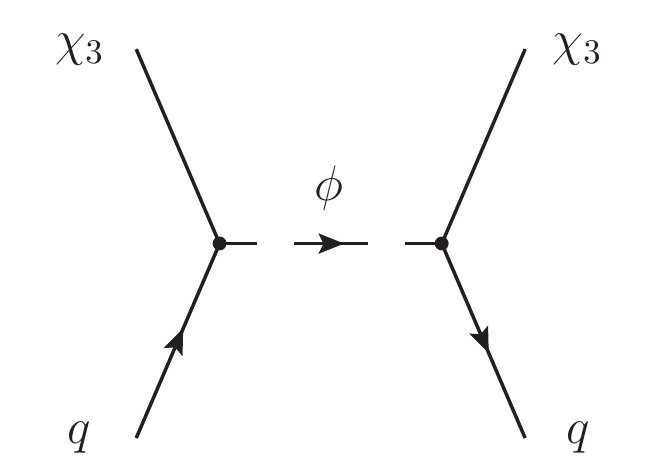}
		\end{subfigure}
		\hfill
		\begin{subfigure}[t]{0.3\textwidth}
		\includegraphics[width=\textwidth]{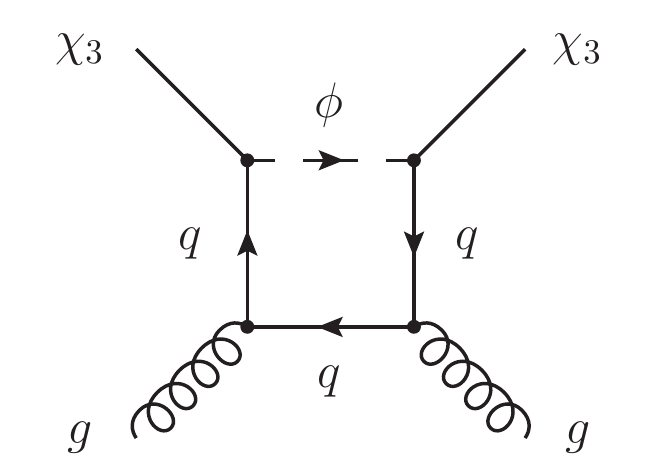}
		\end{subfigure}
		\hfill
		\begin{subfigure}[t]{0.3\textwidth}
		\includegraphics[width=\textwidth]{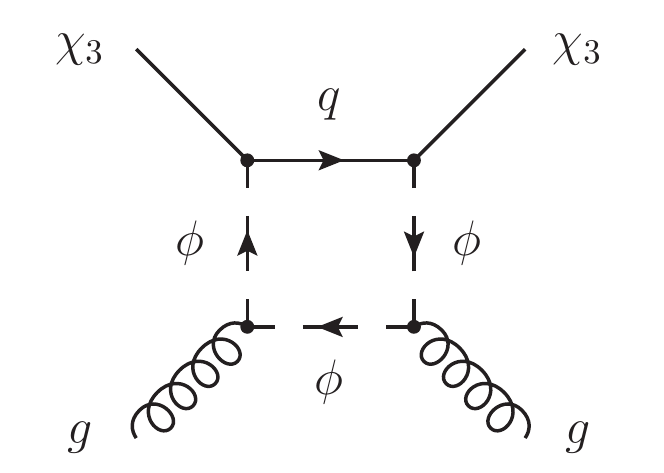}
		\end{subfigure}
		\hfill\\
		\begin{subfigure}[t]{0.3\textwidth}
		\includegraphics[width=\textwidth]{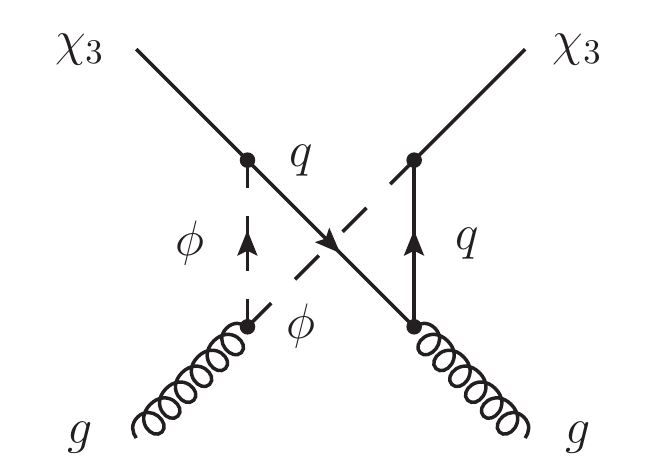}
		\end{subfigure}
		\hspace{1cm}
		\begin{subfigure}[t]{0.3\textwidth}
		\includegraphics[width=\textwidth]{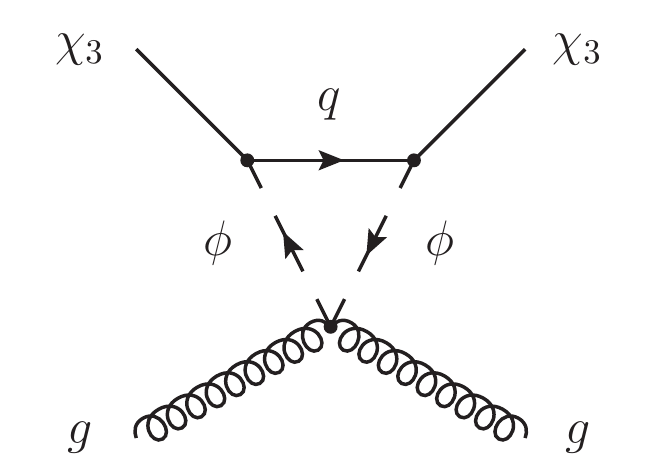}
		\end{subfigure}
		\caption{Tree-level and loop induced Feynman diagrams for DM--nucleon scattering.}
		\label{fig::diagramsdd}
\end{figure}
As the NP only couples to up-type quarks in our model, the expression for $\sigma^N_\text{SD}$ becomes 
 \begin{equation}
\sigma_\text{SD}^N = \frac{3}{16 \pi} \frac{m_N^2 m_\chi^2}{(m_N+m_\chi)^2} \left(\Delta u^N a_u\right)^2\,,
\label{eq::spindep}
\end{equation}
with 
\begin{equation}
a_u = \frac{|\tilde{\lambda}_{u3}|^2}{m_\phi^2-(m_\chi+m_u)^2}\,.
\label{eq::ddsdwilson}
\end{equation}  
For $\sigma_\text{SI}$ we follow the formalism of \cite{Drees:1993bu,ddSIformalism1,ddSIformalism2} and write the effective interaction Lagrangian as
\begin{equation}
\mathcal{L}_\text{SI}^\text{eff}= \sum_{q=u,d,s,c} \mathcal{L}_q^\text{eff}+\mathcal{L}_g^\text{eff}\,.
\end{equation}
Here, $\mathcal{L}_q^\text{eff}$ describes the scattering between DM and quarks, and reads
\begin{equation}
\mathcal{L}_q^\text{eff} = f_q \bar{\chi}\chi\,\mathcal{O}^{(0)}_{q} + \frac{g_q^{(1)}}{m_\chi}\bar{\chi}i(\partial^\mu\gamma^\nu+\partial^\nu\gamma^\mu)\chi \mathcal{O}^{(2)}_{q,\mu\nu}+ \frac{g_q^{(2)}}{m_\chi^2} \bar{\chi}(i\partial^\mu)(i\partial^\nu)\chi \mathcal{O}^{(2)}_{q,\mu\nu}\,,
\end{equation}
while the loop-induced scattering between gluons and DM is described by
\begin{equation}
\mathcal{L}_g^\text{eff} = f_G \bar{\chi}\chi\,\mathcal{O}^{(0)}_{g} + \frac{g_G^{(1)}}{m_\chi}\bar{\chi}i(\partial^\mu\gamma^\nu+\partial^\nu\gamma^\mu)\chi \mathcal{O}^{(2)}_{g,\mu\nu}+ \frac{g_G^{(2)}}{m_\chi^2} \bar{\chi}(i\partial^\mu)(i\partial^\nu)\chi \mathcal{O}^{(2)}_{g,\mu\nu}\,.
\end{equation}
Using the notation of \cite{taitdd} the tensor operators can be written as
\begin{align}
\nonumber
\mathcal{O}^{(2)}_{q,\mu\nu} &= \frac{1}{2} \bar{q}\left(\gamma^{\{\mu}iD_-^{\nu\}}-\frac{g^{\mu\nu}}{4}i\slashed{D}_-\right)\,,\\
\mathcal{O}^{(2)}_{g,\mu\nu} &= -G^{a,\mu\rho}G^{a,\nu}_{\quad\rho}+\frac{g^{\mu\nu}}{4} \left(G^a_{\alpha\beta}\right)^2\,,
\end{align}
and the scalar operators are defined as
\begin{align}
\nonumber
\mathcal{O}^{(0)}_q &= m_q \bar{q}q\,,\\
\mathcal{O}^{(0)}_g &= G^a_{\mu\nu}G^{a,\mu\nu}\,.
\end{align}
In this formalism, the matrix element of the spin-independent scattering process between a DM particle $\chi$ and a nucleon $N$ is given by
\begin{eqnarray}
\nonumber
\frac{f_N}{m_N} &=& \sum_{q=u,d,s,c} f_{Tq} f_q + \frac{3}{4} \left[q(2)+\bar{q}(2)\right]\left(g_q^{(1)}+g_q^{(2)}\right)\\
&&-\frac{8\pi}{9\alpha_s}f_{T_G}f_G+\frac{3}{4} G(2)\left(g_G^{(1)}+g_G^{(2)}\right)\,.
\label{eq::spinindepfN}
\end{eqnarray}
Here, $f_{Tq}$ are the mass fractions of light quarks in the nucleon, and $q(2)$, $\bar q(2)$ and $G(2)$ are the second moments of parton distribution functions of quarks, antiquarks and gluons, respectively.
The Wilson coefficients of the DM--quark interaction read \cite{taitdd}  
\begin{align}
\nonumber
f_q &= \frac{|\tilde{\lambda}_{q3}|^2 m_\chi}{16(m_\phi^2-(m_\chi+m_q)^2)^2}\,,\\
\nonumber
g^{(1)}_q &= \frac{|\tilde{\lambda}_{q3}|^2 m_\chi}{8(m_\phi^2-(m_\chi+m_q)^2)^2}\,,\\
g^{(2)}_q &=0\,.
\end{align}
Note that these coefficients only arise through NLO terms in the expansion of the propagator in the tree-level diagram in Figure \ref{fig::diagramsdd}, and thus they are additionally suppressed by a factor of $(m_\phi^2-(m_\chi+m_q)^2)$ when compared to $a_u$. The Wilson coefficients $f_G$, $g_G^{(1)}$ and $g_G^{(2)}$ of the gluonic operators $\mathcal{O}^{(0)}_{g}$ and $\mathcal{O}^{(2)}_{g,\mu\nu}$ can be found in Appendix \ref{app::ddwilson}.\par
Using the results of \cite{runningdd} we evolve the Wilson coefficients for spin-independent scattering from the new physics scale $m_\phi$ down to the scattering scale $\mu = 2\,\mathrm{GeV}$. The anomalous dimension and matching matrices can be found in \cite{runningdd}. For the running and decoupling of quark masses and the strong coupling $\alpha_s$ we use the \texttt{RunDec} package \cite{rundec}. The numerical values of the hadronic matrix elements as well as the input quark masses for the RG running can be found in the appendix of \cite{taitdd}. 

\subsection{Direct Detection Constraints} 

 In order to determine the constraints from direct detection experiments on the parameter space of our model, we  calculate the spin-dependent as well as the spin-independent scattering cross sections and compare them to the experimental upper bounds. The strongest constraints on spin-dependent scattering are provided by the PICO-60 experiment \cite{pico60} and are obtained for WIMP-proton scattering. For spin-independent scattering the world-leading result is provided by the XENON1T \cite{xenon} experiment.\par
\begin{figure}[!b]
		\centering
		\begin{subfigure}[t]{0.49\textwidth}
		\includegraphics[width=\textwidth]{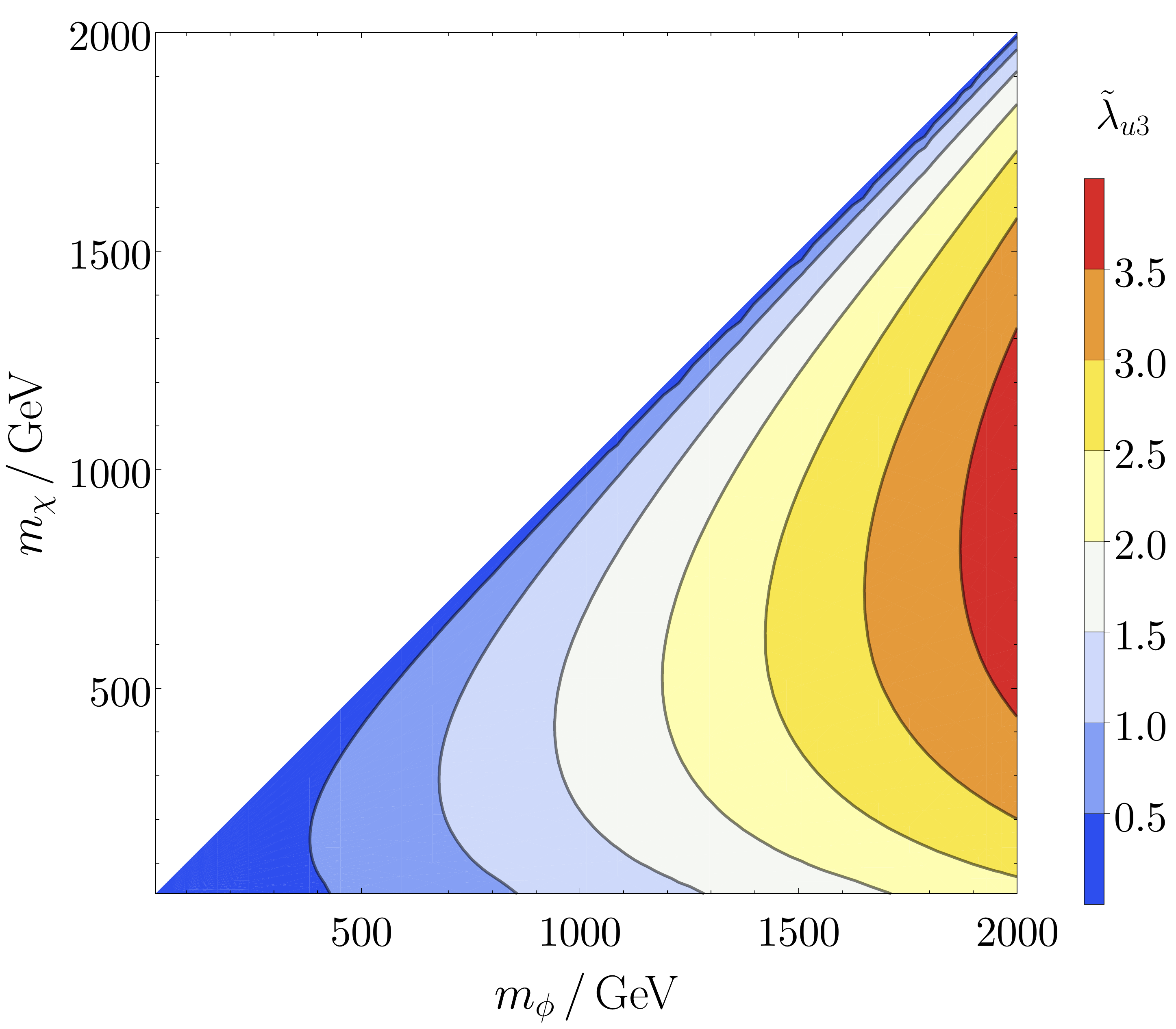}
		\caption{Spin-dependent scattering}
		\label{fig::ddctplta}
		\end{subfigure}
		\hfill
		\begin{subfigure}[t]{0.49\textwidth}
		\includegraphics[width=\textwidth]{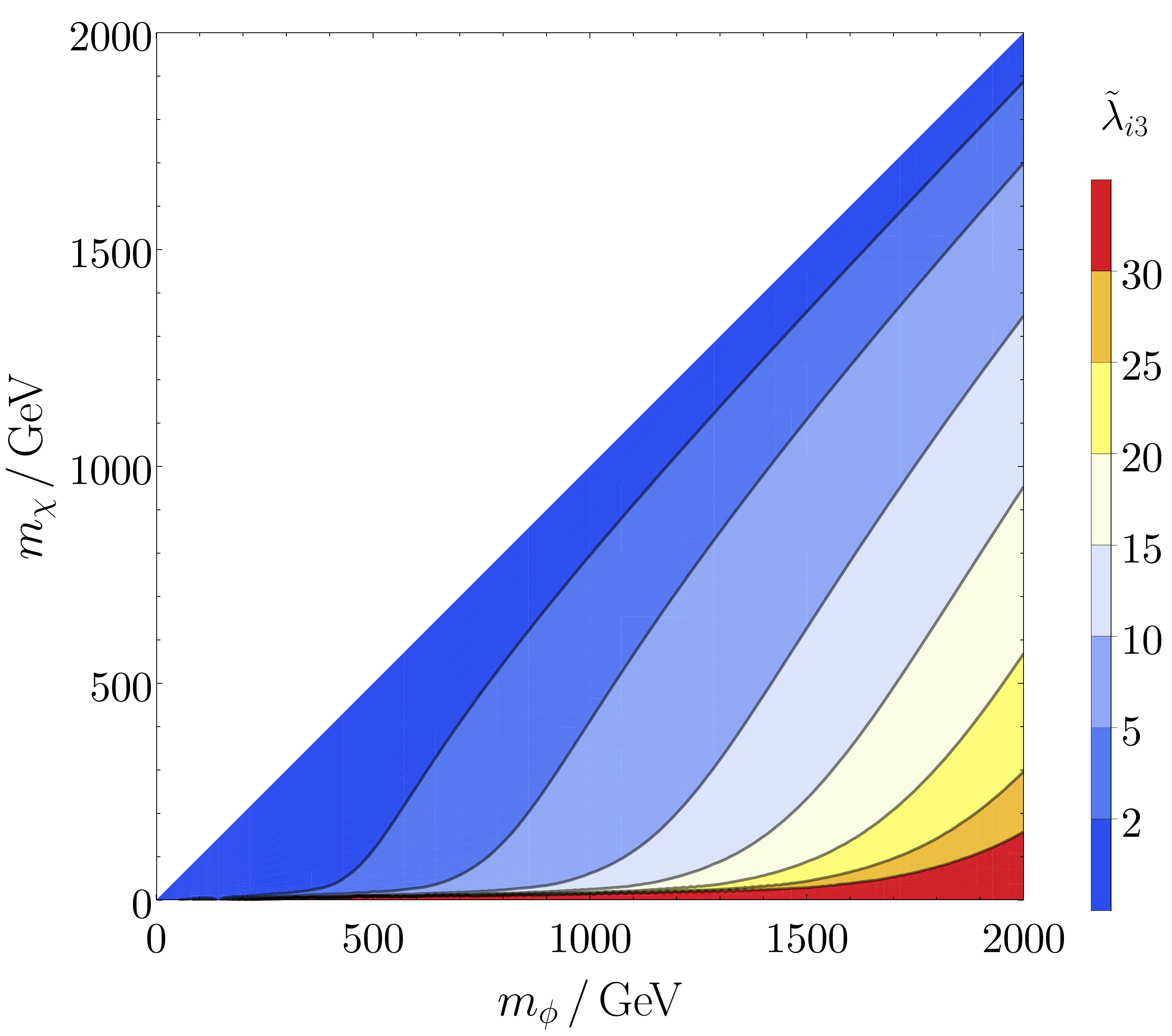}
		\caption{Spin-independent scattering}
		\label{fig::ddctpltb}
		\end{subfigure}
	\caption{Direct detection limits on $\tilde{\lambda}$ from the PICO-60 and XENON1T experiments. For the latter we have set $\tilde{\lambda}_{u3}=\tilde{\lambda}_{c3}=\tilde{\lambda}_{t3}\equiv\tilde{\lambda}_{i3}$.} 
	\label{fig::ddctplt}
	\end{figure}

As the PICO-60 experiment provides limits on spin-dependent WIMP-proton scattering, eq.~\eqref{eq::spindep} becomes \cite{ddsd,taitdd}
\begin{equation}
\sigma_\text{SD}^p = \frac{3}{16 \pi} \frac{m_p^2 m_\chi^2}{(m_p+m_\chi)^2} \left(\Delta u^p a_u\right)^2\,.
\end{equation}
The spin-independent WIMP-nucleon scattering cross section is obtained by summing over all nucleons in the nucleus in eq.\@ \eqref{eq::spinindep} and reads \cite{dmfv,tfdm}
\begin{equation}
\sigma_\text{SI}^N = \frac{4 \mu^2}{\pi A^2} |Z f_p + (A-Z) f_n|^2\,, 
\end{equation}
where the reduced mass is defined as $\mu = m_N m_\chi / (m_N + m_\chi)$. Demanding that these cross sections are below the experimental upper bounds yields limits on the couplings $\tilde{\lambda}_{u3}$, $\tilde{\lambda}_{c3}$ and $\tilde{\lambda}_{t3}$. The results are shown in Figure \ref{fig::ddctplt}.

As we only consider the dominant  tree-level contribution to spin-dependent scattering, the PICO-60 constraints solely apply to $\tilde{\lambda}_{u3}$. In Figure \ref{fig::ddctplta} we see that for large parts of the $m_\phi-m_\chi$ plane the constraints can be completely evaded. For values $m_\phi \GtrSim 1200\,\mathrm{GeV}$ one can always find an $m_\chi$ such that the coupling to up-quarks can grow as large as $\tilde{\lambda}_{u3}=2.0$. As we had restricted the parameters $D_i$ to lie in the range $[0,2]$ to avoid perturbativity issues, this corresponds to the largest possible value the couplings $\tilde{\lambda}_{i3}$ can take. For even larger values $m_\phi \GtrSim 1500\, \mathrm{GeV}$, the constraints only become relevant in the resonance region $m_\chi \approx m_\phi$ and for small DM masses $m_\chi \LessSim 100\, \mathrm{GeV}$. 

In order to analyse the constraints from spin-independent scattering we use the very simple benchmark scenario of a flavour-universal coupling, i.e.\@ we set $\tilde{\lambda}_{u3}=\tilde{\lambda}_{c3}=\tilde{\lambda}_{t3}\equiv\tilde{\lambda}_{i3}$. The results are shown in Figure \ref{fig::ddctpltb}. As can be seen, the constraints are even more lenient in this case. This was to be expected, as the relevant cross section receives contributions from the dimension-seven and dimension-eight operators given above and is thus highly suppressed by the new physics scale $m_\phi$. Only in the resonance region with a mass splitting of at most $10\,\%$ between $m_\phi$ and $m_\chi$ the limits become relevant and force the couplings to lie in the range $\tilde{\lambda}_{i3} \leq 2.0$.\par 

We conclude that the direct detection constraints are less stringent than the ones considered in the previous sections and can even be completely evaded over large parts of the parameter space. Due to the Majorana nature of $\chi$ they are mostly dominated by spin-dependent scattering.

 	\section{Combined Analysis}
	\label{sec::combined}
 	In this section we combine our previous results by analysing the validity of our model within the context of all constraints imposed simultaneously. The remaining allowed regions will then yield a global picture of the viable parameter space. We  further use this section to analyse the flavour of the DM particle.

	\subsection{Combined Constraints}
\begin{figure}[!b]
		\centering
		\begin{subfigure}[t]{0.49\textwidth}
		\includegraphics[width=\textwidth]{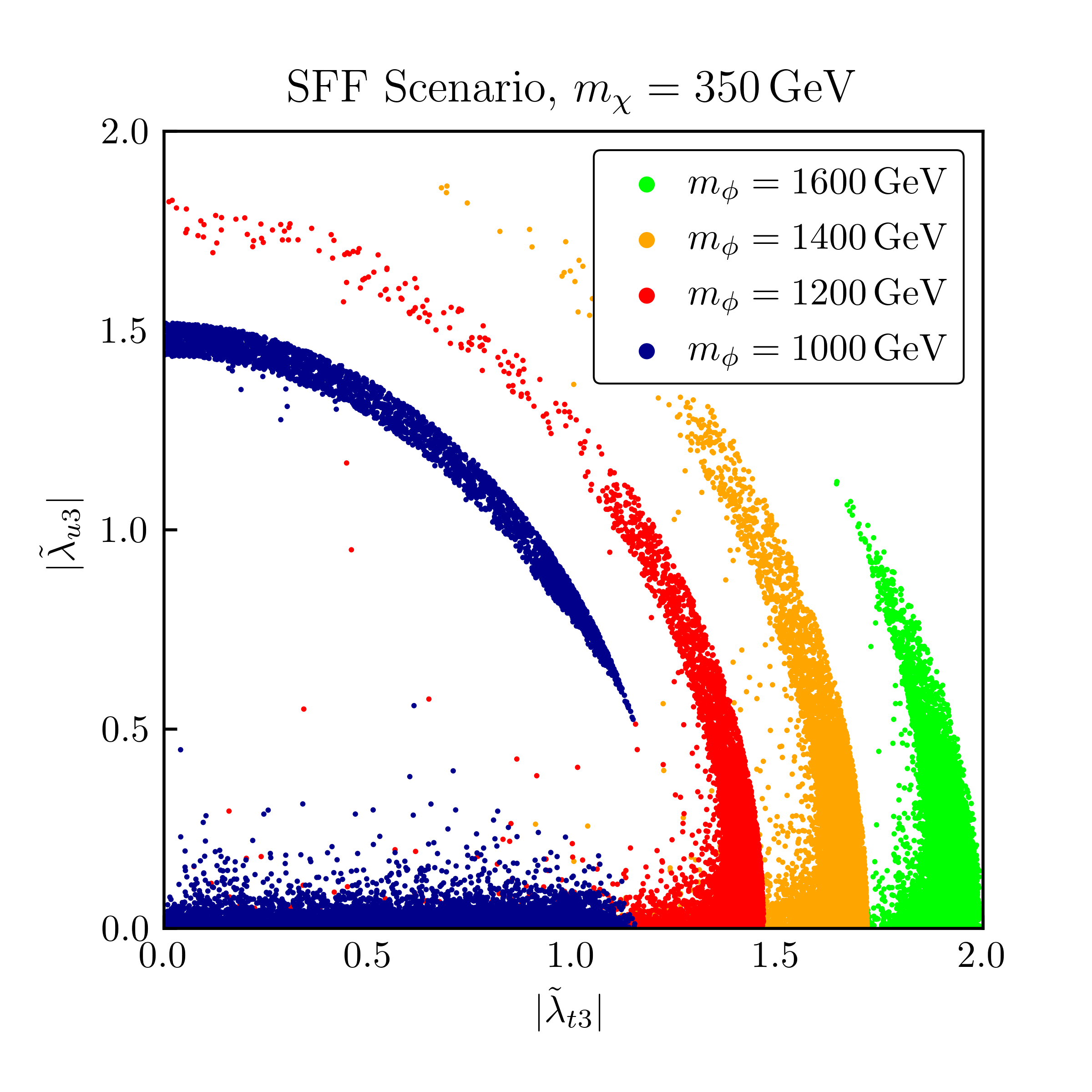}
		\caption{$|\tilde{\lambda}_{t3}|-|\tilde{\lambda}_{u3}|$ plane}
		\label{fig::cbdl33vsl13sff}
		\end{subfigure}
		\hfill
		\begin{subfigure}[t]{0.49\textwidth}
		\includegraphics[width=\textwidth]{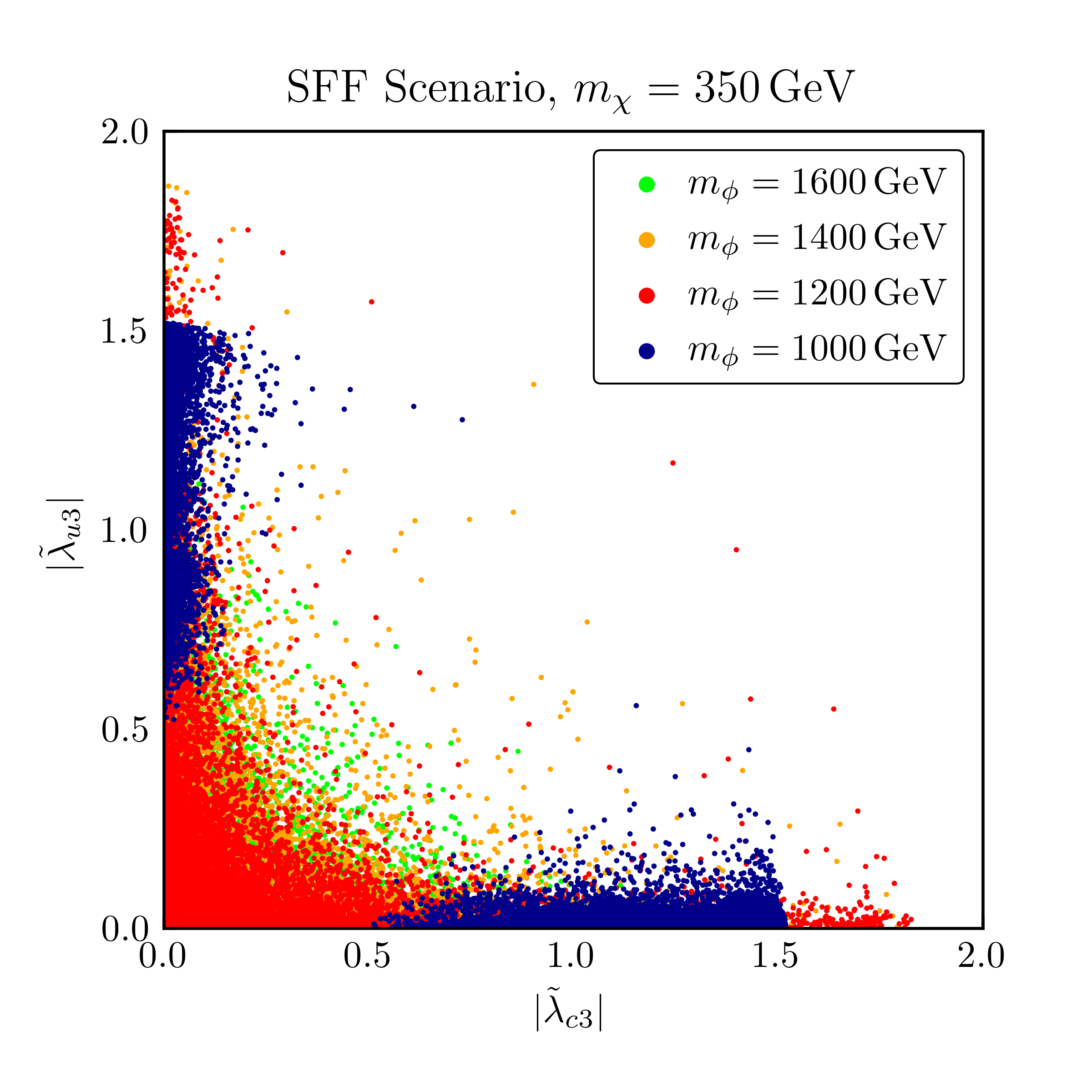}
		\caption{$|\tilde{\lambda}_{c3}|-|\tilde{\lambda}_{u3}|$ plane }
		\label{fig::cbdl23vsl13sff}
		\end{subfigure}
	\caption{Viable couplings $|\tilde{\lambda}_{i3}|$ for $m_\chi =350\,\mathrm{GeV}$ and varying $m_\phi$ within the context of all constraints in the SFF scenario.} 
	\label{fig::cbdsff}
	\end{figure}

\begin{figure}[!b]
		\centering
		\begin{subfigure}[t]{0.49\textwidth}
		\includegraphics[width=\textwidth]{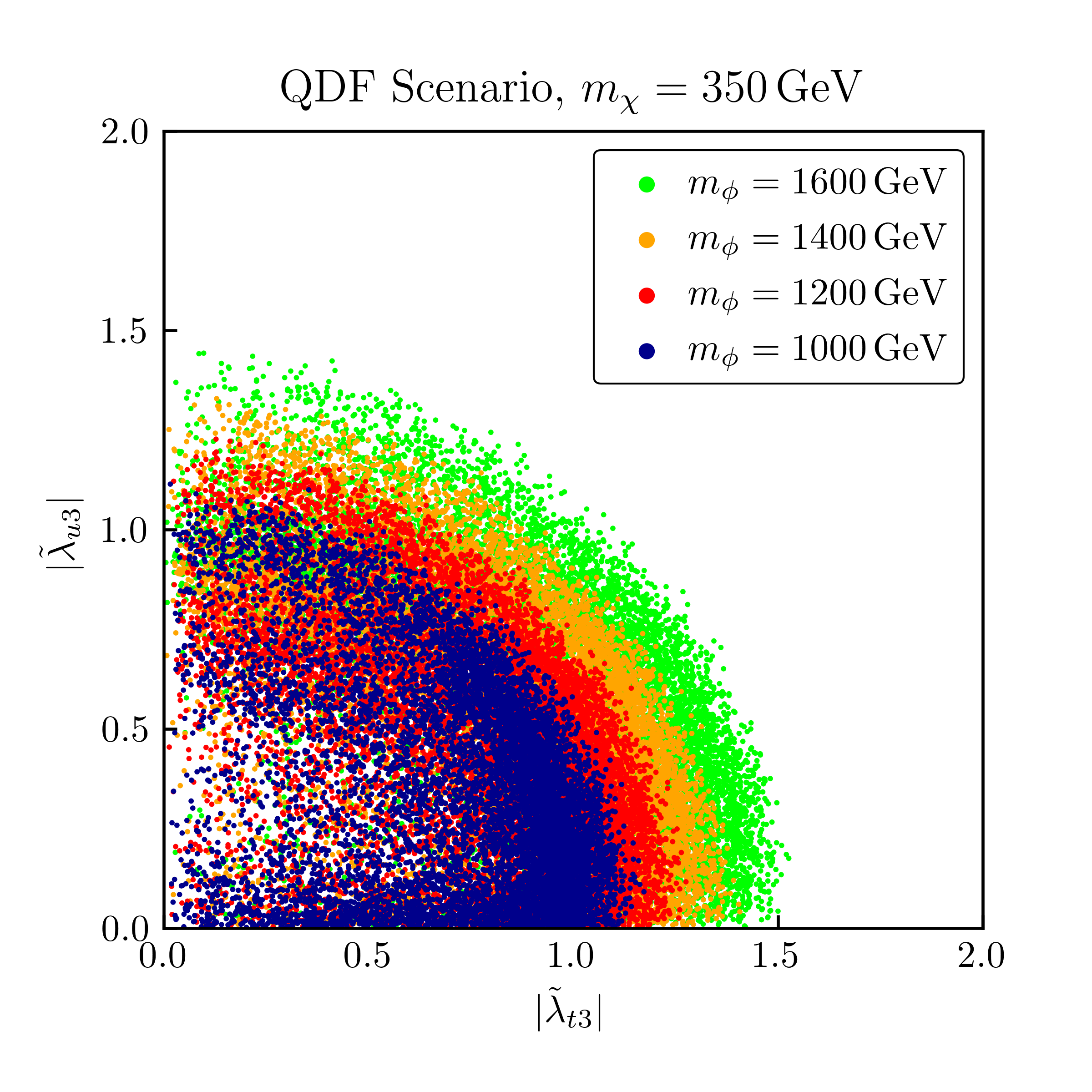}
		\caption{$|\tilde{\lambda}_{t3}|-|\tilde{\lambda}_{u3}|$ plane}
		\label{fig::cbdl33vsl13qdf}
		\end{subfigure}
		\hfill
		\begin{subfigure}[t]{0.49\textwidth}
		\includegraphics[width=\textwidth]{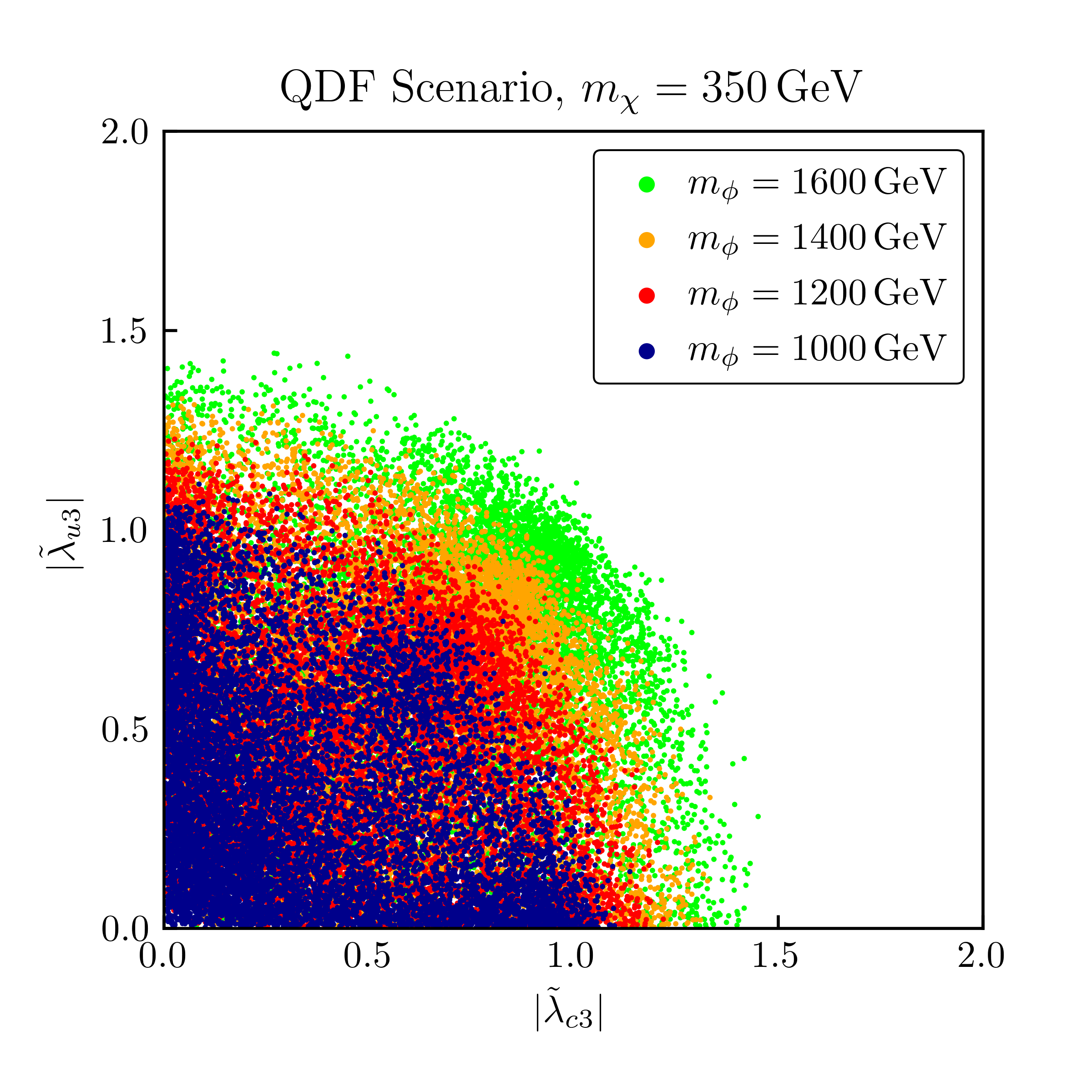}
		\caption{$|\tilde{\lambda}_{c3}|-|\tilde{\lambda}_{u3}|$ plane}
		\label{fig::cbdl23vsl13qdf}
		\end{subfigure}
	\caption{Viable couplings $|\tilde{\lambda}_{i3}|$ for $m_\chi =350\,\mathrm{GeV}$ and varying $m_\phi$ within the context of all constraints in the QDF scenario.} 
	\label{fig::cbdqdf}
	\end{figure}

	The results of the combined application of all constraints are shown in Figure \ref{fig::cbdsff} and Figure \ref{fig::cbdqdf}. Note that the LHC constraints are considered in form of the choices for the masses $m_\phi$ and $m_\chi$. In both freeze-out scenarios the allowed parameter space is mainly determined by the flavour and relic density constraints. The structure of the coupling matrix $\lambda$ is also restricted by the choice of the freeze-out scenario, as the hierarchy in $\lambda$ drives the mass splittings between the dark flavours $\chi_i$.
	
	 In Figure \ref{fig::cbdl33vsl13sff} we show the $|\tilde{\lambda}_{t3}|- |\tilde{\lambda}_{u3}|$ plane for the SFF scenario, i.\,e.\ the coupling strengths of the DM particle to the top and up quark. Likewise, Figure \ref{fig::cbdl23vsl13sff} displays the $|\tilde{\lambda}_{c3}|-|\tilde{\lambda}_{u3}|$ plane.	 
	 As already discussed in Section \ref{sec::relicabundance} the relic abundance constraints reduce the allowed parameter space to the spherical condition
\begin{equation}\label{eq::9d-sphere}
	\sum_{ij}\sum_{kl} |\tilde{\lambda}_{ki}|^2|\tilde{\lambda}_{lj}|^2 \approx \text{const.}\,,
	\end{equation}
for a given pair of $m_\phi$ and $m_\chi$. In general, this condition restricts the couplings $|\tilde{\lambda}_{ij}|$ to lie on a nine-dimensional sphere. However, in the SFF scenario, the sum over the intial state flavours is omitted as only $\chi_3$ is present at freeze-out and the condition in eq.\@ \eqref{eq::9d-sphere} thus reduces to 
\begin{equation}
 |\tilde{\lambda}_{u3}|^2 +|\tilde{\lambda}_{c3}|^2 + |\tilde{\lambda}_{t3}|^2 \approx \text{const.}
\end{equation}
The $D^0-\bar{D}^0$ mixing constraints then further force either $\tilde{\lambda}_{u3}$ or $\tilde{\lambda}_{c3}$ to be small while $\tilde{\lambda}_{t3}$ can be chosen freely, leading to the circular bands of Figure \ref{fig::cbdl33vsl13sff}. The former can also be seen in Figure \ref{fig::cbdl23vsl13sff}, leading to most of the allowed parameter points located close to the axes, with either $|\tilde{\lambda}_{u3}|\simeq 0$ or $|\tilde{\lambda}_{c3}|\simeq 0$. The points with $|\tilde{\lambda}_{u3}|\simeq 0$ are then scattered at the bottom of Figure \ref{fig::cbdl33vsl13sff}, while the points with $|\tilde{\lambda}_{c3}|\simeq 0$ form the circular bands.

The remaining features in Figure \ref{fig::cbdl33vsl13sff} can be understood by considering the  DM annihilation cross section relevant for the thermal freeze-out.
 As the annihilation cross section in eq.\@ \eqref{eq::cxanni} is strongly suppressed by powers of $m_\phi$, low mediator masses require the DM particle to be mainly up- or charm-flavoured in the SFF scenario. This is due to the fact that the velocity suppresion of annihilation channels with up- or charm-flavour in the final state is sufficiently large to compensate the cross section enhancement by the small mediator mass in the denominator.
  In Figure \ref{fig::cbdl33vsl13sff} this can be seen explicitly for a mediator mass of $m_\phi = 1000\,\mathrm{GeV}$. For larger $m_\phi$ the annihilation cross section becomes sufficiently suppressed such that sizeable contributions to annihilation channels with top-flavour in the final state, which are not velocity suppressed, are allowed. 
At even larger values for $m_\phi$ up- and charm-flavoured DM becomes forbidden, as the above-mentioned velocity suppression of annihilations into up- or charm-quarks in combination with a high $m_\phi$-suppression of the annihilation cross section yields a too small relic density. In this case, only top-flavoured DM is viable. 
		
In Figure \ref{fig::cbdqdf} we show analogous plots for the QDF scenario, with fixed $m_\chi=350\,\text{GeV}$ and various values of $m_\phi$. The emerging patterns are much less clear in this case. The reason is that due to the quasi-degeneracy of the DM flavours, the entries in the three columns of $\tilde\lambda$ have to be of similar size. As a result, the plots in Figure \ref{fig::cbdqdf} are two-dimensional projections of  the nine-dimensional coupling parameter space $|\tilde\lambda_{ij}|$. Again we observe the circular bands in the allowed parameter space in Figure \ref{fig::cbdl33vsl13qdf}, stemming from the spherical condition in eq.\ \eqref{eq::9d-sphere}. The varying density of parameter points in  Figure \ref{fig::cbdl33vsl13qdf} suggests that in the QDF scenario, top-flavoured DM is favoured over up-flavoured DM. The symmetry of  Figure \ref{fig::cbdl23vsl13qdf} with respect to interchanging the axes indicates that the constraints do not distinguish between up- and charm-flavour. This was to be expected, as the bounds from direct detection experiments are not relevant for our case of Majorana DM, and the constraints from LHC searches are already accounted for by our choice of mediator and DM masses. Interestingly, the point density in Figure \ref{fig::cbdl23vsl13qdf} is not only high close to the axes but also on the diagonal, indicating that the $D^0-\bar D^0$ mixing constraints are weaker in the QDF scenario. Indeed, the near degeneracy in the $\chi_i$ masses requires a close-to-flavour-universal coupling matrix $\lambda$, which in turn suppresses NP contributions to $D^0-\bar D^0$ mixing.

	\subsection{What's your flavour? Tell me, what's your flavour?}
	
	The discussion in the previous subsection already contains some aspects related to the flavour of the lightest DM particle $\chi_3$. To be more quantitative, we first define our notion of flavour through the following condition: We call the DM particle $\chi_3$ to be $i$-flavoured if 
	\begin{equation}
	|\tilde{\lambda}_{i3}|>|\tilde{\lambda}_{j3}|\,,
	\end{equation}
	with $i \neq j$ and $i,j \in \{u,c,t\}$. In other words, the particle $\chi_3$ has the flavour $i$ if it mainly interacts with the up-type quark of flavour $i$. With this definition at hand, we now summarise our insights on which DM flavour is favoured by the experimental data.
	
	In order to provide quantitative numerical results for the DM flavour analysis we define 
	\begin{align}
	n_i &= \frac{N_i}{N}\,,
	\end{align}
	where $N$ is the total number of allowed parameter points generated in our randomised scan, and $N_i$ the number of allowed points with flavour $i$ for a given constraint. The triple $	n_\text{constr.} = \{n_u, \,n_c, \,n_t\}$ then lists the percentage of viable parameter points with up-, charm- or top-flavoured DM, after imposing that constraint. We stress that our random generation of points prior to the application of the experimental constraints does not favour a specific flavour. The results of this analysis for  $m_\phi=1400\,\text{GeV}$ and several choices for $m_\chi$ are shown in Table \ref{tab::flavour}. Since in the QDF scenario the dependence on $m_\chi$ was found to be much weaker than in the SFF scenario, in {this} case we restrict ourselves to just one value for $m_\chi$.	
		\begin{table}[!t]
	\centering
	\begin{tabular}{cccccc} \toprule
    \multicolumn{6}{c}{\textbf{Flavour of DM}} \\ \bottomrule \toprule
    scenario & $m_\chi\,/\,\mathrm{GeV}$& $n_\text{direct}\, /\,\%$& $n_\text{mixing}\,/\,\%$& $n_\text{relic}\,/\,\%$& $n_\text{combined}\,/\,\%$\\\midrule
   \multirow{4}{*}{SFF} &300 & \multirow{4}{*}{$\{33, \,33, \,34\}$} & \multirow{4}{*}{$\{2, \,4, \,94\}$} & $\{5, \,4, \,91\}$& $\{0, \,0, \,100\}$\\
    &350 & & & $\{23, \,19, \,58\}$& $\{0, \,1, \,99\}$\\
    &400 & & & $\{36, \,33, \,31\}$& $\{1, \,2, \,97\}$\\
    &450 & & & $\{39, \,36, \,25\}$& $\{2, \,3, \,95\}$\\\midrule
    QDF & 350 & $\{33, \,33, \,34\}$& $\{22, \,24, \,54\}$& $\{34, \,34, \,32\}$& $\{23, \,23, \,54\}$\\\bottomrule
	\end{tabular}
	\caption{Numerical results of the flavour analysis for $m_\phi = 1400 \, \mathrm{GeV}$. The constraints from direct detection experiments and $D$ mixing do not exhibit a significant dependence on $m_\chi$, so that in the SFF scenario only one numerical result is shown that applies to all four $m_\chi$ values.}
	\label{tab::flavour}
	\end{table}

	The constraints on the parameter space from direct detection experiments were found to be marginal in Section \ref{sec:dd}, and hence it is not surprising that this constraint has no relevant impact on the preferred flavour of DM neither in the SFF nor in the QDF scenario, quantified by $n_\text{direct}$. The situation is different for 
 the constraints from $D^0-\bar{D}^0$ mixing, as the latter amplitude is sensitive to the coupling of the new particles $\chi_i$ and $\phi$ to up and charm quarks. This especially holds true for the SFF scenario, where the mass splitting  between the different DM flavours requires the couplings of $\chi_3$ to be dominant, and $\chi_3$ having up- or charm-flavour is thus strongly disfavoured. In the QDF scenario the couplings of the heavier flavours $\chi_{1,2}$ are more relevant so that the tendency towards top-flavoured DM implied by $D^0-\bar D^0$ mixing is weaker in this case.
 The relic abundance constraints are blind towards the flavour of the DM particle in the QDF scenario, as the initial-state flavours are summed over in eqs.\@ \eqref{eq::amplit}--\eqref{eq::amplitu}, rendering all nine couplings $|\tilde\lambda_{ij}|$ relevant. This sum is omitted in the SFF scenario, since only the lightest flavour is present at freeze-out, so that only the third column of $\tilde\lambda$ is constrained. Hence, we encounter strong implications for the flavour of the DM particle, as discussed in detail in the previous subsection. One of them is the aforementioned interplay between the velocity suppression of annihilation processes into final states with up- or charm-flavour and the enhancement of the annihilation cross section by small mediator masses. Consequently, low mediator masses require the DM particle to be up- or charm-flavoured in order to compensate for the enhancement through the velocity suppression of the annihilation cross section. As this cross section also grows with the DM mass, we find that for small values of $m_\chi$ top-flavoured DM is required in order to compensate this additional suppresion, as there is no velocity suppresion for annihilations into a top--antitop pair. For growing values of $m_\chi$ up- and charm-flavoured DM become more viable, as the total annihilation cross section is further enhanced and even velocity suppressed annihilations can yield the correct annihilation rate.

Combining the constraints discussed above leads to the distribution of possible DM flavours shown in the last column of Table \ref{tab::flavour}.
	We observe that up- and charm-flavoured DM is largely excluded in the SFF scenario. Only for large DM mass parameters $m_\chi \GtrSim 350\,\mathrm{GeV}$ a tiny part of the parameter space allows for up- or charm-flavoured DM as the relic density constraint on these flavours relaxes here for the reasons explained above. While having a much weaker dependence on the mass choice, the QDF scenario shows a similar behaviour. Here, top-flavoured DM again is favoured, however the DM particle can also have up- or charm-flavour.

\section{Direct CP Violation in Charm Decays}

A central aspect of DMFV models is the structure of the coupling matrix $\lambda$ and its implications on flavour- and CP-violating observables. While in DMFV models with Dirac DM flavour-violating interactions were found to be strongly suppressed by the stringent constraints from neutral meson mixing observables \cite{dmfv,tfdm,Jubb:2017rhm,ldm}, we have seen in Section \ref{sec::flavour} that the Majorana nature of DM implies the presence of additional crossed box diagrams that partially cancel the contributions from the standard box diagrams present in both the Dirac and Majorana case. As a consequence, flavour- and CP-violating interactions are less constrained in the Majorana DM model. The conclusion that other new flavour- and CP-violating effects are suppressed therefore no longer holds in the Majorana model.

 As the new particles in our model couple to up-type quarks, flavour- and CP-violating $D$ meson decays are expected to receive relevant NP contributions. Of particular interest are the  CP asymmetries in $D^0\to K^+K^-$ and $D^0\to \pi^+\pi^-$ decays, whose difference $\Delta A_{CP}$ was measured by the LHCb collaboration, leading to the discovery of CP violation in charm decays \cite{acp}. Notably, the measured value of $\Delta A_{CP}$ is significantly larger than its SM expectation. While the latter is plagued by large hadronic uncertainties, this potential discrepancy raises the need for a possible NP explanation. Thus, in this section we discuss if our model is capable of giving rise to a large $\Delta A_{CP}$.

\label{sec::CPV}
\subsection{Theoretical Approach}

The LHCb collaboration measured the difference  \cite{acp}
\begin{equation}
\Delta A^\text{dir}_{CP,\,\text{LHCb}} = (-0.157 \pm 0.029) \%\,,
\end{equation}
between the time-integrated direct CP asymmetries in $D^0 \rightarrow K^+ K^-$ and $D^0 \rightarrow \pi^+ \pi^-$ decays. The naive SM expectation for this asymmetry can be expressed parametrically as 
\begin{equation}
\Delta A^\text{dir}_{CP,\,\text{SM}} \sim \mathcal{O}((\alpha_s/\pi) (V_{ub}V^*_{cb})/(V_{us}V^*_{cs})) \sim 10^{-4}\,,
\end{equation} 
which is an order of magnitude below the experimental value. A more elaborate SM prediction based on QCD light-cone sum rules finds \cite{acptheo}
\begin{equation}
\Delta A^\text{dir}_{CP,\,\text{SM}} = (0.02 \pm 0.003)\% \,,
\end{equation} 
with a deviation from the data of $4.7\sigma$. This discrepancy suggests that the LHCb result might be a hint at NP. Note, however, that the possibility of a significantly larger $\Delta A^\text{dir}_{CP,\,\text{SM}}$ has been argued for in the literature 
\cite{Brod:2011re,Grossman:2019xcj}.

In investigating the possible size of $\Delta A^\text{dir}_{CP}$ in our model, we  follow the approach in \cite{altman} where the naive QCD factorization results for the relevant hadronic matrix elements have been used. For the final state with $K^+K^-$, for example, it takes the following form
\begin{align}
\bra{K^+K^-}(\bar{u}\,\Gamma_1 \,s)(\bar{s}\,\Gamma_2 \,c) \ket{D^0} &\approx  \bra{K^+}(\bar{u}\,\Gamma_1 \,s)\ket{0}\bra{K^-}(\bar{s}\,\Gamma_2 \,c) \ket{D^0}\,.
\end{align} 
While suffering from large $1/m_c$ corrections, this ansatz enables the calculation of the NP and SM contribution to the CP asymmetry in an effective field theory approach. From \cite{altman} we adopt the expression 
\begin{equation}
\Delta A^\text{dir}_{CP} = A^d_{K^+K^-} - A^d_{\pi^+\pi^-}\,,
\end{equation} 
where the direct CP asymmetry for the final state $f$ is given by 
\begin{equation}
A^d_f = 2 \,r_f \sin\delta_f \sin \phi_f\,,
\end{equation}
under the assumption that $r_f$ is small. Here,  $\delta_f$ and  $\phi_f$ are the differences of strong and weak phases, respectively, of the two interfering decay amplitudes, and $r_f$ is their relative magnitude.
 The general expression of $r_f$ and $\phi_f$ as an expansion of Wilson coefficients for the relevant set of $\Delta F=1$ operators reads \cite{altman}
\begin{align}
r_f e^{i \phi_f} \approx & \left(C_1^{(1) p} +
\frac{C^{(1) p}_2}{N_c} \right)^{-1} \left( \frac{(C_2^{(1)
    p})_\mathrm{NP}}{N_c} + C_4^{(1)} + \frac{C_3^{(1)}}{N_c} -
\frac{C_{10}^{(1)}}{2} - \frac{C_9^{(1)}}{2 N_c} -
\frac{3\alpha_s}{4\pi} \frac{N_c^2-1}{N_c^2} C_{8g}^{(1)}
\right. \nonumber \\
& + \gamma_f \left( C_6^{(1)} + \frac{C_5^{(1)}}{N_c} -
\frac{C_8^{(1)}}{2} - \frac{C_7^{(1)}}{2 N_c} - \frac{\alpha_s}{4\pi}
\frac{N_c^2-1}{N_c^2} C_{8g}^{(1)} \right) + ( C_i^{(1)} \leftrightarrow 
\tilde C_i^{(1)}) \Bigg)\,,
\label{eq::cpvrf}
\end{align}
	where $N_c=3$ is the number of colours and $\gamma_f$ are the chirality factors of the final states $f$, which are approximately 
	\begin{equation}
\gamma_K \approx \frac{2m_K^2}{m_c m_s}\,,	\qquad \gamma_\pi \approx \frac{2m_\pi^2}{m_c(m_d+m_u)}\,.
\end{equation} 
Note that we have dropped the Wilson coefficients of scalar operators in eq.\@ \eqref{eq::cpvrf}, as they are absent in our model and their RG running is decoupled from the $\Delta F=1$ operators included above. A complete list of the operators contributing to the CP asymmetry can be found in \cite{altman}. 

\begin{figure}[b]
	\centering
	\includegraphics[width=0.5\textwidth]{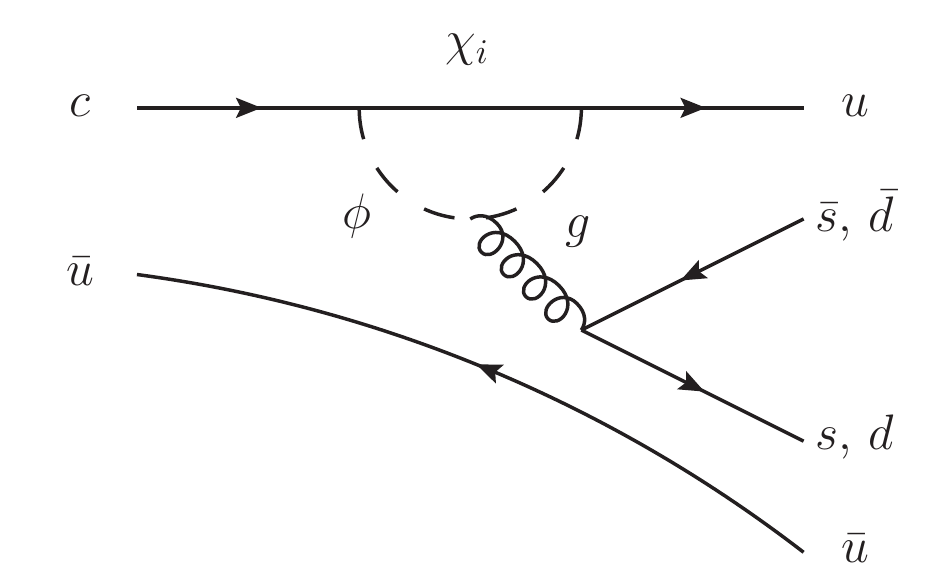}
	\caption{Penguin  diagram contributing to $D^0 \rightarrow K^+K^-$ and $D^0 \rightarrow \pi^+\pi^-$.}
	\label{fig::cpv penguin}
	\end{figure} 
	In our model, the only {sizeable} NP contribution to $\Delta A^\text{dir}_{CP}$ arises through gluon penguins\footnote{Similar contributions from EW penguins are suppressed by a colour factor $1/N_c$ as well as the small hypercharge gauge coupling.} as shown in Figure \ref{fig::cpv penguin}. We determine the following relevant NP contributions
	\begin{align}
	\nonumber
	\tilde{C}_6^{(1)} &= \frac{\alpha_s}{4\pi} \sum_i \tilde{\lambda}_{ui}\tilde{\lambda}_{ci}^*\, \frac{1}{8 m_\phi ^2}\, u(x_i)\,,\\
	\nonumber
	\tilde{C}_3^{(1)} &= \tilde{C}_5^{(1)} = -\frac{1}{N_c}\tilde{C}_4^{(1)} = -\frac{1}{N_c}\tilde{C}_6^{(1)}\,,\\
	\tilde{C}_{8g}^{(1)} &= \sum_i \tilde{\lambda}_{ui}\tilde{\lambda}_{ci}^*\, \frac{1}{4 m_\phi ^2}\, v(x_i)\,,
	\label{eq::cpvwilson}
	\end{align}
	with the loop functions 
	\begin{align}
	\nonumber 
	u(x) &= -\frac{2-7x+11x^2}{36(1-x)^3}-\frac{x^3}{6(1-x)^4}\log(x)\,,\\
	v(x) &= \frac{1-5x-2x^2}{24(1-x)^3}-\frac{x^2}{4(1-x)^4}\log(x)\,,
	\end{align}
and $x=m_\chi^2/m_\phi^2$ \cite{altman}. The Wilson coefficient $C_1^{(1)p}$ is generated at the electroweak scale, where the $W$ boson is integrated out, and reads
\begin{equation}
C_1^{(1)p}=\lambda_p\frac{G_F}{\sqrt{2}}\,,
\end{equation}
with $\lambda_p = V_{cp}V^*_{up}$ and $p=s$ for $f=K$ or $p=d$ for $f=\pi$, respectively. Following \cite{altman, Brod:2011re, strongphase2} we assume $\mathcal{O}(1)$ strong phase differences.

For the numerical analysis we use leading-order renormalization group running to evolve the Wilson coefficients from the NP scale down to the meson scale $\mu \approx m_D$. The anomalous dimensions are adopted from \cite{altman}, and for the running of the quark masses and the strong coupling $\alpha_s$ we again use the \texttt{RunDec} package \cite{rundec}. The values for the CKM elements are obtained from the  UTfit website \cite{utfit}. Due to the large uncertainties stemming from the naive factorization approach we follow \cite{altman, Brod:2011re, strongphase2} and allow for an enhancement factor of two for the ratio $r_f$.

\subsection{Results}

In order to estimate the size of the CP asymmetry $\Delta A^\text{dir}_{CP}$ generated in our model, we use parameter points  that fulfil the constraints discussed in the previous sections at the $2\sigma$ level. We then determine the range of possible $\Delta A^\text{dir}_{CP}$ values spanned by these points, both assuming no enhancement relative to the naive factorization result, and allowing for a factor of two enhancement.

	\begin{figure}[!t]
		\centering
		\begin{subfigure}[t]{0.49\textwidth}
		\includegraphics[width=\textwidth]{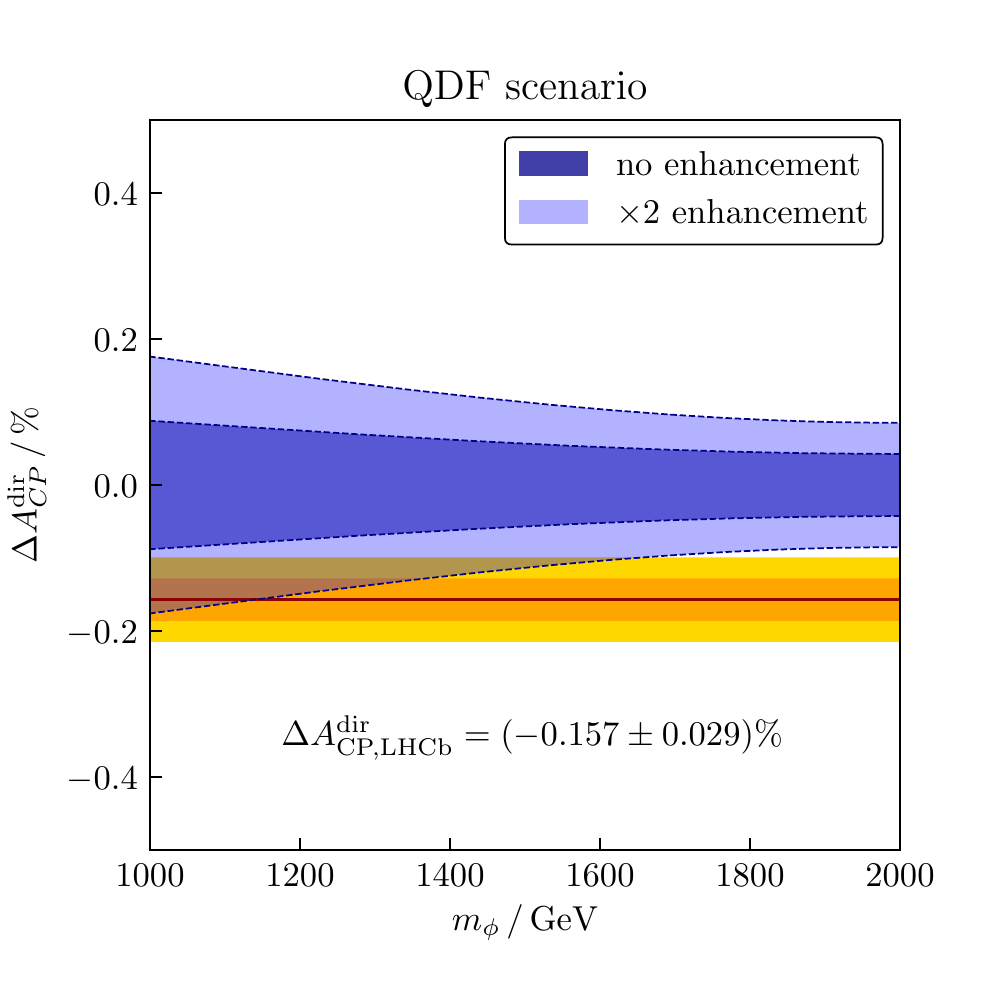}
		\caption{QDF scenario, $m_\chi = 350 \,\mathrm{GeV}$}
		\label{fig::cpvqdf}
		\end{subfigure}
		\hfill
		\begin{subfigure}[t]{0.49\textwidth}
		\includegraphics[width=\textwidth]{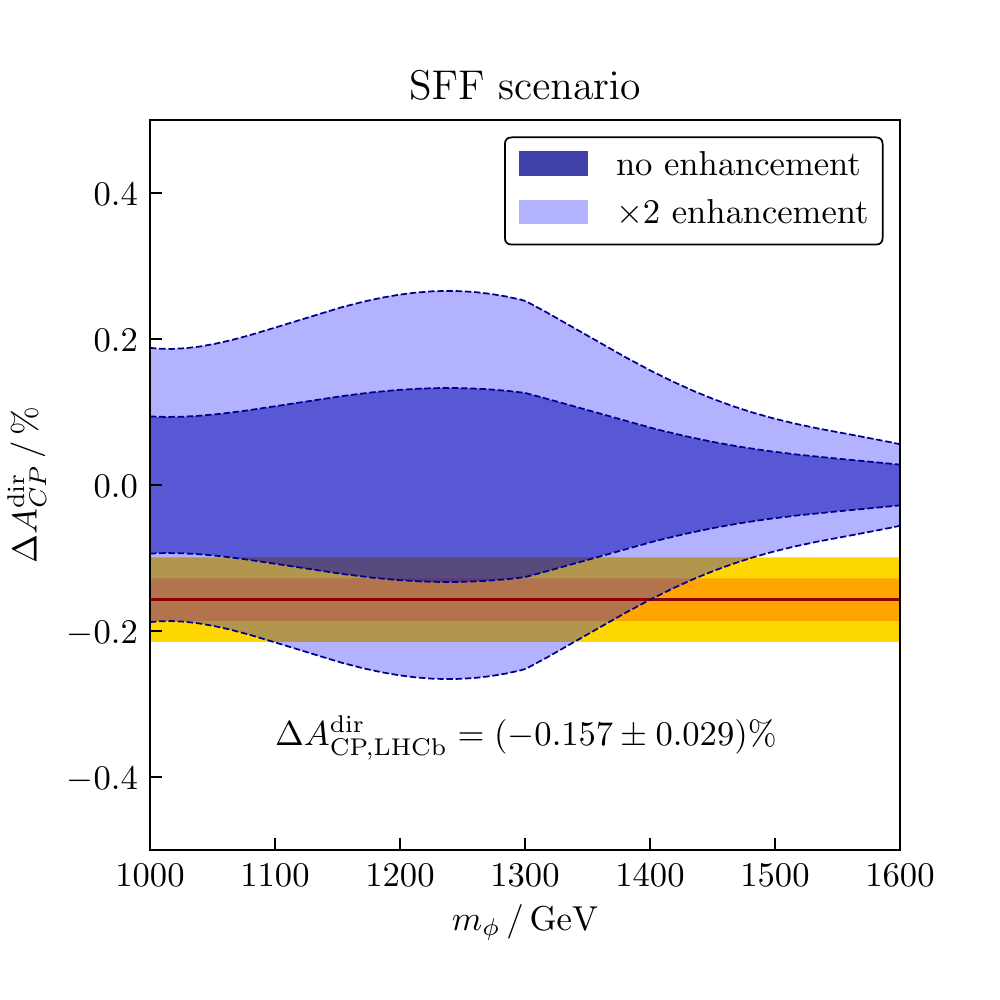}
		\caption{SFF scenario, $m_\chi = 350 \,\mathrm{GeV}$}
		\label{fig::cpvsff}
		\end{subfigure}
	\caption{$\Delta A^\text{dir}_{CP}$ in dependence of $m_\phi$ in the two freeze-out scenarios. The blue contours correspond to the ranges covered by our model, as discussed in the text. The red, orange and yellow bands display the  LHCb measurement with its $1\sigma$ and $2\sigma$ uncertainty bands.} 
	\label{fig::cpv}	
	\end{figure}

The results for the two freeze-out scenarios are gathered in Figure \ref{fig::cpv}.
	In the QDF scenario in Figure \ref{fig::cpvqdf} the $1/m_\phi^2$ suppression of the Wilson coefficients listed in eq.\@ \eqref{eq::cpvwilson} determines the mediator mass dependence of the CP asymmetry $\Delta A^\text{dir}_{CP}$. In the QCD factorization limit, values for $\Delta A^\text{dir}_{CP}\simeq  -0.1\%$  can only be reached for very low mediator masses $m_\phi < 1\,\text{TeV}$, which are excluded by the LHC limits. However, allowing for a factor of two enhancement of $r_f$, the experimental $2\sigma$ region can be reached for mediator masses $m_\phi \LessSim 1600 \, \mathrm{GeV}$.
	As the allowed region grows with decreasing mediator masses $m_\phi$, it  crosses the $1\sigma$ band for $m_\phi \LessSim 1400 \, \mathrm{GeV}$. For smaller masses $m_\phi \LessSim 1200 \, \mathrm{GeV}$, even the central value of the LHCb measurement can be reached. We conclude that the QDF scenario is capable of explaining the large value of $\Delta A^\text{dir}_{CP}$ measured by the LHCb collaboration provided $r_f$ is enhanced relative to its naive QCD factorisation prediction.

	In Figure \ref{fig::cpvsff} we show the results for the SFF scenario. In contrast to the QDF scenario, here $\Delta A^\text{dir}_{CP}$ increases for an increasing mediator mass $m_\phi$ up to a value of $m_\phi \simeq 1300\,\mathrm{GeV}$ and then starts to decrease. This quite counter-intuitive result can be understood well with our results from Section \ref{sec::combined} and Figure \ref{fig::cbdl33vsl13sff} in particular. We had seen there that for a growing mediator mass larger couplings to up- and charm-quarks are allowed, i.e.\@ increasing mediator masses lead to increasing values of at least one of the relevant couplings $|\tilde{\lambda}_{u3}|$ and $|\tilde{\lambda}_{c3}|$. Hence, $\Delta A^\text{dir}_{CP}$ also grows with an increasing value for $m_\phi$ in the SFF scenario. As we had limited the maximum value of the couplings $|\tilde{\lambda}_{i3}|$ and as the relic abundance constraint forces the DM particle to be top-flavoured for large mediator masses, $\Delta A^\text{dir}_{CP}$ encounters a  strong suppression for $m_\phi \GtrSim 1300\,\mathrm{GeV}$. In the SFF scenario the maximal values of $\Delta A^\text{dir}_{CP}$ are larger than in the QDF scenario, as the aforementioned velocity suppression of the annihilation cross section generally allows for larger couplings. Without the factor two enhancement, the experimental $2\sigma$ band is reached between $1100\, \mathrm{GeV} \LessSim m_\phi \LessSim 1350\, \mathrm{GeV}$. 
For values $1200\, \mathrm{GeV} \LessSim m_\phi \LessSim 1300\, \mathrm{GeV}$ the allowed values for $\Delta A^\text{dir}_{CP}$ even reach into the $1\sigma$ band without an enhancement. Including the enhancement factor here yields values for $\Delta A^\text{dir}_{CP}$, such that the central value of the LHCb measurement can be reproduced in the region $m_\phi \LessSim 1400\, \mathrm{GeV}$. Thus, we conclude that the SFF scenario is particularly capable of explaining the large value of $\Delta A^\text{dir}_{CP}$ measured by the LHCb collaboration.

\section{Summary and Outlook}

In this paper we studied a simplified model of flavoured DM 
within the DMFV framework \cite{dmfv} in which the DM relic is formed by the lightest flavour of a Majorana fermion flavour triplet $\chi$ coupling to SM up-type quarks $u_R$ via a coloured scalar mediator $\phi$. The $3\times 3 $ coupling matrix 
 $\lambda$ mediating this interaction constitutes the only new source of flavour and CP violation beyond the SM Yukawa couplings in the DMFV framework. In contrast to the case of Dirac DMFV models investigated in \cite{dmfv,tfdm,Jubb:2017rhm,ldm,Chen:2015jkt}, in the Majorana scenario the underlying flavour symmetry in the dark sector is $O(3)_\chi$. As a consequence, the coupling matrix $\lambda$ contains more physical parameters.
 
 In order to constrain the structure of $\lambda$ and determine the viable parameter space of the model, we investigated the constraints from LHC searches for new particles, $D^0-\bar D^0$ mixing observables, cosmological constraints on the DM relic density, and limits from DM direct detection experiments. Our main findings can be summarised as follows:
 \begin{itemize}
 \item
 The strongest limits from LHC searches on the model are obtained from searches for SUSY squarks, giving rise to the final states $\text{tops}+\slashed{E}_T$ and $jj+\slashed{E}_T$. Recasting the limits obtained by CMS, using $136\,\text{fb}^{-1}$ of run-2 data, we determined the bounds on the masses $m_\chi$ and $m_\phi$, depending on the new coupling parameters.
 We found that the LHC phenomenology is strongly affected by the Majorana nature of $\chi$, as an additional same-sign $\phi\phi$ production channel arises in this case.
 The existing experimental cross section limits can be fulfilled by choosing $0 \leq D_3 \leq 1.5$ and $0 \leq D_1, \, D_2 \leq 0.5$ for mediator masses $m_\phi \GtrSim 1\, \mathrm{TeV}$. While this choice makes a case for top-flavoured DM, it is also possible to fulfil the LHC constraints with larger $D_1$ and $D_2$ but small $m_\chi$ due to the suppression of the same-sign contributions.
 \item
 The Majorana nature of $\chi$, inducing same-sign $\phi\phi$ pair production, gives rise to the new final-state signature of two positively charged top quarks in association with missing transverse energy, $tt+\slashed{E}_T$. In 14\,TeV $pp$ collisions,
we predicted production cross sections of this final state in the multi-fb regime, making it a promising smoking-gun signature for future LHC runs.
 \item
The measurements of $D^0-\bar{D^0}$ mixing observables require the combination of the NP couplings to up and charm quarks to be small. Choosing $0 \leq D_1,D_2 \leq 0.5$ fulfils the constraints while allowing for large mixing angles $\theta_{ij}$ and a freely chosen coupling $D_3$. Relative to the case of Dirac DM,        the Majorana nature of $\chi$ generates an additional mixing diagram with crossed fermion lines. The latter interferes destructively with the contribution from standard box diagrams, leading to relaxed constraints on the flavour- and CP-violating couplings relative to the Dirac model.
\item
The dynamics of the thermal freeze-out of DM depends on the mass hierarchy among the dark flavours $\chi$. Following the studies of flavoured Dirac DM, we investigated two limiting cases: the QDF scenario in which the dark flavours are quasi-degenerate and therefore all three states are present at freeze-out, and the SFF scenario in which the heavier flavours are split significantly from the lightest and thus only the latter contributes to the effective annihilation cross section.
 The relic abundance constraints impose a spherical condition on the elements of the matrix $\lambda$. As we restricted the couplings $D_i$  to lie within the range $\left[0,2\right]$, this poses a lower bound on the DM mass $m_\chi$. In the SFF scenario we encountered a strong $m_\chi$ dependence, a velocity suppression of the annihilation cross section as well as an additional upper bound on the DM mass $m_\chi$.
\item
The constraints from DM direct detection experiments are significantly weaker for Majorana DM than for the Dirac case, as the dominant spin-independent contributions to the DM--nucleus scattering cross section are absent. Consequently, the constraints on the parameter space are rather mild and mainly dominated by contributions from spin-dependent scattering. 
\end{itemize}

In our subsequent combined analysis we investigated the interplay of these constraints
in limiting the viable parameter space for the coupling matrix  $\lambda$. 
In both freeze-out scenarios the interplay between the relic density and $D^0-\bar{D}^0$ mixing constraints mainly determines the allowed regions of parameter space. 
In the QDF scenario the former constraint reduces to a spherical condition on the coupling parameters, without a qualitative dependence on the choice of the masses. 
The SFF condition on the other hand only allows for sharp bands in each plane $|\tilde{\lambda}_{i3}| - |\tilde{\lambda}_{j3}|$, as the couplings of the third DM generation $\chi_3$ are dominant in this scenario. We further identified a strong dependence on the DM mass $m_\chi$ implied by the relic abundance constraint.

We then determined the flavour of the lightest DM field $\chi_3$, i.\,e.\ the quark to which it couples mostly. In the SFF scenario up and charm flavour are forbidden for $m_\chi \LessSim 350\, \mathrm{GeV}$, and $\chi_3$ carries top flavour in most of the allowed parameter space. While the QDF scenario  favours top-flavoured DM, a significant part of the allowed parameter space still corresponds to up and charm flavour.

In the last part of this paper we investigated the possible amount of CP violation in charm decays induced in our model. The CP asymmetry $\Delta A^\text{dir}_{CP}$ in $D^0\to K^+K^-$ and $D^0 \to \pi^+\pi^-$ decays has been measured by the LHCb collaboration with a surprisingly large value, making a  NP contribution plausible. 
Both the QDF and SFF scenarios are capable of significantly enhancing $\Delta A^\text{dir}_{CP}$ w.\@ r.\@ t.\@ its naive SM expectation. While in the QDF scenario a modest deviation from the naive QCD factorization limit is also required to bring our model in agreement with the data, in the SFF scenario values for $\Delta A^\text{dir}_{CP}$ as large as the experimental $1\sigma$ band allows for can be reached even in the strict QCD factorization limit.

In conclusion, the DMFV framework provides an elegant connection of the DM problem to one of the least understood aspects of the SM: flavour. Introducing the DM flavour triplet as Majorana fermions, instead of Dirac fermions as done in the previous literature, turns out to have several phenomenological advantages. On the one hand, the stringent constraints from $D^0-\bar D^0$ mixing and DM direct detection experiments can be softened considerably, opening up additional viable parameter space. Due to the velocity suppression coming from the additional Majorana-specific $u$-channel annihilation diagram, this also holds true for the relic density constraints in the SFF scenario. On the other hand, the Majorana nature of DM leads to the prediction of the rather striking signature of a pair of same-sign top quarks in association with missing transverse energy at the LHC. It can also provide a NP origin for large values of the CP asymmetry $\Delta A^\text{dir}_{CP}$ in charm decays, without the need for large hadronic enhancement effects. Upcoming experimental measurements will hence be able to shed light on the nature of DM and its flavour.

\paragraph*{Acknowledgements} 
We thank the referee who reviewed the first version of our paper and pointed out the missing $u$-channel diagram for the DM annihilation cross section. We further thank
Jan Heisig,
Simon Kast,
Michael Kr\"amer,
Kirtimaan Mohan,
Margarete M\"uhlleitner,
and Mustafa Tabet
for useful discussions. 
This work is supported by the Deutsche Forschungsgemeinschaft (DFG, German Research Foundation) under grant 396021762 -- TRR 257. 
H.A.\@ acknowledges the scholarship and support he receives from the Avicenna-Studienwerk e.V., and the support of
the doctoral school ``Karlsruhe School of Elementary and Astroparticle Physics: Science and Technology (KSETA)''.

\appendix
\section{Partial Wave Expansion Coefficients}
\label{app::partialwave}
For the partial wave expansion of eq.\@ \eqref{eq::cxanni} we find
\begin{align}
\nonumber
a &=\sum_{ijkl}\frac{3 \sqrt{-2 m_k^2 \left(m_l^2+4 m_{\chi}^2\right)+m_k^4+\left(m_l^2-4 m_{\chi}^2\right)^2}}{512 \pi  m_{\chi}^4 \left(m_k^2+m_l^2-2 \left(m_{\chi}^2+m_{\phi}^2\right)\right)^2}\\
\nonumber
   &\times \bigg\{8 m_{\chi}^2 \left(m_k^2+m_l^2\right)
   \text{Re}\left(c^{tu}_{ijkl}\right)-\left(m_k^2-m_l^2\right)^2
   \left(c^u_{ijkl}+c^t_{ijkl}\right)\\
   &-16 m_{\chi}^4 \left(2
   \text{Re}\left(c^{tu}_{ijkl}\right)-c^u_{ijkl}-c^t_{ijkl}\right)\bigg\}\,,\\
	\nonumber \\
\nonumber
b &=\sum_{ijkl}\frac{\sqrt{-2 m_k^2 \left(m_l^2+4 m_{\chi}^2\right)+m_k^4+\left(m_l^2-4 m_{\chi}^2\right)^2}}{4096 \pi  m_{\chi}^4 \left(m_k^2+m_l^2-2 \left(m_{\chi}^2+m_{\phi}^2\right)\right)^4}\\
\nonumber
   &\times\bigg\{-4 \left(-2 m_k^2 \left(m_l^4 \left((c^t_{ijkl}+c^u_{ijkl}) m_{\phi}^2-m_{\chi}^2
   \left(9 (c^t_{ijkl}+c^u_{ijkl})+c^{tu}_{ijkl}\right)\right)\right.\right.\\  
\nonumber 
   &\left.\left. -2 m_l^2 \left(m_{\chi}^4 (c^t_{ijkl}+c^u_{ijkl}-16 c^{tu}_{ijkl})-2 (c^t_{ijkl}+c^u_{ijkl})
   m_{\chi}^2 m_{\phi}^2+(c^t_{ijkl}+c^u_{ijkl}) m_{\phi}^4\right)\right.\right.\\
\nonumber
   &\left.\left. -8 \left(-\left(m_{\chi}^6 (c^t_{ijkl}+c^u_{ijkl}-12
   c^{tu}_{ijkl})\right)+8 (c^t_{ijkl}+c^u_{ijkl}) m_{\chi}^4 m_{\phi}^2+(c^t_{ijkl}+c^u_{ijkl}) m_{\chi}^2 m_{\phi}^4\right)\right. \right.\\
\nonumber
  &\left.\left. +3(c^t_{ijkl}+c^u_{ijkl}) m_l^6\right)-2 m_k^6 \left(m_{\chi}^2 \left(c^t_{ijkl}+c^u_{ijkl}+c^{tu}_{ijkl}\right)+(c^t_{ijkl}+c^u_{ijkl}) \left(3
   m_l^2-m_{\phi}^2\right)\right)\right.\\
\nonumber
   &\left. +2 m_k^4 \left(m_l^2 \left(m_{\chi}^2 \left(9
   (c^t_{ijkl}+c^u_{ijkl})+c^{tu}_{ijkl}\right)-(c^t_{ijkl}+c^u_{ijkl}) m_{\phi}^2\right)+5 (c^t_{ijkl}+c^u_{ijkl}) m_l^4\right.\right.\\
\nonumber
   &\left.\left. -(c^t_{ijkl}+c^u_{ijkl}) \left(9 m_{\chi}^4+14
   m_{\chi}^2 m_{\phi}^2+m_{\phi}^4\right)\right)+2 m_l^6 \left((c^t_{ijkl}+c^u_{ijkl}) m_{\phi}^2\left. -m_{\chi}^2 \left(c^t_{ijkl}+c^u_{ijkl}+c^{tu}_{ijkl}\right)\right)\right.\right.\\
\nonumber
   &\left.+16 m_l^2 m_{\chi}^2 \left(-m_{\chi}^4
   (c^t_{ijkl}+c^u_{ijkl}-12 c^{tu}_{ijkl})+8 (c^t_{ijkl}+c^u_{ijkl}) m_{\chi}^2 m_{\phi}^2\left.+(c^t_{ijkl}+c^u_{ijkl}) m_{\phi}^4\right)\right.\right.\\
\nonumber
   &\left.+32 m_{\chi}^4
   \left(m_{\chi}^4 (c^t_{ijkl}+c^u_{ijkl}-10 c^{tu}_{ijkl})+6 m_{\chi}^2 m_{\phi}^2 (c^t_{ijkl}+c^u_{ijkl}-2 c^{tu}_{ijkl})\right.\right.\\
\nonumber
	&\left.\left. +m_{\phi}^4
   (6 c^{tu}_{ijkl}-7 (c^t_{ijkl}+c^u_{ijkl}))\right)-2 (c^t_{ijkl}+c^u_{ijkl}) m_l^4 \left(9 m_{\chi}^4+14 m_{\chi}^2
   m_{\phi}^2+m_{\phi}^4\right)\right.\\
\nonumber
	&\left.+(c^t_{ijkl}+c^u_{ijkl}) m_l^8\right)+(c^t_{ijkl}+c^u_{ijkl}) m_k^8+9 \left(m_k^2+m_l^2-2 \left(m_{\chi}^2+m_{\phi}^2\right)\right)^2\\
\nonumber
	&\left(-16 m_{\chi}^4 (c^t_{ijkl}+c^u_{ijkl}-2 c^{tu}_{ijkl})+(c^t_{ijkl}+c^u_{ijkl})
   \left(m_k^2-m_l^2\right)^2-8 c^{tu}_{ijkl} m_{\chi}^2
   \left(m_k^2+m_l^2\right)\right)\\
\nonumber
	&+\frac{3 \left(-16 m_{\chi}^2 \left(m_k^2+m_l^2\right)+3
   \left(m_k^2-m_l^2\right)^2+16 m_{\chi}^4\right) \left(m_k^2+m_l^2-2 \left(m_{\chi}^2+m_{\phi}^2\right)\right)^2}{-2
   m_k^2 \left(m_l^2+4 m_{\chi}^2\right)+m_k^4+\left(m_l^2-4 m_{\chi}^2\right)^2}\\
\nonumber
   &\times \left(-16 m_{\chi}^4 (c^t_{ijkl}+c^u_{ijkl}-2 c^{tu}_{ijkl})+(c^t_{ijkl}+c^u_{ijkl})
   \left(m_k^2-m_l^2\right)^2\right.\\
   & -8 c^{tu}_{ijkl} m_{\chi}^2 \left(m_k^2+m_l^2\right)\Big)\bigg\}\,.
\end{align}
\section{Gluonic Wilson Coefficients}
\label{app::ddwilson}
The Wilson coefficents $f_G$, $g_G^{(1)}$ and $g_G^{(2)}$ from eq.\@ \eqref{eq::spinindepfN} read \cite{taitdd}
\begin{align}
f_{G}&=\sum_{i=u,c,t}\alpha_s |\tilde{\lambda}_{i3}|^2
m_{\chi_3}\bigg[ -12 m_i^2 m_\phi^4 m_{\chi_3}^2
(m_i^2-m_\phi^2+m_{\chi_3}^2) \Lambda_i
(m_{\chi_3}^2;m_i,m_\phi)\nonumber \\
&-(m_i-m_\phi-m_{\chi_3}) (m_i+m_\phi-m_{\chi_3})
(m_i-m_\phi+m_{\chi_3})\nonumber \\
&\times  (m_i+m_\phi+m_{\chi_3})\big\{m_i^6-3 m_i^4 (2
m_\phi^2+m_{\chi_3}^2)\nonumber \\
&+m_i^2 (3 m_\phi^4+2 m_\phi^2 m_{\chi_3}^2+3
m_{\chi_3}^4)+(m_\phi^2-m_{\chi_3}^2)^2 
(2 m_\phi^2-m_{\chi_3}^2)\big\}\bigg]\nonumber \\
&\times\frac{(m_i-m_\phi+m_{\chi_3})^{-3} (m_i+m_\phi+m_{\chi_3})^{-3}}{192 \pi  m_\phi^2 (m_i-m_\phi-m_{\chi_3})^3
	(m_i+m_\phi-m_{\chi_3})^3}\,,\\
	\nonumber \\
\frac{g^{(2)}_{G}}{m_{\chi_3}^2}&=\sum_{i=u,c,t}\alpha_s |\tilde{\lambda}_{i3}|^2
\bigg[2 m_{\chi_3}^2 \big\{10 m_{\chi_3}^4
(m_i^6-m_\phi^6)+m_{\chi_3}^8 (m_i^2-5 m_\phi^2)
-5 m_{\chi_3}^2 (m_i^2-m_\phi^2)^3\nonumber \\
&\times (m_i^2+m_\phi^2)+(m_i^2-m_\phi^2)^5 + 2 m_{\chi_3}^6 (-4
m_i^4+2 m_i^2 m_\phi^2+5 m_\phi^4)+m_{\chi_3}^{10}\big\} \nonumber \\
&\times \Lambda_i
(m_{\chi_3}^2;m_i,m_\phi)-(m_i-m_\phi-m_{\chi_3}) (m_i+m_\phi-m_{\chi_3})
(m_i-m_\phi+m_{\chi_3})\nonumber \\
&\times  (m_i+m_\phi+m_{\chi_3}) \bigg\{7 m_{\chi_3}^4
(m_i^4-m_\phi^4)-2 m_{\chi_3}^6 (m_i^2-4 m_\phi^2)-2
m_{\chi_3}^2 (m_i^2-m_\phi^2)^3\nonumber \\
&+2 (m_i^4-2 m_i^2
(m_\phi^2+m_{\chi_3}^2)+(m_\phi^2-m_{\chi_3}^2)^2)^2
\log (\frac{m}{m_\phi})-3 m_{\chi_3}^8\bigg\}\bigg]\nonumber \\
&\times\frac{(m_i-m_\phi+m_{\chi_3})^{-3} (m_i+m_\phi+m_{\chi_3})^{-3}}{48 \pi 
	m_{\chi_3}^5 (m_i-m_\phi-m_{\chi_3})^3 (m_i+m_\phi-m_{\chi_3})^3}\,, \\
		\nonumber \\
\frac{g^{(1)}_{G}}{m_{\chi_3}}&=\sum_{i=u,c,t}\alpha_s |\tilde{\lambda}_{i3}|^2
\bigg[2 m_{\chi_3}^2 (3 m_{\chi_3}^2
(m_\phi^4-m_i^4)+m_{\chi_3}^4 (5
m_i^2+m_\phi^2)+(m_i^2-m_\phi^2)^3-3 m_{\chi_3}^6)\nonumber \\
&\times  \Lambda_i(m_{\chi_3}^2;m_i,m_\phi)+2 (m_i+m_\phi-m_{\chi_3}) (m_i-m_\phi+m_{\chi_3})
(m_i+m_\phi+m_{\chi_3})\nonumber \\
&\bigg\{m_{\chi_3}^2 (m_i-m_\phi-m_{\chi_3}) (m_i^2-m_\phi^2-3
m_{\chi_3}^2)-(m_i+m_\phi-m_{\chi_3}) (m_i-m_\phi+m_{\chi_3})\nonumber \\
&\times(-m_i+m_\phi+m_{\chi_3})^2 (m_i+m_\phi+m_{\chi_3}) \log
(\frac{m_i}{m_\phi})\bigg\}\bigg]\nonumber \\
&\times	\frac{(-m_i+m_\phi+m_{\chi_3})^{-2}
	(m_i+m_\phi+m_{\chi_3})^{-2}}{192 \pi  m_{\chi_3}^4
	(m_i+m_\phi-m_{\chi_3})^2 (m_i-m_\phi+m_{\chi_3})^2}\,,
\end{align}
with
\begin{align}
\Lambda_i(m_{\chi_3}^2;m_i,m_\phi)=
\frac{\lambda_i}{m_{\chi_3}^2} \log \left(\frac{m_i^2+\lambda_i+m_\phi^2-m_{\chi_3}^2}{2 m_i
	m_\phi}\right)\nonumber \\
\lambda_i= \sqrt{m_i^4-2 m_i^2 m_\phi^2-2 m_i^2 m_{\chi_3}^2+m_\phi^4-2 m_\phi^2
	m_{\chi_3}^2+m_{\chi_3}^4}\,.
\end{align}

\bibliography{ref}
\bibliographystyle{JHEP}
\end{document}